\documentclass[portuges,english]{iopart}
\usepackage[T1]{fontenc}
\usepackage[latin9]{inputenc}
\usepackage{geometry}
\usepackage{color}
\usepackage{verbatim}
\usepackage{mathrsfs}
\usepackage{bm}
\usepackage{amssymb}
\usepackage{graphicx}
\usepackage{esint}

\providecommand{\tabularnewline}{\\}

\usepackage{iopams}
\usepackage{setstack}

\usepackage{ae,aecompl}
\usepackage{hyperref}
\usepackage{footmisc}
\newtheorem{theorem}{Theorem}[section]
\newtheorem{proposition}[theorem]{Proposition}

\makeatother

\usepackage{babel}

\begin{document}

\title{Gravitomagnetism in the Lewis cylindrical metrics}

\author{L. Filipe O. Costa$^{1,2}$, José Natário$^1$ and N. O. Santos$^3$}

\address{${^{1}}$ GAMGSD, Departamento de Matemática, Instituto Superior
Técnico, Universidade de Lisboa, 1049-001 Lisboa, Portugal}
\address{${^{2}}$ Centro de Física do Porto --- CFP, Departamento de Física e Astronomia, Universidade do Porto, 4169-007 Porto, Portugal}
\address{${^{3}}$ Sorbonne Université, UPMC Université Paris 06, LERMA, UMRS8112 CNRS, Observatoire de Paris-Meudon, 5, Place Jules Janssen, F-92195 Meudon Cedex, France}

\eads{\mailto{lfilipecosta@tecnico.ulisboa.pt}, \mailto{jnatar@math.ist.utl.pt}, \mailto{Nilton.Santos@obspm.fr}}
\vspace{10pt}

\begin{abstract}
The Lewis solutions describe the exterior gravitational field produced
by infinitely long rotating cylinders, and are useful models for global
gravitational effects. When the metric parameters are real (Weyl class), the exterior metrics
of rotating and static cylinders are locally indistinguishable, but
known to globally differ. The significance of this difference, both
in terms of physical effects (gravitomagnetism) and of the mathematical invariants that detect the rotation,
remain open problems in the literature. 
In this work we show that, by a rigid coordinate rotation, the Weyl
class metric can be put into a ``canonical'' form where the Killing
vector field $\partial_{t}$ is time-like everywhere, and which depends explicitly only on three parameters with a clear physical significance: the Komar mass and angular momentum per unit length, plus the angle deficit. 
This new form of the metric reveals that the two settings differ only at the level of the gravitomagnetic vector potential which, for
a rotating cylinder, cannot be eliminated by any global coordinate
transformation. It manifests itself in the Sagnac and gravitomagnetic
clock effects. The situation is seen to mirror the electromagnetic
field of a rotating charged cylinder, which likewise differs from
the static case only in the vector potential, responsible for the
Aharonov-Bohm effect, formally analogous to the Sagnac effect. The
geometrical distinction between the two solutions is also discussed,
and the notions of local and global staticity revisited. The matching
in canonical form to the van Stockum interior cylinder is also addressed.
\end{abstract}
\vspace{2pc}

\tableofcontents{}

\section{Introduction}

The Lewis metrics \cite{Lewis:1932} are the general stationary solution
of the vacuum Einstein field equations with cylindrical symmetry,
and are usually interpreted as describing the exterior gravitational
field produced by infinitely long rotating cylinders (for a recent
review on cylindrical systems in General Relativity, see \cite{Bronnikov:2019clf}).
They are divided into two sub-classes: the Lewis class and the Weyl
class, the latter corresponding to the case where all the metric parameters
are real. The Weyl class metrics have the same Cartan scalars as in
the special case of a static cylinder (Levi-Civita metric), and so
are locally indistinguishable \cite{SantosGRG1995}; they are known,
however, to have distinct global properties, namely in the matching
to the interior solutions (as the former, but not the latter, can
be matched to rotating interior cylinders). The physical implications
of such difference remain an unanswered question in the literature
\cite{SantosGRG1995,MacCallumSantos1998,GriffithsPodolsky2009}; they
are expected to be manifested as ``gravitomagnetism'', the gravitational
effects generated by the motion of matter (thus known due to their
many analogies with magnetism). From a mathematical point of view,
this distinction also remains an open question, namely whether it
stems from topology \cite{SantosGRG1995,MacCallumSantos1998} or geometry
\cite{Stachel:1981fg}, what are the invariants that detect the rotation,
or what is the nature of the ``transformation\textquotedblright{}
\cite{SantosGRG1995,Bonnor1980,Tipler:1974gt} that is known to relate
the Weyl class rotating and static metrics. The physical significance
of the four Lewis parameters also remains unclear \cite{GriffithsPodolsky2009};
it has been shown in \cite{MacCallumSantos1998} that only three are
independent, but an explicit form of the metric in terms of three
parameters, with a clear physical interpretation, has proved elusive.
Another open question is the rather mysterious ``force'' parallel
to the cylinder's axis found in the literature \cite{HerreraSantosGeoLewis},
which seemingly deflects test particles moving in these spacetimes
axially. In this work we address these questions.

This paper is organized as follows. In the preliminary Section \ref{sec:Preliminaries},
after briefly reviewing some relevant features of stationary spacetimes,
we discuss and formulate, in a suitable framework, the Sagnac effect,
which plays a crucial role in the context of this work. In Sec. \ref{sec:Gravitomagnetism-Levels}
we discuss, in parallel with their electromagnetic analogues, the
different levels of gravitomagnetism, corresponding to different levels
of differentiation of the ``gravitomagnetic vector potential'';
special attention is given to the gravitomagnetic clock effect ---
another important effect in this work --- which is revisited and reinterpreted
in the framework herein. In Sec. \ref{sec:The-electromagnetic-analogue:},
as a preparation for the gravitational problem, we study the electromagnetic
field produced by infinitely long rotating charged cylinders, as viewed
from both static and rotating frames, and the Aharonov-Bohm effect.
In Sec. \ref{sec:lewis metrics} we start by discussing the Lewis
metrics of the Weyl class in their usual form given in the literature,
studying the inertial and tidal fields as measured in the associated
reference frame; we also dissect (Sec. \ref{sub:The-force-parallel})
the origin of the axial \emph{coordinate} acceleration found in the
literature. Subsections \ref{sub:The-canonical-form} and \ref{sub:The-distinction-between}
contain the main results in this paper. In \ref{sub:The-canonical-form}
we show that the usual form of the Weyl class metrics is actually
written in a system of rigidly rotating coordinates; gauging such
rotation away leads to a coordinate system which is inertial at infinity
(thus fixed with respect to the ``distant stars''), the Killing
vector field $\partial_{t}$ is time-like everywhere, and the metric
depends explicitly only on three parameters: the Komar mass and angular
momentum per unit length, plus the angle deficit. We dub such form
of the metric ``canonical''. It makes transparent that the gravitational
fields of (Weyl class) rotating and static cylinders differ only in
the gravitomagnetic potential 1-form $\bm{\mathcal{A}}$ (which is
non-vanishing in the former); the observers at rest measure the same
inertial and tidal fields (Sec. \ref{sub:GEM-fields-Canonical}),
the only distinction being the global effects governed by $\bm{\mathcal{A}}$.
The situation is seen to exactly mirror the electromagnetic fields
of rotating/static charged cylinders. In Sec. \ref{sub:The-distinction-between}
this distinction is explored both on physical grounds, putting forth
(thought) physical apparatuses to reveal it (Sec. \ref{sub:Physical-distinction}),
and on geometrical grounds (Sec. \ref{sub:Mathematical-distinction}).
It turns out to be an archetype of the contrast between globally static,
and locally but non-globally static spacetimes; hence we also revisit
(Secs. \ref{sub:Local-vs-global}-\ref{sub:Holonomy}) the notions
of local and global staticity in the literature, devising equivalent
formulations that are more enlightening in this context. In Sec. \ref{sub:Matching Stockum}
we discuss the matching to the interior van Stockum cylinder. We first
establish the correspondence between the Lewis and van Stockum exterior
solutions, and, using their usual forms in the literature, obtain
the matching to the interior van Stockum solution, using the so-called
``quasi-Maxwell'' formalism. Then, in the same framework, we obtain
the matching in canonical form. Finally, in Sec. \ref{sub:The-Lewis-class},
we briefly discuss the Lewis metrics of the Lewis class, pointing
out their fundamental differences from the Weyl class in the framework
herein.

\subsection{Notation and conventions}

We use the signature $(-+++)$; $\epsilon_{\alpha\beta\gamma\delta}\equiv\sqrt{-g}[\alpha\beta\gamma\delta]$
is the 4-D Levi-Civita tensor, with the orientation $[1230]=1$ (i.e.,
in flat spacetime, $\epsilon_{1230}=1$); \textcolor{black}{Greek
letters $\alpha$, $\beta$, $\gamma$, ... denote 4D spacetime indices,
running 0-3; Roman letters $i,j,k,...$ denote spatial indices, running
1-3}. Our convention for the Riemann tensor is $R_{\ \beta\mu\nu}^{\alpha}=\Gamma_{\beta\nu,\mu}^{\alpha}-\Gamma_{\beta\mu,\nu}^{\alpha}+...$
. $\star$ denotes the Hodge dual (e.g. $\star F_{\alpha\beta}\equiv\epsilon_{\alpha\beta}^{\ \ \ \mu\nu}F_{\mu\nu}/2$,
for a 2-form $F_{\alpha\beta}=F_{[\alpha\beta]}$). The basis vector
corresponding to a coordinate $\phi$ is denoted by $\partial_{\phi}$,
and its $\alpha$-component by $\partial_{\phi}^{\alpha}\equiv\delta_{\phi}^{\alpha}$.

\section{Preliminaries\label{sec:Preliminaries}}

The line element $ds^{2}=g_{\alpha\beta}dx^{\alpha}dx^{\beta}$ of
a stationary spacetime can generically be written as 
\begin{equation}
ds^{2}=-e^{2\Phi}(dt-\mathcal{A}_{i}dx^{i})^{2}+h_{ij}dx^{i}dx^{j}\ ,\label{eq:StatMetric}
\end{equation}
where $e^{2\Phi}=-g_{00}$, $\Phi\equiv\Phi(x^{j})$, $\mathcal{A}_{i}\equiv\mathcal{A}_{i}(x^{j})=-g_{0i}/g_{00}$,
and $h_{ij}\equiv h_{ij}(x^{k})=g_{ij}+e^{2\Phi}\mathcal{A}_{i}\mathcal{A}_{j}$.
Observers whose worldlines are tangent to the timelike Killing vector
field $\partial_{t}$ are \emph{at rest} in the coordinate system
of (\ref{eq:StatMetric}); they are sometimes called ``static''
or ``laboratory'' observers. Their 4-velocity is 
\begin{equation}
u^{\alpha}\equiv u_{{\rm lab}}^{\alpha}=(-g_{00})^{-1/2}\partial_{t}^{\alpha}=e^{-\Phi}\partial_{t}^{\alpha}\equiv e^{-\Phi}\delta_{0}^{\alpha}\ .\label{eq:uLab}
\end{equation}
The quotient of the spacetime by the worldlines of the laboratory
observers yields a 3-D manifold $\Sigma$ in which $h_{ij}$ is a
Riemannian metric, called the spatial or ``orthogonal'' metric \cite{LandauLifshitz,Cattaneo1958,ManyFaces,ZonozBell1998,MenaNatario2008,Zonoz2019}.
It can be identified in spacetime with the projector orthogonal to
$u^{\alpha}$ (\emph{space projector} with respect to $u^{\alpha}$),
\begin{equation}
h_{\alpha\beta}\equiv u_{\alpha}u_{\beta}+g_{\alpha\beta}\ ,\label{eq:SpaceProjector}
\end{equation}
and yields the spatial distances between neighboring laboratory observers,
as measured through Einstein's light signaling procedure\footnote{It is not a metric induced on a hypersurface, since, in general, $u^{\alpha}$
has vorticity, and so is not hypersurface orthogonal. This is the
metric that yields the distance between fixed points in a rotating
frame, such as the terrestrial reference frame (ECEF), where it corresponds
e.g. to the distance measured by radar. It is positive definite since
$h=-ge^{-2\Phi}>0$.} \cite{LandauLifshitz}. In this work we will deal with axistationary
spacetimes, whose line element simplifies to 
\begin{equation}
ds^{2}=-e^{2\Phi}(dt-\mathcal{A}_{\phi}d\phi)^{2}+h_{ij}dx^{i}dx^{j}\ .\label{eq:AxistatMetric}
\end{equation}

\subsection{Stationary observers, angular momentum, and ZAMOs\label{sub:Stationary-observers,-orbital}}

Stationary spacetimes admit a privileged class of observers who see
an unchanging spacetime geometry in their neighborhood, dubbed ``stationary
observers'' \cite{Misner:1974qy,SemerakGRG1998}. Each of their worldlines
is tangent to a time-like Killing vector, forming congruences tangent
to so-called ``quasi-Killing vector fields'' \cite{Iyer1993} $\chi^{\beta}=\partial_{t}^{\beta}+\sum_{n}\alpha_{n}\xi_{(n)}^{\beta}$,
where the $\xi_{(n)}^{\beta}$ are spacelike Killing vectors, and
the coefficients $\alpha_{n}$ are such that $\mathcal{L}_{\chi}\alpha_{n}=0$.
Two classes of stationary observers are especially important in this
work. One are the rest or ``laboratory'' observers, defined in (\ref{eq:uLab}).
In spite of being at rest, their angular momentum is, in general,
non-zero. Take the spacetime to be axisymmetric as in (\ref{eq:AxistatMetric}),
and consider a test particle of 4-momentum $P^{\alpha}=mu^{\alpha}$
and rest mass $m$; the component of its angular momentum along the
symmetry axis is given by \cite{Misner:1974qy,SemerakGRG1998} $P_{\phi}=mu_{\phi}$.
Hence, the laboratory observers have an angular momentum per unit
mass 
\begin{equation}
u_{\phi}=u^{0}g_{0\phi}=\frac{g_{0\phi}}{\sqrt{-g_{00}}}=e^{\Phi}\mathcal{A}_{\phi}\ ,\label{eq:AngMomentumLab}
\end{equation}
which is zero \emph{iff} $g_{0\phi}=0$. Another important class of
stationary observers in axistationary spacetimes are those in circular
motion for which the angular momentum (i.e., $P_{\phi}$) vanishes
--- \emph{the zero angular momentum observers} (ZAMOs). Their 4-velocity,
$u_{{\rm ZAMO}}^{\alpha}=u_{{\rm ZAMO}}^{0}\partial_{0}^{\alpha}+u_{{\rm ZAMO}}^{\phi}\partial_{\phi}^{\alpha}$,
is such that $(u_{{\rm ZAMO}})_{\phi}=0$, i.e., they have angular
velocity 
\begin{equation}
\Omega_{{\rm ZAMO}}\equiv\frac{u_{{\rm ZAMO}}^{\phi}}{u_{{\rm ZAMO}}^{0}}=-\frac{g_{0\phi}}{g_{\phi\phi}}\ .\label{eq:OmegaZamo}
\end{equation}
Thus, $\Omega_{{\rm ZAMO}}=0$ \emph{iff} $g_{0\phi}=0$.

\subsection{Sagnac effect\label{sub:Sagnac-effect}}

A key effect in the context of this work is the Sagnac effect \cite{SagnacI,SagnacII,Laue1920,Post1967,AshtekarMagnon,Chow_et_al1985,Kajari:2009qy,Tartaglia:1998rh,Kajari:2004ms}.
It consists of the difference in arrival times of light-beams propagating
around a closed path in opposite directions. It is a measure of the
\emph{absolute rotation} of an apparatus, i.e., its rotation relative
to the ``spacetime geometry'' \cite{Misner:1974qy}. It was originally
introduced in the context of flat spacetime \cite{SagnacI,SagnacII,Laue1920,Post1967,Chow_et_al1985},
where the time difference is originated by the rotation of the apparatus
with respect to global inertial frames; but, in the presence of a
gravitational field, it arises also in apparatuses which are fixed
relative to the distant stars (i.e., to asymptotic inertial frames);
the effect is in this case assigned to ``frame-dragging''.

In stationary conditions, both effects can be read from the spacetime
metric (\ref{eq:StatMetric}), which encompasses the flat Minkowski
metric expressed in a rotating coordinate system, as well as arbitrary
stationary gravitational fields. Along a photon worldline, $ds^{2}=0$;
by (\ref{eq:StatMetric}), this yields the two solutions $dt=\mathcal{A}_{i}dx^{i}\pm e^{-\Phi}\sqrt{h_{ij}dx^{i}dx^{j}}$.
We are interested in future-oriented worldlines, defined by $k_{\alpha}\partial_{t}^{\alpha}=k_{0}<0$,
where $k^{\alpha}\equiv dx^{\alpha}/d\lambda$ is the vector tangent
to the photon's worldline. Since $k_{0}<0\Leftrightarrow dt>\mathcal{A}_{i}dx^{i}$,
such worldlines correspond to the $+$ solution for $dt$:

\[
dt=\mathcal{A}_{i}dx^{i}+e^{-\Phi}\sqrt{h_{ij}dx^{i}dx^{j}}\equiv\mathcal{A}_{i}dx^{i}+e^{-\Phi}dl\ ,
\]
where $dl\equiv\sqrt{h_{ij}dx^{i}dx^{j}}$ is the spatial distance
element. Consider photons constrained to move within a closed loop
$C$ in the space manifold $\Sigma$ (that is, the photons' worldlines
are such that their projection on the space manifold $\Sigma$ yields
a closed path $C$, see Fig. 2 of \cite{Kajari:2009qy}); for instance,
within an optical fiber loop. Using the + (-) sign to denote the anti-clockwise
(clockwise) directions, the coordinate time it takes for a full loop
is, respectively,
\[
t_{\pm}=\oint_{\pm C}dt=\oint_{C}e^{-\Phi}dl\pm\oint_{C}\mathcal{A}_{i}dx^{i}\ .
\]
Therefore, the Sagnac \emph{coordinate} time delay $\Delta t$ is
\begin{equation}
\Delta t\equiv t_{+}-t_{-}=2\oint_{C}\mathcal{A}_{i}dx^{i}=2\oint_{C}\bm{\mathcal{A}}\ ,\label{eq:SagnacDiffForm}
\end{equation}
where in the last equality we identified (see e.g. \cite{Misner:1974qy})
$\mathcal{A}_{i}dx^{i}$ with the 1-form $\bm{\mathcal{A}}\equiv\mathcal{A}_{i}\mathbf{d}x^{i}$,
where $\mathbf{d}x^{i}$ are basis 1-forms both on the spacetime manifold
and also on the space manifold $\Sigma$ (since $\{x^{i}\}$ is a
coordinate chart on the latter). In Eq. (\ref{eq:SagnacDiffForm})
$\bm{\mathcal{A}}$ is, as usual, understood as its restriction to
the curve $C$, $\bm{\mathcal{A}}|_{C}$. In what follows it will
also be useful to write this result in a different form. Consider
a 2-D submanifold $\mathcal{S}$ on $\Sigma$ with boundary $\partial\mathcal{S}\equiv C$.
Then, by the generalized Stokes theorem, 
\begin{equation}
\Delta t=2\oint_{\partial\mathcal{S}}\bm{\mathcal{A}}=2\int_{\mathcal{S}}\mathbf{d}\bm{\mathcal{A}}=2\int_{\mathcal{S}}(\partial\times\bm{\mathcal{A}})^{k}d\mathcal{S}_{k}\ ,\label{eq:SagnacStokes}
\end{equation}
where $\mathbf{d}\bm{\mathcal{A}}=\mathcal{A}_{j,i}\mathbf{d}x^{i}\wedge\mathbf{d}x^{j}=\mathcal{A}_{[j,i]}\mathbf{d}x^{i}\wedge\mathbf{d}x^{j}$
is the exterior derivative of $\bm{\mathcal{A}}$, and its restriction
to $\mathcal{S}$ is assumed above; $(\partial\times\bm{\mathcal{A}})^{k}\equiv\epsilon^{ijk}\mathcal{A}_{j,i}$
is the vector dual to $\mathcal{A}_{[j,i]}=\epsilon_{ijk}(\partial\times\bm{\mathcal{A}})^{k}/2$,
and $d\mathcal{S}_{k}\equiv\epsilon_{ijk}\mathbf{d}x^{i}\wedge\mathbf{d}x^{j}/2$
is an area element of $\mathcal{S}$ (volume form of $\mathcal{S}$
\cite{Misner:1974qy}). The latter two quantities rely on endowing
the space manifold $\Sigma$ with some metric $(g_{\Sigma})_{ij}$
(even though the integrand is metric independent), with $\epsilon_{ijk}=\sqrt{g_{\Sigma}}[ijk]$
the corresponding Levi-Civita tensor.

The proper time of the laboratory observers (\ref{eq:uLab}) is related
to the coordinate time by $dt/d\tau=u^{0}=(-g_{00})^{-1/2}$; hence,
the Sagnac time delay \emph{as measured by the local laboratory observer}
is 
\begin{eqnarray}
\Delta\tau & = & \sqrt{-g_{00}}\Delta t=e^{\Phi}\Delta t\ .\label{eq:PropertimeCoordinateTime}
\end{eqnarray}

\subsubsection{Axistationary case, circular loop around the axis\label{sub:circular-loop-around}}

\begin{figure}
\includegraphics[width=1\columnwidth]{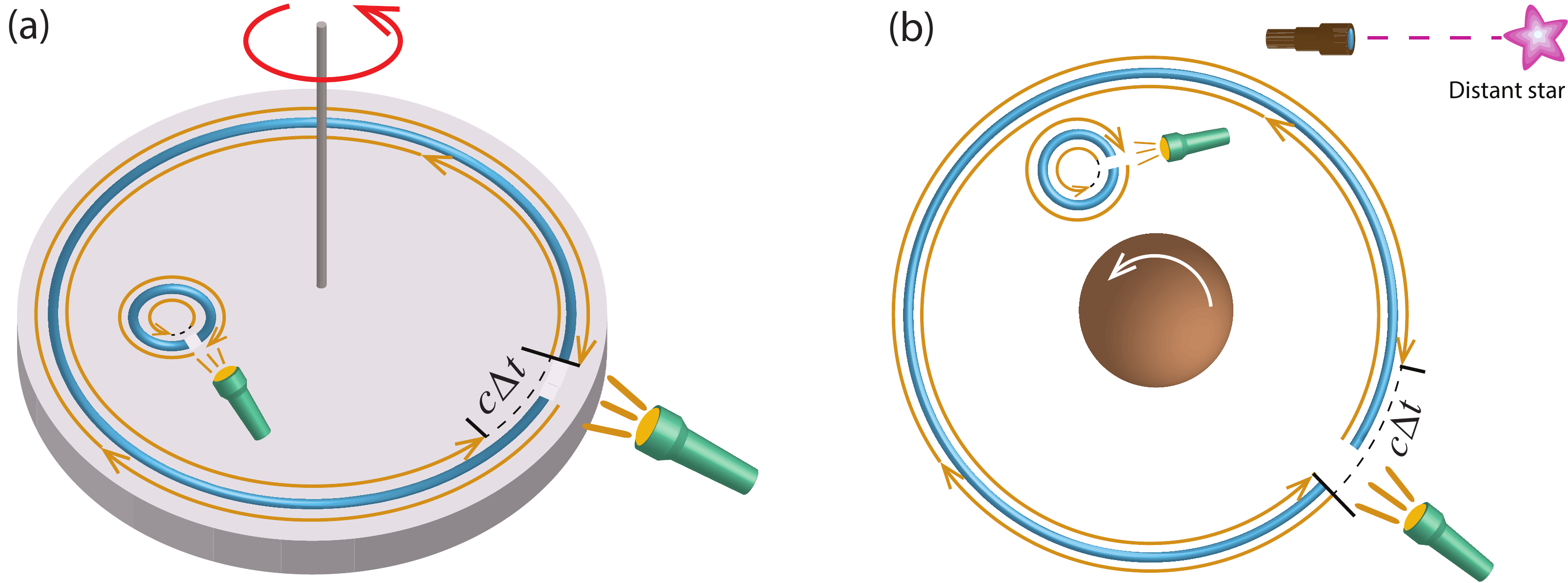}\protect\protect\protect\caption{\label{fig:Sagnac}(a) Sagnac effect in special relativity: a flashlight
sends light beams propagating in opposite directions along optical
fiber loops attached to a rotating platform; they take different times
to complete the loop, the co-rotating beam taking longer. (b) General
relativistic Sagnac effect (``frame dragging''): optical fiber loops
fixed with respect to the ``distant stars'' (i.e., to the asymptotic
inertial frame at infinity), placed around, or in the vicinity, of
a spinning object. Again, counter-propagating light beams take different
times to complete the loops. In both (a) and (b) the coordinate time
difference $\Delta t$ of arrival is twice the circulation of the
gravitomagnetic potential 1-form $\bm{\mathcal{A}}$ {[}cf. Eq. (\ref{eq:SagnacDiffForm}){]};
that amounts to the component $\mathcal{A}_{\phi}$ governing the
effect for the circular loops around the axis, and (approximately)
its curl $\partial\times\bm{\mathcal{A}}$ (times the enclosed area)
for the small loops (optical gyroscopes).}
\end{figure}

Consider an axistationary metric (\ref{eq:AxistatMetric}), and a
circular optical fiber loop centered at the symmetry axis, as depicted
in Fig. \ref{fig:Sagnac}. From Eq. (\ref{eq:SagnacDiffForm}), counter-propagating
light beams complete such loop with a coordinate time difference,

\begin{equation}
\Delta t=2\oint_{C}\mathcal{A}_{\phi}d\phi=2\mathcal{A}_{\phi}\int_{0}^{2\pi}d\phi=4\pi\mathcal{A}_{\phi}\ .\label{eq:DtBigLoop}
\end{equation}
In terms of the proper time of the local laboratory observer (\ref{eq:uLab}),
the difference is $\Delta\tau=\sqrt{-g_{00}}\Delta t=4\pi u_{\phi}$.
That is, it is, up to a $4\pi$ factor, the angular momentum per unit
mass of the apparatus (or, equivalently, of the laboratory observers
attached to it), cf. Sec. \ref{sub:Stationary-observers,-orbital}.
Hence, in such an apparatus, a Sagnac effect arises \emph{iff} its
angular momentum is non-zero. Notice that this singles out the zero
angular momentum observers (ZAMOs) as those which regard the $\pm\phi$
directions as geometrically equivalent; for this reason they are said
to be those that do not rotate with respect to ``the local spacetime
geometry'' \cite{Misner:1974qy}.

\emph{Physical interpretation.---} In the flat spacetime case in Fig.
\ref{fig:Sagnac} (a), the physical interpretation of the Sagnac effect
is simple, from the point of view of an inertial frame: the beams
undergo different paths in their round trips. The co-rotating one
undergoes a longer path, comparing to the case that the apparatus
does not rotate, because the arrival point is ``running away'' from
the beam during the trip, thus taking longer to complete the loop
(since the speed of light is the same). Conversely, the counter-rotating
one undergoes a shorter path, since the arrival point is approaching
the beam during the trip. This provides an intuitive argument for
understanding the general relativistic Sagnac effect as well. Consider
the gravitational field of a spinning body, as depicted in Fig. \ref{fig:Sagnac}(b).
As is well known, in such a field the observers (or objects) with
zero angular momentum actually have, from the point of view of a star-fixed
coordinate system, a non-vanishing angular velocity $\Omega_{{\rm ZAMO}}$,
Eq. (\ref{eq:OmegaZamo}). For the far field of a \emph{finite}, isolated
spinning source with angular momentum $J$ (see e.g. \cite{CiufoliniWheeler}),
$\mathcal{A}_{\phi}\simeq-2J/r$ and $\Omega_{{\rm ZAMO}}\simeq2J/r^{3}$,
in the same sense as the source. Thus, by being at rest with respect
to the distant stars, the large optical fiber loop in Fig. \ref{fig:Sagnac}(b)
is in fact rotating with respect to ``the local geometry'' (i.e.,
to the ZAMOs), with angular velocity $-\Omega_{{\rm ZAMO}}$, in the
sense \emph{opposite} to the source's rotation. Therefore, beams counter-rotating
with the source should take longer to complete the loop, comparing
to the co-rotating ones. The difference is given by Eq. (\ref{eq:DtBigLoop}):
$t_{-}-t_{+}=-\Delta t\simeq8\pi J/r$.

\subsubsection{Small loop --- optical gyroscope\label{sub:optical gyroscope}}

Consider a small loop centered at some point (call it $x_{O}^{\alpha}$)
at rest in the coordinate system of (\ref{eq:StatMetric}), as depicted
in Fig. \ref{fig:Sagnac}. Making a Taylor expansion, around $x_{O}^{\alpha}$,
of the components $(\partial\times\bm{\mathcal{A}})^{k}$, and keeping
only the lowest order terms, it follows, from Eq. (\ref{eq:SagnacStokes}),

\begin{equation}
\Delta t\approx2(\partial\times\bm{\mathcal{A}})^{k}|_{O}\int_{\mathcal{S}}d\mathcal{S}_{k}=2(\partial\times\bm{\mathcal{A}})^{k}|_{O}({\rm Area}_{\mathcal{S}})_{k}\ ,\label{eq:DtGyro}
\end{equation}
where ${\rm Area}_{\mathcal{S}}^{k}$ is the ``area vector'' of
the small loop (i.e., a vector approximately normal to $\mathcal{S}$
at $x_{O}^{\alpha}$, whose magnitude ${\rm Area}_{\mathcal{S}}$
approximately equals the enclosed area\footnote{Here, unlike in the exact Eq. (\ref{eq:SagnacStokes}), the surface
$\mathcal{S}$ is not arbitrary. In flat spacetime the loop is assumed
flat, so that ${\rm Area}_{\mathcal{S}}^{k}$ is normal to its plane,
and ${\rm Area}_{\mathcal{S}}$ \emph{exactly} the enclosed area.
In a curved spacetime the approximation is acceptable as long as the
loop and $\mathcal{S}$ are nearly flat (ideally, when they are the
image, by the exponential map, of a plane loop in the tangent space
at $x_{O}^{\alpha}$).}). Hence, for such setting, the Sagnac effect is governed by \emph{the
curl} of $\bm{\mathcal{A}}$. Although $\Delta t$ itself does not
depend on it, both the loop area and $(\partial\times\bm{\mathcal{A}})^{k}|_{O}$
require defining a metric on the space manifold $\Sigma$. The usual
notion of area relies on the measurement of distances between observers,
and so the most natural metric to use is the ``orthogonal'' metric
$h_{ij}$ defined above, which yields the distances as measured through
Einstein's light signaling procedure. With such choice\footnote{Had one chosen some other metric $(g_{\Sigma})_{ij}$ on $\Sigma$,
an extra factor $\sqrt{h/g_{\Sigma}}$ would arise in expressions
(\ref{eq:DeltatVort}). }, it follows that $(\partial\times\bm{\mathcal{A}})^{k}|_{O}=2e^{-\Phi}\omega^{k}|_{O}$,
where 
\begin{equation}
\omega^{\alpha}=\frac{1}{2}\epsilon^{\alpha\beta\gamma\delta}u_{\gamma;\beta}u_{\delta}\label{eq:Vorticity}
\end{equation}
is the vorticity of the observers (\ref{eq:uLab}), at rest in the
coordinate system of (\ref{eq:StatMetric}). Therefore, 
\begin{equation}
\Delta t\approx4e^{-\Phi}\omega^{k}|_{O}({\rm Area}_{\mathcal{S}})_{k}\ ;\qquad\Delta\tau\approx4\omega^{k}|_{O}({\rm Area}_{\mathcal{S}})_{k}\ .\label{eq:DeltatVort}
\end{equation}
Hence, the Sagnac effect in such a small loop is a measure of the
vorticity of the observers that are at rest with respect to the apparatus.
It represents the local absolute rotation of such observers, i.e.,
their rotation with respect to the ``local compass of inertia''
(e.g. \cite{MassaZordan,RindlerPerlick,CiufoliniWheeler,Analogies}).
Let us make this notion more precise. The local compass of inertia
is mathematically defined by a system of axis undergoing \emph{Fermi-Walker
transport} (e.g. \cite{Misner:1974qy}), and materialized physically
by the spin axes of guiding gyroscopes. The vorticity $\omega^{\alpha}$
corresponds to the angular velocity of rotation of the connecting
vectors between neighboring observers with respect to axes Fermi-Walker
transported along the observer congruence\footnote{The Fermi-Walker derivative, whose vanishing defines the Fermi-Walker
transport law \cite{Misner:1974qy}, reads, for a vector $\eta^{\mu}$,
\[
\frac{D_{F}\eta^{\mu}}{d\tau}=\eta_{\ ;\nu}^{\mu}u^{\nu}-2u^{[\mu}a^{\nu]}\eta_{\nu}
\]
(where $a^{\mu}\equiv u_{\ ;\nu}^{\mu}u^{\nu}$). If $\eta^{\mu}$
is a connecting vector, $\mathcal{L}_{u}\eta^{\nu}=0\Rightarrow\eta_{\ ;\nu}^{\mu}u^{\nu}=u_{\ ;\nu}^{\mu}\eta^{\nu}$;
since, for a rigid congruence, $u_{\mu;\nu}=-a_{\mu}u_{\nu}-\epsilon_{\mu\nu\alpha\beta}\omega^{\alpha}u^{\beta}$
(e.g. \cite{Hawking:1973uf,ManyFaces,Invariants}), it follows that
$D_{F}\eta^{\mu}/d\tau=\epsilon_{\ \alpha\nu\beta}^{\mu}\omega^{\alpha}\eta^{\nu}u^{\beta}-u^{\mu}a^{\nu}\eta_{\nu}$,
whose space components (orthogonal to $u^{\nu}$) read $D_{F}\vec{\eta}/d\tau=\vec{\omega}\times\vec{\eta}$,
manifesting that $\vec{\eta}$ indeed rotates with respect to Fermi-Walker
transport with angular velocity $\vec{\omega}$. } \cite{RindlerPerlick,Iyer1993,CiufoliniWheeler}. The Sagnac effect
in the small optical fiber loop is thus a probe for such rotation,
and is for this reason called an \emph{optical gyroscope}.

\emph{Physical interpretation.---} Concerning the small loop placed
in the turntable of Fig. \ref{fig:Sagnac}(a), essentially the same
principle as for the large loop (Sec. \ref{sub:circular-loop-around})
explains that the beam propagating in the same sense as the turntable's
rotation takes longer to complete the loop. Consider now the small
loop in Fig. \ref{fig:Sagnac}(b). A well known facet of frame-dragging
is that, close to a spinning source, the compass of inertia rotates
with respect to inertial frames at infinity (i.e., to the star-fixed
frame). For the far field of a finite isolated source, the corresponding
angular velocity is, in the equatorial plane, $\simeq-\vec{J}/r^{3}$
(e.g. \cite{CiufoliniWheeler,Misner:1974qy}), in the sense \emph{opposite}
to the source's rotation. By being fixed with respect to the distant
stars, the small loop in Fig. \ref{fig:Sagnac}(b) is thus rotating
with respect to the compass of inertia, with angular velocity $\vec{\omega}\simeq\vec{J}/r^{3}$.
Therefore, \emph{contrary to the situation for the large loop}, beams
propagating in the same sense as the source's rotation take longer
to complete the loop.

\subsection{Closed forms, exact forms, and Stokes theorem\label{sub:Closed-forms,-exact}}

\begin{figure}
\includegraphics[width=0.45\columnwidth]{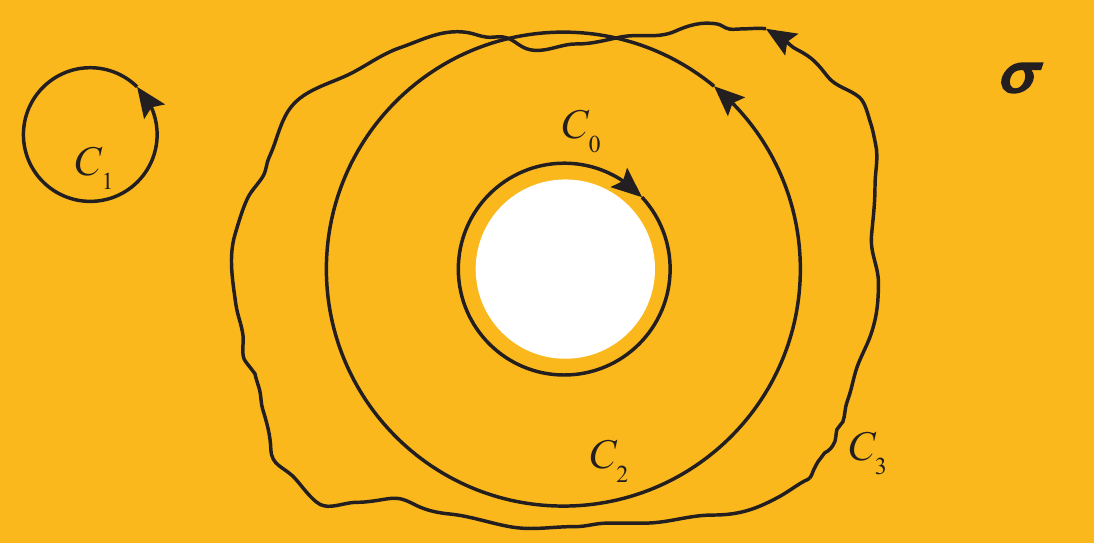}\protect\protect\protect\caption{\label{fig:Forms}A closed 1-form $\bm{\sigma}$ in the punctured
plane $\mathbb{R}^{2}\backslash\{0\}$. By Stokes' theorem, $\oint_{C_{1}}\bm{\sigma}=0$
and $\oint_{C_{2}}\bm{\sigma}=\oint_{C_{3}}\bm{\sigma}$.}
\end{figure}

A 1-form $\bm{\sigma}$ is closed if $\mathbf{d}\bm{\sigma}=0$; it
is moreover exact if $\bm{\sigma}=\mathbf{d}\varphi$, for some smooth
(single-valued) function $\varphi$. \emph{Locally}, the two conditions
are equivalent, but globally it is not so. Exact forms have a vanishing
circulation $\oint_{C}\bm{\sigma}$ around \emph{any} closed curve
$C$. In simply connected regions, every closed form is exact; multiply
connected regions allow for the existence of closed but non-exact
forms. Consider a closed form $\bm{\sigma}$ in a manifold with topology
$\mathbb{R}^{2}\backslash\{0\}$, as illustrated in Fig. \ref{fig:Forms}.
The loop $C_{1}$ lies in a simply connected region (so that $C_{1}$
can be shrunk to a point); by the Stokes theorem, $\oint_{C_{1}}\bm{\sigma}=\int_{\mathcal{S}_{1}}\mathbf{d}\bm{\sigma}=0$,
where $\mathcal{S}_{1}$ is a compact 2-D manifold bounded by $C_{1}$
($C_{1}=\partial\mathcal{S}_{1}$). Loops $C_{2}$ and $C_{3}$ enclose
a multiply connected region. The disjoint unions of curves $C_{2}\sqcup C_{0}$
and $C_{3}\sqcup C_{0}$ form boundaries of compact 2-D manifolds,
to which the Stokes theorem can be applied. The theorem demands in
this case that $\int_{C_{2}\sqcup C_{0}}\bm{\sigma}=0=\int_{C_{3}\sqcup C_{0}}\bm{\sigma}$,
i.e., 

\[
\int_{C_{2}\sqcup C_{0}}\bm{\sigma}\equiv\oint_{C_{2}}\bm{\sigma}+\oint_{C_{0}}\bm{\sigma}\ =\ \int_{C_{3}\sqcup C_{0}}\bm{\sigma}\ \ \ \Rightarrow\ \ \ \oint_{C_{2}}\bm{\sigma}=\oint_{C_{3}}\bm{\sigma}\ .
\]
So, the circulation of $\bm{\sigma}$ vanishes along any loop not
enclosing the origin (``hole''), and has the same value for any
loop enclosing it. When $\oint_{C_{2}}\bm{\sigma}=\oint_{C_{3}}\bm{\sigma}\ne0$,
the form $\bm{\sigma}$ is non-exact; an example is the 1-form $\bm{\sigma}=\mathbf{d}\phi$.

\subsection{Komar Integrals\label{sub:Komar-Integrals}}

In stationary spacetimes admitting Killing vectors fields $\xi^{\alpha}$,
and for a compact spacelike hypersurface (i.e., 3-volume) $\mathcal{V}$
with boundary $\partial\mathcal{V}$, the Komar integrals are defined
as \cite{Wald:1984,Townsend,HansenWinicour1975,Komar1959,NatarioMathRel}
\begin{equation}
Q_{\xi}(\mathcal{V})=-\frac{K}{16\pi}\int_{\partial\mathcal{V}}\star\mathbf{d}\bm{\xi}\ ,\label{eq:Komar}
\end{equation}
where $(\star\mathbf{d}\bm{\xi})_{\alpha\beta}\equiv\xi_{\nu;\mu}\epsilon_{\ \ \alpha\beta}^{\mu\nu}$
is the 2-form dual to $\mathbf{d}\bm{\xi}$, and $K$ a dimensionless
constant specific to each $\xi^{\alpha}$. Since $\mathcal{V}$ is
compact, an application of the Stokes theorem leads to the equivalent
expressions 
\begin{equation}
Q_{\xi}(\mathcal{V})=-\frac{K}{16\pi}\int_{\mathcal{V}}\mathbf{d}(\star\mathbf{d}\bm{\xi})=\frac{K}{8\pi}\int_{\mathcal{V}}R_{\ \beta}^{\alpha}\xi^{\beta}d\mathcal{V}_{\alpha}=-K\int_{\mathcal{V}}\left[T_{\ \beta}^{\alpha}-\frac{1}{2}g_{\ \beta}^{\alpha}T_{\ \gamma}^{\gamma}\right]\xi^{\beta}n_{\alpha}d\mathcal{V},\label{eq:KomarRicci}
\end{equation}
where $d\mathcal{V}_{\alpha}=-n_{\alpha}d\mathcal{V}=\epsilon_{\alpha\mu\nu\lambda}\mathbf{d}x^{\mu}\wedge\mathbf{d}x^{\nu}\wedge\mathbf{d}x^{\lambda}/6$
is the volume element 1-form of $\mathcal{V}$, $n^{\alpha}$ is the
future-pointing unit vector normal to $\mathcal{V}$, and we used
the well known relation for Killing vectors fields $\xi_{\ ;\delta\alpha}^{\alpha}=R_{\beta\delta}\xi^{\beta}$
to notice that $\mathbf{d}(\star\mathbf{d}\bm{\xi})=-2R_{\alpha\beta}\xi^{\beta}d\mathcal{V}^{\alpha}$.
Observe that this expression implies that, in \emph{vacuum} ($R_{\mu\nu}=0$),
$\star\mathbf{d}\bm{\xi}$ is a \emph{closed 2-form}. Via the Stokes
theorem, this means that $Q_{\xi}(\mathcal{V})=0$ for any compact
hypersurface $\mathcal{V}$ not enclosing sources, and has the same
value for any compact $\mathcal{V}$ enclosing an isolated system
(see Sec. \ref{sub:Closed-forms,-exact} above). Due to this hypersurface
independence, $Q_{\xi}(\mathcal{V})$ is said to be \emph{conserved}.

In an asymptotically flat axistationary spacetime, \emph{and} in a
suitable coordinate system \cite{Wald:1984,Whittaker1935,HansenWinicour1975},
where the Killing vector field $\partial_{t}^{\alpha}=\xi^{\alpha}$
is time-like and tangent to inertial observers at infinity (corresponding
to the source's asymptotic inertial ``rest'' frame), and is moreover
normalized so that $\xi^{\alpha}\xi_{\alpha}\stackrel{r\rightarrow\infty}{\rightarrow}-1$,
the quantity $M=Q_{\xi}(\mathcal{V})$, with $K=-2$, has the physical
meaning of ``active gravitational mass,'' or total mass/energy present
in the spacetime \cite{Wald:1984,Whittaker1935,Komar1959,HansenWinicour1975}.
Similarly, $J=Q_{\zeta}(\mathcal{V})$, for $\zeta^{\alpha}=\partial_{\phi}^{\alpha}$
and $K=1$, is the angular momentum present in the spacetime. Other
coordinate systems/Killing vectors can be considered; in that case,
however, the interpretation of quantities such as mass or angular
momentum is in general not appropriate. Consider, e.g., a rigidly
rotating coordinate system $\{x^{\alpha'}\}$, obtained from the asymptotically
inertial coordinate system $\{x^{\alpha}\}$ by the transformation
$\phi'=\phi-\Omega't$, $x^{\alpha'\ne\phi'}=x^{\alpha}$. In terms
of the new Killing vector field $\partial'_{t}=\partial_{t}+\Omega'\partial_{\phi}$,
one has 
\begin{equation}
M'=\frac{1}{8\pi}\int_{\partial\mathcal{V}}\star\mathbf{d}\bm{\xi}'=\frac{1}{8\pi}\int_{\partial\mathcal{V}}\star\mathbf{d}(\bm{\xi}+\Omega'\bm{\zeta})=M-2\Omega'J\ ,\label{eq:Mprime}
\end{equation}
i.e., $M'$ is a mixture of the mass $M$ and the angular momentum
$J$ of the spacetime (as computed in the asymptotically inertial
frame). The latter in this case stays the same, $J'=J$, as $\partial'_{\phi'}=\partial_{\phi}$.

\section{Gravitomagnetism and its different levels\label{sec:Gravitomagnetism-Levels}}

The gravitational effects generated by the motion of matter, or, more
precisely, by mass/energy currents, are known as ``gravitomagnetism\textquotedblright .
The reason for the denomination is its many analogies with magnetism
(generated by charge currents). To make them apparent, consider a
stationary metric with line element written in the form (\ref{eq:StatMetric}),
and let, as in (\ref{eq:uLab}), $u^{\alpha}$ be the 4-velocity of
the laboratory observers, and $U^{\alpha}=dx^{\alpha}/d\tau$ the
4-velocity of a test point particle in geodesic motion. The space
components of the geodesic equation $DU^{\alpha}/d\tau=0$ yield\footnote{\label{fn:Christoffel}The relevant Christoffel symbols read $\Gamma_{00}^{i}=-e^{2\Phi}G^{i}$,
$\Gamma_{j0}^{i}=e^{2\Phi}\mathcal{A}_{j}G^{i}-e^{\Phi}H_{\ j}^{i}/2$,
and $\Gamma_{jk}^{i}=\Gamma(h)_{jk}^{i}-e^{\Phi}\mathcal{A}_{(k}H_{j)}^{\ i}-e^{2\Phi}G^{i}\mathcal{A}_{j}\mathcal{A}_{k}$,
where $H_{ij}\equiv e^{\Phi}[\mathcal{A}_{j,i}-\mathcal{A}_{i,j}]$. } \cite{LandauLifshitz,ZonozBell1998,NatarioQM2007,Analogies,Zonoz2019}
\begin{equation}
\frac{\tilde{D}\vec{U}}{d\tau}=\gamma\left[\gamma\vec{G}+\vec{U}\times\vec{H}\right]\label{eq:QMGeo}
\end{equation}
where $\gamma=-U^{\alpha}u_{\alpha}=e^{\Phi}(U^{0}-U^{i}\mathcal{A}_{i})$
is the Lorentz factor between $U^{\alpha}$ and $u^{\alpha}$, 
\begin{equation}
\left[\frac{\tilde{D}\vec{U}}{d\tau}\right]^{i}=\frac{dU^{i}}{d\tau}+\Gamma(h)_{jk}^{i}U^{j}U^{k}\ ;\qquad\Gamma(h)_{jk}^{i}=\frac{1}{2}h^{il}\left(h_{lj,k}+h_{lk,j}-h_{jk,l}\right)\label{eq:3DAccel}
\end{equation}
is a covariant derivative with respect to the spatial metric $h_{ij}$,
with $\Gamma(h)_{jk}^{i}$ the corresponding Christoffel symbols,
and 
\begin{equation}
\vec{G}=-\tilde{\nabla}\Phi\ ;\qquad\quad\vec{H}=e^{\Phi}\tilde{\nabla}\times\vec{\mathcal{A}}\ ,\label{eq:GEMFieldsQM}
\end{equation}
are vector fields living on the space manifold $\Sigma$ with metric
$h_{ij}$, dubbed, respectively, ``gravitoelectric'' and ``gravitomagnetic''
fields. These play in Eq. (\ref{eq:QMGeo}) roles analogous to those
of the electric ($\vec{E}$) and magnetic ($\vec{B}$) fields in the
Lorentz force equation, $DU^{i}/d\tau=(q/m)[\gamma\vec{E}+\vec{U}\times\vec{B}]^{i}$.
Here $\tilde{\nabla}$ denotes covariant differentiation with respect
to the spatial metric $h_{ij}$ {[}i.e., the Levi-Civita connection
of $(\Sigma,h)${]}. Notice that Eq. (\ref{eq:3DAccel}) is the standard
covariant expression for the 3-D acceleration (e.g.~Eq.~(6.9) in
\cite{kay1988schaum}); equation (\ref{eq:QMGeo}) describes the acceleration
of the curve obtained by projecting the time-like geodesic onto the
space manifold $(\Sigma,h)$, and $\vec{U}$ is its tangent vector
{[}identified in spacetime with the projection of $U^{\alpha}$ onto
$(\Sigma,h)$: $(\vec{U})^{\alpha}=h_{\ \beta}^{\alpha}U^{\beta}$,
cf. Eq. (\ref{eq:SpaceProjector}){]}. The physical interpretation
of Eq. (\ref{eq:QMGeo}) is that, from the point of view of the laboratory
observers, the spatial trajectory will seem to be accelerated, as
if acted by fictitious forces---\emph{inertial} \emph{forces}. These
arise from the fact that the laboratory frame is not inertial; in
fact, $\vec{G}$ and $\vec{H}$ are identified in spacetime, respectively,
with minus the acceleration and twice the vorticity (\ref{eq:Vorticity})
of the laboratory observers: 
\begin{equation}
G^{\alpha}=-\nabla_{\mathbf{u}}u^{\alpha}\equiv-u_{\ ;\beta}^{\alpha}u^{\beta}\;;\qquad H^{\alpha}=\epsilon^{\alpha\beta\gamma\delta}u_{\gamma;\beta}u_{\delta}\;.\label{eq:GEM Fields Cov}
\end{equation}
One may cast $\vec{G}$ as a relativistic generalization of the Newtonian
gravitational field (embodying it as a limiting case), and $\vec{H}$
as a generalization of the Coriolis field \cite{Costa:2015hlh}. Equations
(\ref{eq:QMGeo})-(\ref{eq:GEMFieldsQM}) apply to stationary spacetimes
(being part of the so-called 1+3 ``quasi-Maxwell'' formalism \cite{LandauLifshitz,ZonozBell1998,NatarioQM2007,Analogies,Nouri-Zonoz:2013rfa,Zonoz2019});
formulations for arbitrary spacetimes are given in \cite{Analogies,ManyFaces,Cattaneo1958,MassaII}.

Since $\vec{B}=\nabla\times\vec{A}$ and, in stationary settings,
$\vec{E}=-\nabla\phi$, Eqs. (\ref{eq:GEMFieldsQM}) suggest also
an analogy between the electric potential $\phi$ and the ``Newtonian''
potential $\Phi$, and between the magnetic potential vector $\vec{A}$
and the vector $\vec{\mathcal{A}}$ (which, as seen in Sec. \ref{sub:Sagnac-effect},
governs the Sagnac effect); for this reason $\vec{\mathcal{A}}$ is
dubbed \emph{gravitomagnetic vector potential.}

Other realizations of the analogy exist, namely in the equations of
motion for a ``gyroscope'' (i.e., a spinning pole-dipole particle)
in a gravitational field and a magnetic dipole in a electromagnetic
field. According to the Mathisson-Papapetrou equations \cite{Mathisson:1937zz,Papapetrou:1951pa,Dixon:1970zza,Dixon1964,Gralla:2010xg,Costa:2012cy},
under the Mathisson-Pirani spin condition \cite{Mathisson:1937zz,Pirani:1956tn},
the spin vector of a gyroscope of 4-velocity $U^{\alpha}$ evolves
as $DS^{\alpha}/d\tau=S^{\mu}a_{\mu}U^{\alpha}$; here $a^{\alpha}\equiv DU^{\alpha}/d\tau$
and $S^{\alpha}$ is the spin vector, which is spatial with respect
to $U^{\alpha}$ ($S^{\alpha}U_{\alpha}=0$). For a gyroscope whose
center of mass is at rest in the coordinate system of (\ref{eq:StatMetric}),
$U^{\alpha}=u^{\alpha}$ {[}see Eq. (\ref{eq:uLab}){]}, and the space
part of the spin evolution equation reads (see footnote on page \pageref{fn:Christoffel},
and notice that $S^{\alpha}u_{\alpha}=0\Rightarrow S^{0}=S^{i}\mathcal{A}_{i}$)
\begin{equation}
\frac{d\vec{S}}{d\tau}=\frac{1}{2}\vec{S}\times\vec{H}\ ,\label{eq:SpinPrec}
\end{equation}
which is analogous to the precession of a magnetic dipole in a magnetic
field, $D\vec{S}/d\tau=\vec{\mu}\times\vec{B}$. Another effect directly
governed by the gravitomagnetic field $\vec{H}$ is the Sagnac time
delay in an optical gyroscope, as follows from Eqs. (\ref{eq:DeltatVort}),
\begin{equation}
\Delta t\approx2e^{-\Phi}\vec{H}\cdot\vec{{\rm A}}\!{\rm rea}_{\mathcal{S}}\ ;\qquad\Delta\tau\approx2\vec{H}\cdot\vec{{\rm A}}\!{\rm rea}_{\mathcal{S}}\ .\label{eq:DeltatH}
\end{equation}

When the electromagnetic field is non-homogeneous, a force is exerted
on a magnetic dipole, covariantly described by $DP^{\alpha}/d\tau=B^{\beta\alpha}\mu_{\beta}$
\cite{Costa:2012cy,Dixon1964}, where $\mu^{\beta}$ is the magnetic
dipole moment 4-vector, and $B_{\alpha\beta}=\star F_{\alpha\mu;\beta}U^{\mu}$
($F^{\alpha\beta}\equiv$ Faraday tensor, $\star\equiv$ Hodge dual)
is the ``magnetic tidal tensor'' as measured by the particle. A
covariant force is likewise exerted on a gyroscope in a gravitational
field (the spin-curvature force \cite{Mathisson:1937zz,Papapetrou:1951pa,Dixon:1970zza,Dixon1964,Gralla:2010xg,Costa:2012cy}),
which (again, under the Mathisson-Pirani spin condition) takes a remarkably
similar form \cite{Costa:2012cy,CostaHerdeiro} 
\begin{equation}
\frac{DP^{\alpha}}{d\tau}=-\mathbb{H}^{\beta\alpha}S_{\beta};\qquad\quad\mathbb{H}_{\alpha\beta}\equiv\star R_{\alpha\mu\beta\nu}U^{\mu}U^{\nu}=\frac{1}{2}\epsilon_{\alpha\mu}^{\ \ \ \lambda\tau}R_{\lambda\tau\beta\nu}U^{\mu}U^{\nu}.\label{eq:SpinCurvature}
\end{equation}
Here $\mathbb{H}_{\alpha\beta}$ is the ``gravitomagnetic tidal tensor''
(or ``magnetic part'' of the Riemann tensor \cite{Dadhich:1999eh})
as measured by the particle, playing a role analogous to that of $B_{\alpha\beta}$
in electromagnetism. For a congruence of observers at rest in a stationary
field in the form (\ref{eq:StatMetric}), the relation between these
tidal tensors and the magnetic/gravitomagnetic fields is \cite{Analogies}
\begin{eqnarray}
B_{ij} & = & \tilde{\nabla}_{j}B_{i}+\frac{1}{2}\left[\vec{E}\cdot\vec{H}h_{ij}-E_{j}H_{i}\right]\ ;\label{eq:BijGEM}\\
\mathbb{H}_{ij} & = & -\frac{1}{2}\left[\tilde{\nabla}_{j}H_{i}+(\vec{G}\cdot\vec{H})h_{ij}-2G_{j}H_{i}\right]\ .\label{eq:HijGEM}
\end{eqnarray}
In a locally inertial frame (and rectangular coordinates) $B_{ij}=B_{i,j}$,
and the force on a comoving magnetic dipole reduces to the textbook
expression $DP^{i}/d\tau=B^{j,i}u_{j}\equiv\nabla(\vec{\mu}\cdot\vec{B})$.
Moreover, in the linear regime, $\mathbb{H}_{ij}\approx H_{i,j}$,
and so the force (\ref{eq:SpinCurvature}) on a gyroscope at rest
yields $D\vec{P}/d\tau\approx\nabla(\vec{S}\cdot\vec{H})/2$. The
tidal tensors $B_{\alpha\beta}$ and $\mathbb{H}_{\alpha\beta}$ are
essentially quantities one order higher in differentiation, comparing
to the corresponding fields $B^{\alpha}$ and $H^{\alpha}$. The gravitoelectric
counterpart of $\mathbb{H}_{\alpha\beta}$ is the tidal tensor $\mathbb{E}_{\alpha\beta}\equiv R_{\alpha\mu\beta\nu}U^{\mu}U^{\nu}$
(``electric part'' of the Riemann tensor) \cite{CostaHerdeiro,Analogies},
which governs the geodesic deviation equation $D^{2}\delta x^{\alpha}/d\tau=-\mathbb{E}_{\ \beta}^{\alpha}\delta x^{\beta}$.
In vacuum, these tensors together fully determine the Riemann (i.e.
Weyl) tensor, cf. the decomposition (30) of \cite{MaartensBasset1997},
and hence the tidal forces felt by any set of test particles/observers.
\begin{table*}
\begin{tabular}{|c|l||c|l|}
\hline 
\multicolumn{2}{|c||}{\raisebox{3.5ex}{}\raisebox{0.5ex}{\textbf{Levels of Magnetism}}} & \multicolumn{2}{c|}{\raisebox{0.5ex}{\textbf{Levels of Gravitomagnetism}}}\tabularnewline
\hline 
\hline 
\raisebox{4ex}{}\raisebox{0.7ex}{Governing Field}  & \raisebox{0.7ex}{Physical effect}  & \raisebox{0.7ex}{Governing Field}  & \raisebox{0.7ex}{Physical effect} \tabularnewline
\hline 
\raisebox{9ex}{}\raisebox{2.5ex}{%
\begin{tabular}{c}
$\vec{A}$\tabularnewline
{\footnotesize{}{}{}(magnetic }\tabularnewline
\raisebox{0.5ex}{{\footnotesize{}{}{}vector potential)}}\tabularnewline
\end{tabular}}  & \raisebox{3ex}{%
\begin{tabular}{l}
$\bullet$ Aharonov-Bohm effect\tabularnewline
~~~~(quantum theory)\tabularnewline
\end{tabular}}  & \raisebox{2.5ex}{%
\begin{tabular}{c}
$\vec{\mathcal{A}}$\tabularnewline
{\footnotesize{}{}{}(gravitomagnetic }\tabularnewline
\raisebox{0.5ex}{{\footnotesize{}{}{}vector potential)}}\tabularnewline
\end{tabular}}  & \raisebox{3ex}{%
\begin{tabular}{l}
$\bullet$ Sagnac effect\tabularnewline
\raisebox{2.5ex}{}$\bullet$ part of gravitomagnetic\tabularnewline
\raisebox{0.5ex}{~~~ ``clock effect''}\tabularnewline
\end{tabular}}\tabularnewline
\hline 
\raisebox{17ex}{}\raisebox{3.5ex}{%
\begin{tabular}{c}
\raisebox{0.1ex}{$\vec{B}$}\tabularnewline
{\footnotesize{}{}{}(magnetic field }\tabularnewline
{\footnotesize{}{}{}$=\nabla\times\vec{A}$)}\tabularnewline
\end{tabular}}  & \raisebox{2.5ex}{%
\begin{tabular}{l}
$\bullet$ magnetic force \tabularnewline
\raisebox{0.5ex}{~~~$q\vec{U}\times\vec{B}$}\tabularnewline
\raisebox{3.8ex}{}$\bullet$ dipole precession\tabularnewline
\raisebox{0.5ex}{~~~$D\vec{S}/d\tau=\vec{\mu}\times\vec{B}$}\tabularnewline
\raisebox{3.8ex}{}$\bullet$ magnetic ``clock\tabularnewline
\raisebox{1ex}{~~~effect''}\tabularnewline
\end{tabular}}  & \raisebox{3.5ex}{%
\begin{tabular}{c}
\raisebox{0.1ex}{$\vec{H}$}\tabularnewline
{\footnotesize{}{}{}(gravitomagnetic }\tabularnewline
{\footnotesize{}{}{}field $=e^{\phi}\nabla\times\vec{\mathcal{A}}$)}\tabularnewline
\end{tabular}}  & \raisebox{2.5ex}{%
\begin{tabular}{l}
$\bullet$ gravitomagnetic force \tabularnewline
\raisebox{0.5ex}{~~~$m\gamma\vec{U}\times\vec{H}$}\tabularnewline
$\bullet$ gyroscope precession\tabularnewline
\raisebox{0.5ex}{~~~$d\vec{S}/d\tau=\vec{S}\times\vec{H}/2$}\tabularnewline
$\bullet$ Sagnac effect in \tabularnewline
\raisebox{0.5ex}{~~~light gyroscope}\tabularnewline
$\bullet$ part of gravitomagnetic\tabularnewline
\raisebox{0.5ex}{~~~ ``clock effect''}\tabularnewline
\end{tabular}}\tabularnewline
\hline 
\raisebox{9.5ex}{}\raisebox{3.3ex}{%
\begin{tabular}{c}
\raisebox{0.1ex}{$B_{\alpha\beta}$}\tabularnewline
{\footnotesize{}{}{}(magnetic tidal}\tabularnewline
\raisebox{0.5ex}{{\footnotesize{}{}{}tensor $\sim\partial_{i}\partial_{j}A_{k}$)}}\tabularnewline
\end{tabular}}  & \raisebox{3.3ex}{%
\begin{tabular}{l}
$\bullet$ Force on mag. dipole\tabularnewline
\raisebox{2.4ex}{}~~~{\small{}{}{}$DP^{\alpha}/d\tau=B^{\beta\alpha}\mu_{\beta}$}\tabularnewline
\end{tabular}}  & \raisebox{3.3ex}{%
\begin{tabular}{c}
$\mathbb{H}_{\alpha\beta}$\tabularnewline
{\footnotesize{}{}{}(gravitomagnetic}\tabularnewline
\raisebox{0.5ex}{{\footnotesize{}{}{}tidal tensor $\sim\partial_{i}\partial_{j}\mathcal{A}_{k}$)}}\tabularnewline
\end{tabular}}  & \raisebox{3.5ex}{%
\begin{tabular}{l}
$\bullet$ Force on gyroscope\tabularnewline
\raisebox{2.4ex}{}~~~{\small{}{}{}$DP^{\alpha}/d\tau=-\mathbb{H}^{\beta\alpha}S_{\beta}$}\tabularnewline
\end{tabular}}\tabularnewline
\hline 
\end{tabular}\protect\protect\protect\caption{\label{tab:Levels}Levels of magnetism and gravitomagnetism, corresponding
to different levels of differentiation of, respectively, the magnetic
($\vec{A}$) and gravitomagnetic ($\vec{\mathcal{A}}$) vector potentials.}
\end{table*}

The analogy between magnetic and gravitomagnetic effects can thus
be cast into the three distinct levels in Table \ref{tab:Levels},
corresponding to three different levels of differentiation of the
gravitomagnetic vector potential $\vec{\mathcal{A}}$.

We close this overview with a note on the so-called ``frame dragging'';
in the literature this denomination is used for two main kinds of
effects: 
\begin{description}
\item [{(i)}] One, the fact that near a moving source (e.g. a rotating
body) the compass of inertia, and thus the locally inertial frames,
rotate with respect to inertial frames at infinity (i.e. to the ``distant
stars''). Or, conversely, near a rotating source a frame anchored
to the distant stars in fact rotates with respect to the local compass
of inertia \cite{LenseThirringTranslated,EinsteinLetterMach,EinsteinMeaning,Schiff_PNAS1960,CohenBrill1968,CiufoliniWheeler,Misner:1974qy,Iyer1993,Costa:2015hlh},
and observers at rest therein have non-vanishing vorticity \cite{Herrera_FrameDragging2006,Herrera_FrameDragging_and_SE2007}.
This is manifest in that, relative to such frame, gyroscopes precess
{[}as described by Eq. (\ref{eq:SpinPrec}){]}, and Coriolis (i.e.
gravitomagnetic) forces arise {[}cf. Eq. (\ref{eq:QMGeo}){]}, causing
e.g. orbits of test bodies to precess (Lense-Thirring orbital precession
\cite{LenseThirringTranslated,CiufoliniPavlisNature2004,CiufoliniNature2007}),
and the plane of a Foucault pendulum to rotate \cite{Misner:1974qy}. 
\item [{(ii)}] The other, the fact that, close to a rotating source, the
orbits of zero angular momentum (e.g. the ZAMOs of Sec. \ref{sub:Stationary-observers,-orbital})
have non-zero angular velocity as seen from infinity (or, conversely,
objects with zero angular velocity have non-zero angular momentum)
\cite{Misner:1974qy,BlackSaturn,CizekSemerakDisks2017}. Associated
to this, in axistationary spacetimes, a system of axes carried by
the ZAMOs and spatially locked to the background symmetries (dubbed,
somewhat misleadingly \cite{Misner:1974qy,Analogies}, ``locally
non-rotating frames'' \cite{BaardenPressTeukolsky,Bardeen1970ApJ}),
rotates with respect to comoving gyroscopes \cite{Semerak_CQG1996_GyroscopePrec}. 
\end{description}
We point out, in view of the above, that the phenomena in (i) and
(ii) have distinct origins, corresponding to two different levels
of gravitomagnetism, the former being governed by $\vec{H}$, and
the latter by $\vec{\mathcal{A}}$. The effects are independent: in
fact, as we shall see, there exist solutions for which $\vec{H}$
vanishes whilst $\vec{\mathcal{A}}$ is non-zero, of which the metric
in Sec. \ref{sub:The-metric-in-3-parameters} is an example.

\subsection{The gravitomagnetic clock effect\label{sub:The-gravitomagnetic-clock}}

When a body rotates, the period of co- and counter-rotating geodesics
around it differs in general; such effect has been dubbed \cite{COHENMashhoon1993,BonnorClockEffect,BiniJantzenMashhoon_Clock1,MaartensMashhoonMatravers_Holonomy}
gravitomagnetic ``clock effect''. Let $U^{\alpha}\equiv dx^{\alpha}/d\tau=U^{0}\delta_{0}^{\alpha}+U^{\phi}\delta_{\phi}^{\alpha}$
be the 4-velocity of a test particle describing a circular geodesic
in an axistationary spacetime, and $\mathcal{L}=g_{\mu\nu}U^{\mu}U^{\nu}/2$
the corresponding Lagrangian. The angular velocity $\Omega_{{\rm geo}}\equiv d\phi/dt=U^{\phi}/U^{0}$
of the circular geodesics is readily obtained from the Euler-Lagrange
equations, 
\begin{equation}
\frac{d}{d\tau}\left(\frac{\partial\mathcal{L}}{\partial U^{\alpha}}\right)-\frac{\partial\mathcal{L}}{\partial x^{\alpha}}=0\ ,\label{eq:EL}
\end{equation}
which reduce to 
\begin{equation}
g_{\phi\phi,r}\Omega_{{\rm geo}}^{2}+2g_{0\phi,r}\Omega_{{\rm geo}}+g_{00,r}=0\ .\label{eq:circ-geo}
\end{equation}
Solving this equation yields 
\begin{equation}
\Omega_{{\rm geo}\pm}=\frac{-g_{0\phi,r}\pm\sqrt{g_{0\phi,r}^{2}-g_{\phi\phi,r}g_{00,r}}}{g_{\phi\phi,r}}\label{eq:OmegaGeo}
\end{equation}
the $+$ ($-$) sign corresponding, for $g_{\phi\phi,r}>0$ and $g_{00,r}<0$
(i.e., attractive $\vec{G}$), to prograde (retrograde) geodesics,
i.e., positive (negative) $\phi$ directions. The orbital period is,
in \emph{coordinate} time, $t_{{\rm geo}}=2\pi/|\Omega_{{\rm geo}}|$;
hence, the difference between the periods of prograde and retrograde
geodesics reads 
\[
\Delta t_{{\rm geo}}=2\pi\left(\frac{1}{\Omega_{{\rm geo+}}}+\frac{1}{\Omega_{{\rm geo-}}}\right)=-4\pi\frac{g_{0\phi,r}}{g_{00,r}}\ .
\]
Since $g_{0\phi}=-g_{00}\mathcal{A}_{\phi}$, this result can be re-written
as 
\begin{equation}
\Delta t_{{\rm geo}}=4\pi\mathcal{A}_{\phi}-2\pi\frac{\mathcal{A}_{\phi,r}}{G_{r}}=4\pi\mathcal{A}_{\phi}-2\pi\frac{\star H_{r\phi}}{G_{r}e^{\Phi}}\ ,\label{eq:GMClock}
\end{equation}
where $\star H_{jk}\equiv\epsilon_{ijk}H^{i}$ is the 2-form dual
to the gravitomagnetic field $\vec{H}$. In cylindrical coordinates
one can substitute $\star H_{r\phi}=\sqrt{h}H^{z}$; in spherical
coordinates, $\star\!H_{r\phi}=-\sqrt{h}H^{\theta}$. Hence, the gravitomagnetic
clock effect consists of the sum of two contributions of different
origin: the ``global'' Sagnac effect around the source, Eq. (\ref{eq:DtBigLoop}),
which is governed by $\mathcal{A}_{\phi}$, plus a term governed by
the gravitomagnetic field $\vec{H}$. The physical interpretation
of the latter is as follows: for circular orbits, the gravitomagnetic
force $\gamma\vec{U}\times\vec{H}$ in Eq. (\ref{eq:QMGeo}) is radial
(since $\vec{H}$ is parallel to the axis, and $\vec{U}=U^{\phi}\partial_{\phi}$),
being attractive or repulsive depending on the $\pm\phi$ direction
of the orbit. Namely, it is attractive when the test body counter-rotates
with the central source, and repulsive when it co-rotates. This highlights
the fact that (anti-) parallel mass/energy currents have a repulsive
(attractive) gravitomagnetic interaction, which is opposite to the
situation in electromagnetism, where (anti-) parallel charge currents
attract (repel) (see \cite{schutz_2003} and Sec. 13.6 of \cite{Feynman:1963uxa},
respectively, for enlightening analogous explanations of these relativistic
effects). In fact, the second term has an exact electromagnetic analogue,
as we shall now show.

\subsubsection*{Electromagnetic analogue}

Consider, in flat spacetime, a charged test particle of charge $q$
and mass $m$ in a circular orbit around a spinning charged body,
and let $\mathcal{L}=mg_{\mu\nu}U^{\mu}U^{\nu}/2+qg_{\mu\nu}U^{\mu}A^{\nu}$,
with $A^{\mu}=(\phi,\vec{A})$, be the corresponding Lagrangian. The
Euler-Lagrange equations (\ref{eq:EL}) yield, for a circular orbit,
\[
qE_{r}+q\Omega_{{\rm orb}}A_{\phi,r}+\frac{1}{2}U^{0}mg_{\phi\phi,r}\Omega_{{\rm orb}}^{2}=0\ ,
\]
where $\Omega_{{\rm orb}}\equiv d\phi/dt=U^{\phi}/U^{0}$, leading
to 
\[
\Omega_{{\rm orb\pm}}=\frac{-qA_{\phi,r}\pm\sqrt{q^{2}A_{\phi,r}^{2}-2U^{0}qmg_{\phi\phi,r}E_{r}}}{U^{0}mg_{\phi\phi,r}}\ .
\]
Thus, for $qE_{r}<0$ (attractive electric force), the difference
between the periods of prograde and retrograde orbits is 
\begin{equation}
\Delta t_{{\rm orb}}=2\pi\left(\frac{1}{\Omega_{{\rm orb+}}}+\frac{1}{\Omega_{{\rm orb-}}}\right)=-\frac{2\pi A_{\phi,r}}{E_{r}}=-2\pi\frac{\star B_{r\phi}}{E_{r}}\ ,\label{eq:EMClock}
\end{equation}
where $\star B_{jk}\equiv\epsilon_{ijk}B^{i}$ is the 2-form dual
to the magnetic field $\vec{B}$ (not to be confused with the magnetic
tidal tensor $B_{\alpha\beta}$). In cylindrical coordinates, $\star B_{r\phi}=B^{z}$;
in spherical coordinates, $\star B_{r\phi}=-B^{\theta}$. Notice the
analogy with Eq. (\ref{eq:GMClock}), identifying $\{\vec{E},\ \vec{B}\}\leftrightarrow\{\vec{G},\ \vec{H}\}$.
We can thus say that the gravitomagnetic clock effect in Eq. (\ref{eq:GMClock})
consists of a term with a direct electromagnetic analogue, plus a
term (the Sagnac time delay $4\pi\mathcal{A}_{\phi}$) that has no
electromagnetic counterpart in Eq. (\ref{eq:EMClock}).

\subsubsection*{Observer-independent ``two-clock'' effect\label{sub:Observer-independent-two-clock}}

The time delay (\ref{eq:GMClock}) corresponds to orbital periods
as seen by the laboratory observers (\ref{eq:uLab}), and measured
in \emph{coordinate} time {[}which can be converted into observer's
proper time via Eq. (\ref{eq:PropertimeCoordinateTime}){]}. Other
observers, rotating with respect to the laboratory observers, will
measure different periods, since, from their point of view, the closing
of the orbits occurs at different points. The effect can also be formulated
in terms of the orbital proper times \cite{COHENMashhoon1993,Bonnor_ClockEffect1999,SantosGEM}
(``two clock effect''); for a discussion of such alternative formulations
and their relationships, we refer to \cite{BiniJantzenMashhoon_Clock1}.
An \emph{observer independent} clock effect can however be derived,
based on the proper times ($\tau_{+}$ and $\tau_{-}$) measured by
each orbiting clock between the events where they meet \cite{Tartaglia_ClockI_CQG2000}.
Consider two oppositely rotating circular geodesics at some fixed
$r$, and set a starting meeting point at $\phi_{+}=\phi_{-}=0$,
$t=0$. The next meeting point is defined by $\phi_{+}=2\pi+\phi_{-}$.
Since $\phi_{\pm}=\Omega_{{\rm geo}\pm}t$, the meeting point occurs
at a coordinate time $t=2\pi/(\Omega_{{\rm geo}+}-\Omega_{{\rm geo}-})$.
Since $dt/d\tau_{\pm}=U_{\pm}^{0}$, and 
\begin{equation}
U_{\pm}^{0}=\left[-g_{00}-2\Omega_{{\rm geo}\pm}g_{0\phi}-\Omega_{{\rm geo}\pm}^{2}g_{\phi\phi}\right]^{-1/2}\label{eq:gammaPM}
\end{equation}
is constant along a circular orbit, it follows that $\tau_{\pm}=t/U_{\pm}^{0}$,
thus 
\begin{equation}
\tau_{\pm}=\frac{2\pi(U_{\pm}^{0})^{-1}}{\Omega_{{\rm geo}+}-\Omega_{{\rm geo}-}}\ ;\qquad\quad\Delta\tau\equiv\tau_{+}-\tau_{-}=2\pi\frac{(U_{+}^{0})^{-1}-(U_{-}^{0})^{-1}}{\Omega_{{\rm geo}+}-\Omega_{{\rm geo}-}}\ .\label{eq:InvClockEffect}
\end{equation}

\section{The electromagnetic analogue: the field of an infinite rotating charged
cylinder\label{sec:The-electromagnetic-analogue:}}

Consider, in flat spacetime, a charged, infinitely long rotating cylinder
along the $z-$axis. Its exterior electromagnetic field is described
(cf. e.g. \cite{Feynman:1963uxa}) by the 4-potential 1-form $A_{\alpha}=(-\phi,\mathbf{A})$,
\begin{equation}
\phi=-2\lambda\ln(r/r_{0})\ ;\qquad\mathbf{A}=A_{\phi}\mathbf{d}\phi=2\mathfrak{m}\mathbf{d}\phi\qquad(\vec{A}=\frac{2\mathfrak{m}}{r^{2}}\partial_{\phi})\ ,\label{eq:EMphi_and_A}
\end{equation}
where $\phi\equiv\phi(r)$ is the electric potential, $r_{0}$ is
an arbitrary constant, $\vec{A}$ is the (3-D) magnetic vector potential
(with associated 1-form $\mathbf{A}$), and $\lambda$ and $\mathfrak{m}$
are, respectively, the charge and magnetic moment per unit $z-$length.
The corresponding electric and magnetic fields read 
\begin{equation}
\vec{E}=-\nabla\phi=\frac{2\lambda}{r}\partial_{r}\ ;\qquad\vec{B}=\nabla\times\vec{A}=0\ .\label{eq:EB}
\end{equation}
The magnetic tidal tensor $B_{\alpha\beta}$ also vanishes trivially
since $\vec{H}=0$ for an inertial frame, cf. Eq. (\ref{eq:BijGEM}).
Hence, the electromagnetic field of a rotating cylinder differs from
that of a static one \emph{only in the vector potential} $\vec{A}$,
which vanishes in the latter case ($\mathfrak{m}=0$).

\subsection{Aharonov-Bohm effect\label{sub:Aharonov-Bohm-effect}}

\begin{figure}
\includegraphics[width=0.85\textwidth]{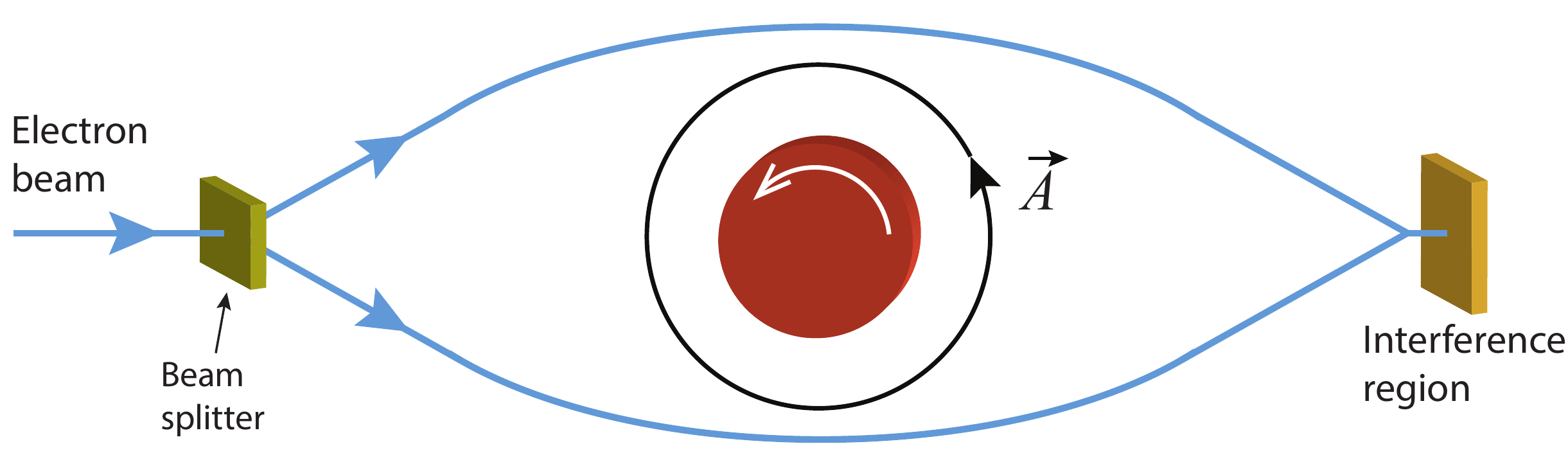}

\protect\protect\protect\caption{\label{fig:Aharonov}Simplified scheme for the Aharonov-Bohm effect
around a rotating charged cylinder. }
\end{figure}

Classically, the physics in the exterior field of a rotating cylinder
are the same as for a static one, since $\vec{A}$ itself plays no
role in any physical process (only its curl $\vec{B}$). In other
words, \emph{classically}, an irrotational vector potential $\vec{A}$
is pure gauge. Quantum theory, however, changes the picture, since
$\vec{A}$ intervenes physically in the so-called Aharonov-Bohm effect
\cite{AharonovBohm}. This effect can be described as follows. The
wave function of a particle of charge $q$ moving in a stationary
electromagnetic field along a spatial path $C$ acquires a phase shift
given by $\varphi=q/\hbar\int_{C}\mathbf{A}\equiv q/\hbar\int_{C}\vec{A}\cdot d\vec{l}$
\cite{AharonovBohm}. Now consider, as in Fig. \ref{fig:Aharonov},
a beam of electrons which is split into two, each passing around a
rotating charged cylinder but on opposite sides (while avoiding it).
Since $\vec{A}$ circulates around the cylinder, that will lead to
a phase difference between the beams, which manifests itself in a
shift of the interference pattern when the beams are recombined. 

Let $\varphi_{+}$ ($\varphi_{-}$) denote the phase shift induced
in the beam propagating in the same (opposite) sense of the cylinder's
rotation. Since the field lines of $\vec{A}$ are circles around the
cylinder, the phase shifts in the two paths are of equal magnitude
but opposite sign: $\varphi_{+}=-\varphi_{-}$. The two paths together
form a closed loop; since $\nabla\times\vec{A}=0$ outside the cylinder,
by the Stokes theorem $\int_{C}\mathbf{A}$ is the same for every
closed spatial loop $C$ enclosing the cylinder (in particular, a
circular one); hence, the phase difference between the two paths,
$\Delta\varphi=\varphi_{+}-\varphi_{-}$, equals the phase shift along
any circular loop $C$ enclosing the cylinder 
\begin{equation}
\Delta\varphi=\frac{q}{\hbar}\oint_{C}\mathbf{A}=\frac{q}{\hbar}A_{\phi}\int_{0}^{2\pi}d\phi=\frac{2\pi q}{\hbar}A_{\phi}\ .\label{eq:dphiAharonov}
\end{equation}
Notice the formal analogy with the expression (\ref{eq:DtBigLoop})
for the Sagnac effect on a circular loop around the axis of an axistationary
metric: $\Delta t$ therein corresponds to a difference in beam arrival
times for one full loop; for a half loop {[}as is the case in Eq.
(\ref{eq:dphiAharonov}){]} the time difference is $\Delta t/2$,
corresponding to a phase difference $\Delta\varphi=(E/\hbar)\Delta t/2=(2\pi E/\hbar)\mathcal{A}_{\phi}$,
where $E$ denotes the photon's energy. Identifying $\{E,\mathcal{A}_{\phi}\}\leftrightarrow\{q,A_{\phi}\}$,
this \emph{exactly} matches (\ref{eq:dphiAharonov}).

For comparison with the gravitational analogue below, it is worth
observing the following. The fact that $\int_{C}\mathbf{A}\ne0$ for
loops $C$ enclosing the cylinder is, in connection with Stokes' theorem,
assigned to the fact that, within the cylinder, $\nabla\times\vec{A}=\vec{B}\ne0$.
However, one could as well restrict our analysis to the region outside
the cylinder (as originally done in \cite{AharonovBohm}); that is,
cut the cylinder out of the space manifold, and consider the field
$\vec{A}$ defined in the multiply connected region thereby obtained.
The fact that $\int_{C}\mathbf{A}\ne0$, in spite of $\mathbf{d}\mathbf{A}=0$
everywhere, is then explained through the fact that $C$ lies in a
region which is not simply connected \cite{Stachel:1981fg}, where
$\mathbf{A}$ is a closed but non-exact form: in this case Stokes'
theorem does not require its circulation to vanish, but only to have
the same value $2\pi A_{\phi}$ for any closed loop $C$ around the
cylindrical hole, cf. Sec. \ref{sub:Closed-forms,-exact}.

\subsection{Rotating frame\label{sub:EMRotating-frame}}

Consider the coordinate system $\{x^{\bar{\alpha}}\}$, obtained from
the globally inertial coordinate system $\{x^{\alpha}\}$ by the transformation
$x^{\bar{i}\ne\bar{\phi}}=x^{i}$, $\bar{\phi}=\phi-\Omega t$, corresponding
to a reference frame rotating with angular velocity $\Omega$ about
the cylinder's axis. The Minkowski metric in such coordinates reads
\begin{equation}
ds^{2}=(-1+\Omega^{2}r^{2})dt^{2}+dr^{2}+2\Omega r^{2}dtd\bar{\phi}+dz^{2}+r^{2}d\bar{\phi}^{2}\ .\label{eq:RotatingMetric}
\end{equation}
The 4-potential 1-form $A_{\bar{\alpha}}=\Lambda_{\ \bar{\alpha}}^{\beta}A_{\beta}$
($\Lambda_{\ \bar{\alpha}}^{\beta}\equiv\partial x^{\beta}/\partial x^{\bar{\alpha}}$)
becomes, in such coordinates, $A_{\bar{\alpha}\ne\bar{t}}=A_{\alpha}$,
$A_{\bar{t}}=-\phi+A_{\phi}\Omega$. The electric and magnetic fields
are given by $\bar{E}^{\bar{\alpha}}=F^{\bar{\alpha}\bar{\beta}}\bar{u}_{\bar{\beta}}$
and $\bar{B}^{\bar{\alpha}}=\star F^{\bar{\alpha}\bar{\beta}}\bar{u}_{\bar{\beta}}$,
where $F_{\alpha\beta}\equiv2A_{[\beta,\alpha]}=2A_{[\beta;\alpha]}$
is the Faraday 2-form and $\bar{u}^{\bar{\alpha}}=(-g_{\bar{0}\bar{0}})^{-1/2}\delta_{\bar{0}}^{\bar{\alpha}}$
is the 4-velocity of the observers at rest in the rotating coordinates;
they read 
\[
\vec{\bar{E}}=\frac{2\lambda}{r\sqrt{1-r^{2}\Omega^{2}}}\partial_{r}\ ;\qquad\vec{\bar{B}}=\frac{2\Omega\lambda}{\sqrt{1-r^{2}\Omega^{2}}}\partial_{z}
\]
($\bar{E}^{\bar{0}}=\bar{B}^{\bar{0}}=0$). So now a non-vanishing
magnetic $\vec{\bar{B}}$ field arises.

Finally, observe that the curves of constant spatial coordinates $x^{\bar{i}}$,
tangent to the Killing vector field $\partial_{\bar{t}}$, cease to
be time-like for $r>1/\Omega$, since $g_{\bar{0}\bar{0}}>0$ therein.
Hence, no observers $\bar{u}^{\bar{\alpha}}$, at rest with respect
to such frame, can exist past that value of $r$ (they would exceed
the speed of light).

\section{Gravitational field of infinite rotating cylinders: the Lewis metrics\label{sec:lewis metrics}}

The exterior gravitational field of an infinitely long, rotating or
non-rotating cylinder is generically described by the Lewis metric
\cite{SantosGRG1995} 
\begin{equation}
ds^{2}=-fdt^{2}+2kdtd\phi+r^{(n^{2}-1)/2}(dr^{2}+dz^{2})+ld\phi^{2}\ ;\label{eq:LewisMetric}
\end{equation}
\begin{equation}
f=ar^{1-n}-\frac{c^{2}r^{n+1}}{n^{2}a}\ ;\qquad k=-Cf\ ;\qquad l=\frac{r^{2}}{f}-C^{2}f\ ;\qquad C=\frac{cr^{n+1}}{naf}+b\ .\label{eq:LewisFunctions}
\end{equation}
Here $a$, $b$, $c$, and $n$ are constants, which can be real or
complex, corresponding, respectively, to the Weyl or Lewis classes
of solutions. The constant $n$, in particular, is real for the Weyl
class, and purely imaginary for the Lewis class \cite{StephaniExact,SantosCQG1995}.
This is the most general stationary solution of the vacuum Einstein
field equations with cylindrical symmetry. It encompasses, as special
cases, the van Stockum \cite{Stockum1938} exterior solutions for
the field produced by a rigidly rotating dust cylinder, and the static
Levi-Civita metric, corresponding to a non-rotating cylinder.

\emph{Curvature invariants.---} As is well known (e.g. \cite{StephaniExact,Invariants}),
in vacuum there are four independent scalar invariants one can construct
from the Riemann tensor (which equals the Weyl tensor): two quadratic,
namely the Kretchmann and Chern-Pontryagin invariants, which read,
for the metric (\ref{eq:LewisMetric}),

\begin{equation}
\mathbf{R}\cdot\mathbf{R}\equiv R_{\alpha\beta\gamma\delta}R^{\alpha\beta\gamma\delta}=\frac{1}{4}(n^{2}-1)^{2}(3+n^{2})r^{-3-n^{2}};\qquad\star\!\mathbf{R}\cdot\mathbf{R}\equiv\star R_{\alpha\beta\gamma\delta}R^{\alpha\beta\gamma\delta}=0\ ,\label{eq:QuadInvariants}
\end{equation}
plus the two cubic invariants 
\begin{eqnarray}
\mathbb{A} & \equiv & -\frac{1}{16}R_{\ \ \lambda\mu}^{\alpha\beta}R_{\ \ \ \rho\sigma}^{\lambda\mu}R_{\ \ \alpha\beta}^{\rho\sigma}=\frac{3}{256}(n^{2}-1)^{4}r^{-3(n^{2}+3)/2}\ ;\label{eq:AInv}\\
\mathbb{B} & \equiv & \frac{1}{16}R_{\ \ \lambda\mu}^{\alpha\beta}R_{\ \ \ \rho\sigma}^{\lambda\mu}\star\!R_{\ \ \alpha\beta}^{\rho\sigma}=0\ .\label{eq:BInv}
\end{eqnarray}

\subsection{Gravitoelectromagnetic (GEM) fields and tidal tensors}

The metric (\ref{eq:LewisMetric})-(\ref{eq:LewisFunctions}) can
be put in the form (\ref{eq:AxistatMetric}), with 
\begin{equation}
e^{2\Phi}=f;\qquad\mathcal{A}_{\phi}=-\frac{cr^{n+1}}{naf}-b\ ;\qquad h_{rr}=h_{zz}=r^{(n^{2}-1)/2};\qquad h_{\phi\phi}=r^{2}e^{-2\Phi},\label{eq:LewisQM1}
\end{equation}
and $h_{ij}=0$ for $i\ne j$. The gravitoelectric and gravitomagnetic
fields read, cf. Eqs. (\ref{eq:GEMFieldsQM}),
\begin{equation}
G_{i}=\frac{a^{2}(n-1)n^{2}+c^{2}(1+n)r^{2n}}{2r(a^{2}n^{2}-c^{2}r^{2n})}\delta_{i}^{r};\qquad\vec{H}=\frac{2acn^{2}r^{-(n-1)^{2}/2}}{c^{2}r^{2n}-a^{2}n^{2}}\partial_{z}\ .\label{eq:GEMLewis}
\end{equation}
Equation (\ref{eq:QMGeo}) then tells us that test particles in geodesic
motion are, from the point of view of the reference frame associated
to the coordinate system of (\ref{eq:LewisMetric}), acted upon by
two inertial forces: a radial force (per unit mass) $\gamma^{2}\vec{G}$,
which can be attractive or repulsive (depending on $n$, $a$, and
$c$), and a gravitomagnetic force (per unit mass) $\gamma\vec{U}\times\vec{H}$,
likewise lying on the plane orthogonal to the cylinder. A consequence
of the latter is that test particles dropped from rest are deflected
sideways, instead of moving radially. The non-vanishing $\vec{H}$
means also that gyroscopes precess relative to the frame associated
to the coordinates of (\ref{eq:LewisMetric}), cf. Eq. (\ref{eq:SpinPrec}).

The non-vanishing components of the gravitoelectric $\mathbb{E}_{\alpha\beta}\equiv R_{\alpha\mu\beta\nu}u^{\mu}u^{\nu}$
and gravitomagnetic $\mathbb{H}_{\alpha\beta}\equiv\star R_{\alpha\mu\beta\nu}u^{\mu}u^{\nu}$
tidal tensors as measured by the observers at rest in the coordinates
of (\ref{eq:LewisMetric}) read 
\begin{eqnarray}
\mathbb{E}_{rr} & = & -(1-n^{2})\frac{a^{2}n^{2}(n+1)+c^{2}(n-1)r^{2n}}{8r^{2}(a^{2}n^{2}-c^{2}r^{2n})}\ ;\qquad\mathbb{E}_{\phi\phi}=(1-n^{2})\frac{an^{2}r^{-(n-1)^{2}/2}}{4(a^{2}n^{2}-c^{2}r^{2n})}\ ;\nonumber \\
\mathbb{E}_{zz} & = & (1-n^{2})\frac{a^{2}n^{2}(n-1)+c^{2}(n+1)r^{2n}}{8r^{2}(a^{2}n^{2}-c^{2}r^{2n})}\ ,\label{eq:LewisEab}
\end{eqnarray}
\begin{equation}
\mathbb{H}_{rz}=\mathbb{H}_{zr}=-\frac{(1-n^{2})acn^{2}r^{n-2}}{4(c^{2}r^{2n}-a^{2}n^{2})}\ .\label{eq:LewisHab}
\end{equation}
The fact that $\mathbb{H}_{\alpha\beta}\ne0$ means that a spin-curvature
force (\ref{eq:SpinCurvature}) is exerted on gyroscopes at rest in
the coordinates of (\ref{eq:LewisMetric}).

Comparing with the electromagnetic analogue, we observe that both
the inertial and tidal fields (in particular the non-vanishing $\vec{H}$
and $\mathbb{H}_{\alpha\beta}$) are in contrast with the electromagnetic
field of a rotating cylinder as measured in the inertial rest frame,
discussed in Sec. \ref{sec:The-electromagnetic-analogue:}, resembling
more the situation in a rotating frame, discussed in Sec. \ref{sub:EMRotating-frame}.

\subsubsection{The Levi-Civita static cylinder\label{sub:The-Levi-Civita-static}}

It is known \cite{SantosGRG1995,SantosCQG1995} that when $n$ is
real (Weyl class), and\footnote{We shall see in Sec. \ref{sub:The-canonical-form} that the condition
$c=0$ is actually not necessary.} $b=0=c$, the metric (\ref{eq:LewisMetric}) becomes 
\begin{equation}
ds^{2}=-ar^{1-n}dt^{2}+r^{(n^{2}-1)/2}(dr^{2}+dz^{2})+\frac{r^{1+n}}{a}d\phi^{2}\ ,\label{eq:Levi-Civita}
\end{equation}
which is the static Levi-Civita metric, corresponding to a non-rotating
cylinder. Imposing $a>0$ (so that $t$ remains the time coordinate,
and $\phi$ the spacelike periodic coordinate) yields, identifying
the appropriate constants, a dimensionless version of the original
line element in \cite{Levi-Civita2011}. Further re-scaling the time
coordinate $t\rightarrow a^{-1/2}t$, so that $g_{00}=-r^{1-n}$,
yields the line element in the coordinate system in \cite{MacCallumEdnote,SantosGRG1995,Bronnikov:2019clf}.
For this metric we have {[}cf. Eqs. (\ref{eq:StatMetric}), (\ref{eq:GEMFieldsQM}),
(\ref{eq:HijGEM}){]} $\mathcal{A}_{i}=\vec{H}=\mathbb{H}_{\alpha\beta}=0$,
and 
\begin{equation}
\Phi=\frac{1-n}{2}\ln(r)+K;\qquad G_{i}=-\Phi_{,i}=-\frac{1-n}{2r}\delta_{i}^{r}\ ,\label{eq:LeviCivitaGEM}
\end{equation}
where $K$ is an arbitrary constant (depending on the choice of units
for $t$). That is, the Newtonian potential $\Phi$ and the gravitoelectric
field 1-form $G_{i}$ exactly match, with the identification $(1-n)/4\leftrightarrow\lambda$,
\emph{minus} the electrostatic potential $\phi$ and electric field
1-form $E_{i}$ of the electromagnetic analogue in Sec. \ref{sec:The-electromagnetic-analogue:},
cf. Eqs. (\ref{eq:EMphi_and_A})-(\ref{eq:EB}). This analogy suggests
identifying the quantity $(1-n)/4$ with the source's mass density
per unit $z-$length, in agreement with earlier interpretations (e.g.
\cite{Bonnor1992,SantosGRG1995,SantosGEM,Bronnikov:2019clf}). The
speed of the circular geodesics with respect\footnote{\label{fn:Relative-velocity}The relative velocity $v^{\alpha}$ of
a test particle of 4-velocity $U^{\alpha}$ with respect to an observer
of 4-velocity $u^{\alpha}$ is given by \cite{ManyFaces,Invariants,Costa:2012cy}
$U^{\alpha}=\gamma(u^{\alpha}+v^{\alpha})$, with $\gamma\equiv-u^{\alpha}U_{\alpha}=(1-v^{2})^{-1/2}$.
Relative to a ``laboratory'' observer at rest in the given coordinate
system, $u^{\alpha}=(-g_{00})^{-1/2}\delta_{0}^{\alpha}$, its magnitude
$v$ is simply given by $(1-v^{2})^{-1/2}=-u^{0}U_{0}$.} to the coordinate system in (\ref{eq:Levi-Civita}) is 
\begin{equation}
v_{{\rm geo}}=\sqrt{\frac{2}{1+n}-1}\label{eq:vgeoLC}
\end{equation}
(cf. e.g. \cite{Bronnikov:2019clf}), which is independent of $r$
(like in the Newtonian/electric analogues, albeit with a different
value). These are possible only when $0\le n<1$ \cite{GautreauHoffman69,GriffithsPodolsky2009,BonnorMartins1991,Bonnor1992,SantosLC,Bronnikov:2019clf},
since $\vec{G}$ is attractive only for $n<1$, and their speed becomes
superluminal for $n<0$.

\subsubsection{The ``force'' parallel to the axis\label{sub:The-force-parallel}}

In some literature \cite{HerreraSantosGeoLewis,Bronnikov:2019clf}
it was found that test particles in geodesic motion appeared to be
deflected by a rather mysterious ``force'' parallel to the cylinder's
axis. Let us examine the origin of such effect. It follows from Eqs.
(\ref{eq:QMGeo}) and (\ref{eq:GEMLewis}) that, in the reference
frame associated to the coordinate system of (\ref{eq:LewisMetric}),
the only inertial forces acting on a test particle in geodesic motion
are the radial gravitoelectric force $m\vec{G}$ and the gravitomagnetic
force $m\gamma\vec{U}\times\vec{H}$ (with $\vec{H}$ parallel to
the axis), both always orthogonal to the cylinder's axis. It is thus
clear that no axial inertial force exists. {[}In other words, the
3-D curve obtained by projecting the geodesic onto the space manifold
$(\Sigma,h)$ has no axial acceleration, cf. Sec. \ref{sec:Gravitomagnetism-Levels}{]}.
The axial component of Eq. (\ref{eq:QMGeo}) reads 
\begin{equation}
\frac{\tilde{D}U^{z}}{d\tau}=0\ \Leftrightarrow\ \frac{dU^{z}}{d\tau}=-2\Gamma(h)_{rz}^{z}U^{r}U^{z}=\frac{1-n^{2}}{2r}U^{r}U^{z}\ ,\label{eq:AxialAccel}
\end{equation}
which is Eq. (74) of \cite{HerreraSantosGeoLewis}. That is, the coordinate
``acceleration'' $d^{2}z/d\tau^{2}\equiv dU^{z}/d\tau$ is down
to the fact that the Christoffel symbol $\Gamma(h)_{rz}^{z}$ of the
spatial metric $h_{ij}$ is non-zero. In particular, if $U^{z}\equiv dz/d\tau$
is initially zero, it actually remains zero along the motion. The
question then arises on whether the effect is due to the curvature
of the space manifold $(\Sigma,h)$ or to a coordinate artifact, as
both are generically encoded in $\Gamma(h)_{jk}^{i}$. The distinction
between the two is not clear in general (an example of a pure coordinate
artifact is the variation of $U^{i}$ when one describes geodesic
motion in flat spacetime using a non-Cartesian coordinate system,
e.g. spherical coordinates). In the present case, however, the translational
Killing vector $\partial_{z}$ gives us a notion of fixed axial direction;
on the other hand, the dependence of $g_{zz}$ on $r$ (i.e., the
fact that the magnitude $\sqrt{g_{zz}}$ of the basis vector $\partial_{z}$
is not constant along the particle's trajectory) causes the coordinate
acceleration $dU^{z}/d\tau$ to include a trivial coordinate artifact.
Such effect is gauged away by switching to an orthonormal tetrad frame
$\mathbf{e}_{\hat{\alpha}}$ such that $\mathbf{e}_{z}=(g_{zz})^{-1/2}\partial_{z}$,
where the axial component of the 4-velocity reads $U^{\hat{z}}=U^{z}\sqrt{g_{zz}}$.
It evolves as {[}using (\ref{eq:AxialAccel}){]} $dU^{\hat{z}}/d\tau=(1-n^{2})r^{(n^{2}-5)/4}U^{r}U^{z}/4$,
which corresponds to the axial component of the geodesic equation
written in such tetrad, $DU^{\hat{\alpha}}/d\tau=0$. Hence, even
in an orthonormal frame $\mathbf{e}_{\hat{\alpha}}$, $U^{\hat{z}}$
is not constant; in other words, the axial vector component itself,
$U^{z}\partial_{z}$ (not just the coordinate component $U^{z}$),
varies along the particle's motion. This is a consequence of the curvature
of the space manifold. We conclude thus that $dU^{z}/d\tau$ in Eq.
(\ref{eq:AxialAccel}), interpreted in \cite{HerreraSantosGeoLewis,Bronnikov:2019clf}
as an axial ``force'', consists of a combination of a coordinate
artifact caused by the variation of the basis vector $\partial_{z}$
along the particle's trajectory, with a physical effect due to the
curvature of the space manifold $(\Sigma,h)$.

\subsection{The canonical form of the Weyl class\label{sub:The-canonical-form}}

The Weyl class corresponds to all parameters in Eqs. (\ref{eq:LewisMetric})-(\ref{eq:LewisFunctions})
being real. We observe that, for $r^{2n}>a^{2}n^{2}/c^{2}$, the Killing
vector field $\partial_{t}$ ceases to be time-like; thus no physical
observers $u^{\alpha}=f^{-1/2}\partial_{t}^{\alpha}$, at rest in
the coordinates of (\ref{eq:LewisMetric}), can exist past that value
of $r$. This resembles the situation for a rotating frame in flat
spacetime, see Sec. \ref{sub:EMRotating-frame}.\textcolor{blue}{{}
}Moreover, the non-vanishing gravitomagnetic field $\vec{H}$ and
tidal tensor $\mathbb{H}_{\alpha\beta}$, Eqs. (\ref{eq:GEMLewis})
and (\ref{eq:LewisHab}), contrast with the electromagnetic problem
of a charged rotating cylinder as viewed from the inertial rest frame
(Sec. \ref{sec:The-electromagnetic-analogue:}); they resemble instead
the corresponding electromagnetic fields when measured in a rotating
frame (Sec. \ref{sub:EMRotating-frame}). This makes one wonder whether
these two features might be mere artifacts of some trivial rotation
of the coordinate system in which the metric, in its usual form (\ref{eq:LewisMetric})-(\ref{eq:LewisFunctions}),
is written. In what follows we shall argue this to be the case.

For the Weyl class, we have the invariant structure, cf. Eqs. (\ref{eq:QuadInvariants})-(\ref{eq:BInv}):
$\mathbf{R}\cdot\mathbf{R}>0$, $\star\!\mathbf{R}\cdot\mathbf{R}=0$,
$\mathbb{B}=0$, 
\begin{equation}
\mathbb{M}\equiv\frac{I^{3}}{(\mathbb{A}-i\mathbb{B})^{2}}-6=\frac{2n^{2}(n^{2}-9)^{2}}{9(n^{2}-1)^{2}}\ge0\quad\mbox{(real)},\label{eq:MInvariant}
\end{equation}
where $I\equiv(\mathbf{R}\cdot\mathbf{R}-i\star\!\mathbf{R}\cdot\mathbf{R})/8$.
We shall see below (Secs. \ref{sub:Komar-Canonical} and \ref{sub:GEM-fields-Canonical})
that, in order for the cylinder's Komar mass per unit length to be
positive and its gravitational field attractive, while at the same
time allowing for circular geodesics, it is necessary that $n^{2}<1$:
for larger values of $|n|$ the physical significance of the solutions
is unclear already in the static Levi-Civita special case, no longer
representing the gravitational field of a cylindrical source \cite{GriffithsPodolsky2009,MacCallumEdnote,Bonnor1992,GautreauHoffman69,BonnorMartins1991,SantosLC,BonnorDavidson1992,Bronnikov:2019clf,BonnorStaticCylinder1999,Philbin_1996}.
Moreover, for $n=0$ the metric coefficients diverge. We consider
thus the range $0<n^{2}<1$ for solutions of physical interest to
the problem at hand. In this case, $\mathbb{M}>0$; this, together
with the above relations on the quadratic invariants, implies that
the spacetime is purely ``electric'' \cite{McIntosh_1994,Invariants,StephaniExact,WyllemanBergh2006},
i.e., everywhere there are observers for which the magnetic part of
the Weyl tensor ($=\mathbb{H}_{\alpha\beta}$, in vacuum) vanishes.
They imply also that the Petrov type is I, and thus at each point
the observer measuring $\mathbb{H}_{\alpha\beta}=0$ is unique \cite{Invariants}.
Let us find such observer congruence. The nontrivial components of
the gravitomagnetic tidal tensor as measured by an observer of arbitrary
4-velocity $U^{\alpha}=(U^{t},U^{r},U^{\phi},U^{z})$ read ($\mathbb{H}_{\alpha\beta}=\mathbb{H}_{\beta\alpha})$:
\begin{eqnarray*}
\mathbb{H}_{tt} & = & \frac{\alpha cU^{r}U^{z}}{2r};\qquad\mathbb{H}_{tr}=\frac{\alpha(\beta U^{\phi}-2cU^{t})U^{z}}{8r};\qquad\mathbb{H}_{t\phi}=\frac{\alpha(2bc-n)U^{r}U^{z}}{4r};\\
\mathbb{H}_{tz} & = & \frac{\alpha\left[\chi U^{\phi}-2cU^{t}\right]U^{r}}{8r};\qquad\mathbb{H}_{r\phi}=\frac{\alpha\left[2b\xi U^{\phi}+\chi U^{t}\right]U^{z}}{8r};\qquad\mathbb{H}_{\phi\phi}=-\frac{\alpha b\xi U^{r}U^{z}}{2r};\\
\mathbb{H}_{rz} & = & \frac{\alpha(bU^{\phi}+U^{t})(cU^{t}-\xi U^{\phi})}{4r};\qquad\mathbb{H}_{\phi z}=\frac{\alpha\left[2b\xi U^{\phi}+\beta U^{t}\right]U^{r}}{8r},
\end{eqnarray*}
where $\alpha\equiv1-n^{2}$, $\beta\equiv n-2bc-3$, $\chi\equiv n-2bc+3$
and $\xi\equiv n-bc$. The condition $\mathbb{H}_{\alpha\beta}=0$
implies $U^{r}=U^{z}=0$, plus one of the following conditions:
\begin{equation}
\frac{U^{\phi}}{U^{t}}=-\frac{1}{b}\quad({\rm i})\qquad{\rm or}\qquad\frac{U^{\phi}}{U^{t}}=\frac{c}{n-bc}\ .\quad({\rm ii})\label{eq:vphi}
\end{equation}
Notice that these conditions cannot hold simultaneously for 4-velocities
$U^{\alpha}=U^{t}\partial_{t}^{\alpha}+U^{\phi}\partial_{\phi}^{\alpha}$:
condition (\ref{eq:vphi}i) leads to a vector which is time-like \emph{iff}
$a<0$, whereas (\ref{eq:vphi}ii) leads to a vector which is time-like
\emph{iff} $a>0$. Hence, for any given values of the parameters ($a\ne0$,
$b$, $c$, $n$), there is one, and only one, congruence of observers
for which $\mathbb{H}_{\alpha\beta}=0$; such congruence has 4-velocity
\begin{equation}
U^{\alpha}=U^{t}(\partial_{t}^{\alpha}+\Omega\partial_{\phi}^{\alpha})\ ;\qquad\Omega=\frac{c}{n-bc}\ \ {\rm for}\ a>0\ ;\qquad\Omega=-\frac{1}{b}\ \ {\rm for}\ a<0\ ,\label{eq:PrincipalObs}
\end{equation}
consisting of observers \emph{rigidly} rotating around the cylinder
with \emph{constant} angular velocity\footnote{The angular velocities (\ref{eq:vphi}) coincide with those for which
gyroscopes do not precess, previously found in \cite{HerreraSantosGravitomagnetic};
however, the significance of such result remained unclear then. } $\Omega\equiv U^{\phi}/U^{t}\equiv d\phi/dt$.

Since $\Omega$ is constant, by using the coordinate transformation
\begin{equation}
\bar{\phi}=\phi-\Omega t\ ;\qquad x^{\bar{\alpha}\ne\bar{\phi}}=x^{\alpha}\label{eq:TransformCanonical}
\end{equation}
one can switch to a coordinate system where the observers (\ref{eq:PrincipalObs})
are at rest. In such a coordinate system, the metric (\ref{eq:LewisMetric})
becomes 
\begin{equation}
ds^{2}=-\bar{f}(dt+\bar{C}d\bar{\phi})^{2}+r^{(n^{2}-1)/2}(dr^{2}+dz^{2})+\frac{r^{2}}{\bar{f}}d\bar{\phi}^{2}\ ,\label{eq:Canonical}
\end{equation}
with, for $\Omega=c/(n-bc)$, 
\begin{equation}
\bar{f}=\frac{r^{1-n}}{\alpha}\ ;\qquad\alpha=\frac{\bar{C}^{2}}{ab^{2}}\ ;\qquad\bar{C}=b\frac{n-bc}{n}\ ,\label{eq:CanonicalFunctions}
\end{equation}
and, for $\Omega=-1/b$, 
\begin{equation}
\bar{f}=\frac{r^{1+n}}{\alpha}\ ;\qquad\alpha=-ab^{2}\ ;\qquad\bar{C}=-b\frac{n-bc}{n}\ .\label{eq:Canonical2Functions}
\end{equation}
The special cases $bc=n$ and $b=0$, which are excluded from, respectively,
the former and the latter transformations, both lead to the Levi-Civita
line-element. That it is so for $b=0$ can be immediately seen by
substituting in (\ref{eq:Canonical})-(\ref{eq:CanonicalFunctions}),
yielding directly (\ref{eq:Levi-Civita}). This has previously been
noticed in \cite{SantosGEM,HerreraSantosGravitomagnetic}, by a different
route. That $n=bc$ also leads to the Levi-Civita metric (which seems
to have gone unnoticed in the literature) can be seen by substituting
$n\rightarrow bc$ in the expression for $\bar{C}$ in (\ref{eq:Canonical2Functions})
and: (i), for $a<0$, by substituting in the remainder $n\equiv-m$,
yielding (\ref{eq:Levi-Civita}) in a different notation (with $\{m,\alpha^{-1}\}$
in the place of $\{n,a\}$); (ii) for $a>0$, by substituting $\alpha=-|\alpha|$,
yielding (\ref{eq:Levi-Civita}) with $t$ and $\phi'$ swapped ($t$
the angular coordinate and $\phi'$ the temporal coordinate, and $|\alpha|$
in the place of $a$).

Notice the simplicity of these forms of the metric, comparing to the
usual form (\ref{eq:LewisMetric})-(\ref{eq:LewisFunctions}). In
particular, we remark that $\bar{C}$ is a constant (we shall see
below the importance of this result), and that these metrics depend
only on three effective parameters: $\alpha$, $n$, and $\bar{C}$.
This makes explicit the assertion in \cite{MacCallumSantos1998} that
the four parameters ($a$, $b$, $c$, $n$) in the usual form of
the metric are not independent. Observe moreover that, contrary to
the situation in the usual form, the Killing vector $\partial_{t}$
is \emph{everywhere time-like}, that is, $g_{00}<0$ for all $r$
{[}for $a>0$ in (\ref{eq:CanonicalFunctions}), and $a<0$ in (\ref{eq:Canonical2Functions}){]}.
Therefore, physical observers $u^{\alpha}=\bar{f}^{-1/2}\partial_{t}^{\alpha}$,
at rest in the coordinates of (\ref{eq:Canonical}), exist everywhere
(even for arbitrarily large $r$).

\subsubsection{Komar Integrals\label{sub:Komar-Canonical}}

Infinite cylinders are not isolated sources, hence a conserved \emph{total}
mass or angular momentum (which is infinite) cannot be defined for
bounded hypersurfaces. In these systems one can define instead a mass
and angular momentum per unit length, obeying conservation laws analogous
to those of $M$ and $J$ for finite sources. In order to effectively
suppress the irrelevant $z-$coordinate from the problem, consider
simply connected tubes $\mathcal{V}$ parallel to the $z-$axis, of
unit $z-$length and arbitrary section. Let $\partial\mathcal{V}=\mathcal{S}\cup\mathcal{B}_{1}\cup\mathcal{B}_{2}$
be the boundary of such tubes, where $\mathcal{S}$ is the tube's
lateral surface, parameterized by $\{\phi,z\}$, and $\mathcal{B}_{1}$
and $\mathcal{B}_{2}$ its bases (of disjoint union $\mathcal{B}_{1}\sqcup\mathcal{B}_{2}$),
lying in planes orthogonal to the $z-$axis and parametrized by $\{r,\phi\}$.
Since, by the equation $\mathbf{d}(\star\mathbf{d}\bm{\xi})=-2R_{\alpha\beta}\xi^{\beta}d\mathcal{V}^{\alpha}$
(see Sec. \ref{sub:Komar-Integrals}), $\star\mathbf{d}\bm{\xi}$
is a closed form outside the cylinder, by the Stokes theorem the Komar
integrals (\ref{eq:Komar}) vanish for all $\mathcal{V}$ exterior
to the cylinder, and are the same for all $\mathcal{V}$ enclosing
it. They are thus conserved quantities for such tubes. We can write
\begin{equation}
Q_{\xi}(\mathcal{V})=-\frac{K}{16\pi}\int_{\partial\mathcal{V}}\star\mathbf{d}\bm{\xi}=-\frac{K}{16\pi}\left[\int_{\mathcal{B}_{1}\sqcup\mathcal{B}_{2}}(\star d\xi)_{r\phi}drd\phi+\int_{\mathcal{S}}(\star d\xi)_{\phi z}d\phi dz\right]\ .\label{eq:KomarCylinder}
\end{equation}
In the coordinates of (\ref{eq:Canonical})-(\ref{eq:CanonicalFunctions})
for $a>0$, or (\ref{eq:Canonical2Functions}) for $a<0$, $\xi^{\alpha}=\partial_{t}^{\alpha}$
is everywhere time-like {[}contrary to the the usual form of the metric
(\ref{eq:LewisMetric})-(\ref{eq:LewisFunctions}){]}; it is actually
tangent to inertial observers at infinity, as we shall see below (Sec.
\ref{sub:GEM-fields-Canonical}). Hence, following the discussion
in Sec. \ref{sub:Komar-Integrals}, we argue that the corresponding
Komar integral has the physical interpretation of mass per unit length
($\lambda_{{\rm m}}$). Let us consider first the form (\ref{eq:Canonical})-(\ref{eq:CanonicalFunctions}).
Since the only non-trivial component of $\star\mathbf{d}\bm{\xi}$
is $(\star d\xi)_{\phi z}=1-n$, it follows that (with $K\rightarrow-2$)
\begin{equation}
\lambda_{{\rm m}}=Q_{\partial_{t}}(\mathcal{V})=\frac{1}{8\pi}\int_{\mathcal{S}}(\star d\xi)_{\phi z}d\phi dz=\frac{1-n}{8\pi}\int_{z=0}^{1}dz\int_{0}^{2\pi}d\phi=\frac{1-n}{4}\ .\label{eq:MassCylinder}
\end{equation}
It formally matches the Komar mass per unit length of the metric (\ref{eq:Levi-Civita})
for the Levi-Civita static cylinder. Actually, the fact that $\partial_{t}$
is everywhere time-like, and the reference frame asymptotically inertial,
puts the Weyl class metrics in equal footing with the Levi-Civita
solution, for which integral definitions of mass and angular momentum
have been put forth \cite{Bonnor1979_linemass,Israel1977,Marder1958,WangSilvaSantos1997},
and which amount to Komar integrals (or approximations to it, case
of the Hansen-Winicour \cite{HansenWinicour1975} integral in \cite{WangSilvaSantos1997}).

Equation (\ref{eq:MassCylinder}) has the interpretation of Komar
mass per unit length for $a>0$ {[}case in which $\partial_{t}$ in
(\ref{eq:Canonical})-(\ref{eq:CanonicalFunctions}) is time-like{]}.
Had we considered instead the form (\ref{eq:Canonical2Functions}),
we would obtain $\lambda_{{\rm m}}=(1+n)/4$, i.e., a similar expression
but with the sign of $n$ changed; this is the quantity that should
be interpreted as the Komar mass for $a<0$ {[}case in which $\partial_{t}$
in (\ref{eq:Canonical}), (\ref{eq:Canonical2Functions}) is time-like{]}.
In either case, $\lambda_{{\rm m}}>0$ for attractive gravitational
field, as we shall see in Sec. \ref{sub:GEM-fields-Canonical} below.

A subtlety concerning this result must however be addressed. Rescaling,
in (\ref{eq:Canonical}), $t=\mbox{\ensuremath{\kappa}}\tilde{t}$,
for some constant $\kappa$, yields an equivalent metric form with
a Killing vector $\partial_{\tilde{t}}=\kappa\partial_{t}$ tangent
to the same congruence of rest observers $u^{\alpha}$; however, $Q_{\partial_{\tilde{t}}}(\mathcal{V})=\kappa\lambda_{{\rm m}}$
no longer yields the correct mass per unit length. For the asymptotically
flat spacetimes of isolated sources, the arbitrariness in the normalization
of $\xi^{\alpha}$ is naturally eliminated by demanding $\xi^{\alpha}\xi_{\alpha}\stackrel{r\rightarrow\infty}{=}-1$,
i.e., by choosing coordinates such that $g_{00}\stackrel{r\rightarrow\infty}{=}-1$.
This is not possible, however, in the cylindrical metrics (\ref{eq:Canonical})-(\ref{eq:Canonical2Functions}),
where $g_{00}\stackrel{r\rightarrow\infty}{=}-\infty$. An alternative
route is as follows. Consider, for a moment, the spacetime to be globally
static (see Sec. \ref{sub:Local-vs-global}), so that $\xi^{\alpha}=\partial_{t}^{\alpha}$
is hypersurface orthogonal, and $\mathcal{V}$ lies on such hypersurfaces.
Using $\epsilon_{\ \ \alpha\beta}^{\mu\nu}\mathbf{d}x^{\alpha}\wedge\mathbf{d}x^{\beta}=-2d\mathcal{S}^{\mu\nu}$
\cite{Misner:1974qy}, where $d\mathcal{S}^{\mu\nu}$ is the area
2-form, equation (\ref{eq:Komar}), for $K=-2$, can be written as
(cf. \cite{Wald:1984,Townsend,NatarioMathRel}) 
\begin{equation}
Q_{\partial_{t}}(\mathcal{V})=-\frac{1}{8\pi}\int_{\partial\mathcal{V}}\xi_{\nu;\mu}d\mathcal{S}^{\mu\nu}=\frac{1}{4\pi}\int_{\partial\mathcal{V}}\xi_{\nu;\mu}n^{\nu}u^{\mu}d\mathcal{S}=-\frac{1}{4\pi}\int_{\partial\mathcal{V}}\sqrt{-g_{00}}G_{\nu}n^{\nu}d\mathcal{S}\ ,\label{eq:KomarG}
\end{equation}
where $u^{\alpha}=(-g_{00})^{-1/2}\partial_{t}^{\alpha}$ {[}cf. Eq.
(\ref{eq:uLab}){]}, $n^{\alpha}$ is the unit (outward pointing)
normal to $\partial\mathcal{V}$ which is orthogonal to $\xi^{\alpha}$
(so that $2n^{[\mu}u^{\nu]}$ is the normal bivector to $\partial\mathcal{V}$
\cite{Wald:1984}), $d\mathcal{S}$ is the area element on $\partial\mathcal{V}$,
and $G^{\alpha}$ is the gravitoelectric field as given in Eq. (\ref{eq:GEM Fields Cov}).
Equation (\ref{eq:KomarG}) is the relativistic generalization of
Gauss' theorem $M=-(1/4\pi)\int_{\mathcal{S}}\vec{G}_{{\rm N}}\cdot\vec{n}d\mathcal{S}$
($\vec{G}_{{\rm N}}=-\nabla\Phi_{{\rm N}}\equiv$ Newtonian gravitational
field) \cite{Wald:1984,Whittaker1935,NatarioMathRel}; in fact, for
an isolated source, $\Phi\stackrel{r\rightarrow\infty}{=}-Q_{\partial_{t}}(\mathcal{V})/r$,
yielding the ``Newtonian'' potential of the Komar mass $M=Q_{\partial_{t}}(\mathcal{V})$.
One can thus equivalently say that $\xi^{\alpha}$ is normalized so
that the Komar mass matches the ``active'' mass one infers from
$\Phi$ or $G_{i}=-\Phi_{,i}$ (namely their asymptotic behavior).
This is a criterion that translates to the case of infinite cylinders:
as we shall see in Sec. \ref{sub:GEM-fields-Canonical} below, $\lambda_{{\rm m}}$
matches precisely the mass per unit length inferred from $\Phi$ and
$G_{i}$, based now on their \emph{exact} behavior, as well as from
the comparison with the Newtonian (and electromagnetic) analogues.

The angular momentum per unit length, $j$, follows from substituting
$\xi^{\alpha}\rightarrow\zeta^{\alpha}=\partial_{\phi}^{\alpha}$
and $K\rightarrow1$ in Eq. (\ref{eq:KomarCylinder}). It is the same
for (\ref{eq:CanonicalFunctions}) or (\ref{eq:Canonical2Functions}),
as well as for the original form of the metric (\ref{eq:LewisMetric})-(\ref{eq:LewisFunctions}),
since $\partial_{\phi}=\partial_{\bar{\phi}}$ remains the same in
all cases. The non-trivial components of $\star\mathbf{d}\bm{\zeta}$
are $(\star d\zeta)_{zt}=1+n$ and $(\star d\zeta)_{\phi z}=2b(bc-n)$,
and so 
\begin{equation}
j=Q_{\partial_{\phi}}(\mathcal{V})=-\frac{1}{16\pi}\int_{\mathcal{S}}(\star d\zeta)_{\phi z}d\phi dz=\frac{1}{4}b(n-bc)\ .\label{eq:AngularMomentumCylinder}
\end{equation}

Had one chosen instead the Killing vector $\partial_{t}$ of the metric
in the usual coordinates (\ref{eq:LewisMetric})-(\ref{eq:LewisFunctions}),
one would obtain $\lambda'_{{\rm m}}=(1-n+2bc)/4=\lambda_{{\rm m}}-2\Omega'j$,
with $\Omega'=-\Omega$ the angular velocity of such coordinate system
relative to the star-fixed coordinates of (\ref{eq:Canonical}), given
by either of Eqs. (\ref{eq:PrincipalObs}), according to $\pm a>0$,
and $\lambda_{{\rm m}}=(1\mp n)/4$. The integral $\lambda'_{{\rm m}}$
no longer matches that of the Levi-Civita static cylinder; that $\lambda'_{{\rm m}}$
indeed should not be interpreted as the cylinder's Komar mass per
unit length is made evident by the fact that for $r^{2n}>a^{2}n^{2}/c^{2}$
such Killing vector field is not even time-like.

\subsubsection{The metric in terms of physical parameters --- ``canonical'' form
of the metric\label{sub:The-metric-in-3-parameters}}

The metric forms (\ref{eq:CanonicalFunctions}) and (\ref{eq:Canonical2Functions})
are actually two \emph{equivalent} facets of a more fundamental result.
As seen in Sec. \ref{sub:Komar-Canonical} above, in the case of (\ref{eq:CanonicalFunctions})
we have $n=1-4\lambda_{{\rm m}}$, whereas for (\ref{eq:Canonical2Functions})
we have $n=4\lambda_{{\rm m}}-1$; that is, in terms of the Komar
mass per unit length associated to the time-like Killing vector $\partial_{t}$
of the corresponding coordinate system, the expression for parameter
$n$ in (\ref{eq:CanonicalFunctions}) is the exact symmetrical of
that in (\ref{eq:Canonical2Functions}). Hence, in both cases, we
have $\bar{f}=r^{4\lambda_{{\rm m}}}/\alpha$, cf. Eqs. (\ref{eq:CanonicalFunctions})-(\ref{eq:Canonical2Functions}),
and $r^{(n^{2}-1)/2}=r^{4\lambda_{{\rm m}}(2\lambda_{{\rm m}}-1)}$.
Notice, moreover, using (\ref{eq:AngularMomentumCylinder}), that
one can write, in (\ref{eq:CanonicalFunctions}), $\bar{C}=-4j/n$,
and, in (\ref{eq:Canonical2Functions}), $\bar{C}=4j/n$; hence, in
both cases, we end up likewise with the same expression for $\bar{C}$
in terms of $\lambda_{{\rm m}}$ and $j$: $\bar{C}=4j/(1-4\lambda_{{\rm m}})$.
Therefore, we can write the single expression 
\begin{equation}
ds^{2}=-\frac{r^{4\lambda_{{\rm m}}}}{\alpha}\left(dt-\frac{j}{\lambda_{{\rm m}}-1/4}d\bar{\phi}\right)^{2}+r^{4\lambda_{{\rm m}}(2\lambda_{{\rm m}}-1)}(dr^{2}+dz^{2})+\alpha r^{2(1-2\lambda_{{\rm m}})}d\bar{\phi}^{2}\ ,\label{eq:MetricKomar}
\end{equation}
encompassing both the metrics forms (\ref{eq:Canonical})-(\ref{eq:CanonicalFunctions})
and (\ref{eq:Canonical2Functions}). This is an irreducible, fully
general expression for the Lewis metric of the Weyl class. The fact
that it can be written in the forms (\ref{eq:CanonicalFunctions})
or (\ref{eq:Canonical2Functions}), reflects the existing redundancy
in the original four parameters: in fact, two sets of parameters $(a_{1},b_{1},c_{1},n_{1})$
and $(a_{2},b_{2},c_{2},n_{2})$, with $a_{1}>0$ and $a_{2}<0$,
such that the values of $(\lambda_{{\rm m}},j,\alpha)$ are the same
in both cases, necessarily represent the same solution, since they
can both be written in the same form (\ref{eq:MetricKomar}). Its
degree of generality is such that, swapping the time and angular coordinates,
$t\leftrightarrow\phi$, in the original metric (\ref{eq:LewisMetric})-(\ref{eq:LewisFunctions}),
again leads (through entirely analogous steps) to the metric form
(\ref{eq:MetricKomar}). We argue Eq. (\ref{eq:MetricKomar}) to be
the most natural, or \emph{canonical}, form for the metric of a rotating
cylinder of the Weyl class for, in addition to the above, the following
reasons: 
\begin{itemize}
\item the Killing vector $\partial_{t}$ is (for $\alpha>0$) everywhere
time-like (i.e., $g_{00}<0$ for all $r$), therefore physical observers
$u^{\alpha}=(-g_{00})^{-1/2}\partial_{t}^{\alpha}$, at rest in the
coordinates of (\ref{eq:MetricKomar}), exist everywhere. 
\item The associated reference frame is \emph{asymptotically} inertial,
and thus fixed with respect to the ``distant stars'' (see Sec. \ref{sub:GEM-fields-Canonical}
below). 
\item A conserved Komar mass per unit length ($\lambda_{{\rm m}}$) can
be defined from $\partial_{t}$ on arbitrary spatial tubes (even at
$r\rightarrow\infty$) which matches its expected value from the gravitational
field $\vec{G}$ and potential $\Phi$, and also that of the Levi-Civita
static cylinder (Secs. \ref{sub:GEM-fields-Canonical} and \ref{sub:Komar-Canonical}). 
\item It is irreducibly given in terms of three parameters with a clear
physical interpretation: the Komar mass ($\lambda_{{\rm m}}$) and
angular momentum ($j$) per unit length, plus the parameter $\alpha$
governing the angle deficit of the spatial metric $h_{ij}$ {[}cf.
Eq. (\ref{eq:AxistatMetric}){]}. 
\item The GEM fields are strikingly similar to the electromagnetic analogue
--- the electromagnetic fields of a rotating cylinder, from the point
of view of the inertial rest frame (namely $\bm{\mathcal{A}}=\mathcal{A}_{\bar{\phi}}\mathbf{d}\bar{\phi}$;
$\mathcal{A}_{\bar{\phi}}\equiv$constant, $\vec{H}=\mathbb{H}_{\alpha\beta}=0$,
and $\Phi$ and $G_{,i}$ match the electromagnetic counterparts identifying
the Komar mass per unit length $\lambda_{{\rm m}}$ with the charge
per unit length $\lambda$, see Sec. \ref{sub:GEM-fields-Canonical}). 
\item The GEM inertial fields and tidal tensors are the \emph{same }as those
of the Levi-Civita static cylinder; hence the dynamics of test particles
is, with respect to the coordinate system in (\ref{eq:MetricKomar}),
the same as in the static metric (\ref{eq:Levi-Civita}), see Sec.
\ref{sub:GEM-fields-Canonical} below (just like the electromagnetic
forces produced by a charged spinning cylinder are the same as by
a static one). 
\item It is obtained from a simple rigid rotation of coordinates, Eq. (\ref{eq:TransformCanonical}),
which is a well-defined \emph{global} coordinate transformation associated
to a Killing vector field. 
\item It makes immediately transparent the locally static but globally stationary
nature of the metric (Sec. \ref{sub:Local-vs-global} below). 
\item It evinces that the vanishing of the Komar angular momentum $j$ is
the necessary and \emph{sufficient} condition for the metric to reduce
to the Levi-Civita static one (\ref{eq:Levi-Civita}).
\end{itemize}
We thus suggest that the Lewis metric in its usual form (\ref{eq:LewisMetric})-(\ref{eq:LewisFunctions})
possesses a trivial coordinate rotation {[}of angular velocity $-\Omega$,
equivalently given by either of Eqs. (\ref{eq:PrincipalObs}){]},
which has apparently gone unnoticed in the literature, causing $\partial_{t}$
to fail to be time-like everywhere, and the GEM fields to be very
different from the electromagnetic analogue in an inertial frame,
being instead more similar to the situation in a rotating frame in
flat spacetime.

\subsubsection{GEM fields and tidal tensors\label{sub:GEM-fields-Canonical}}

For $\alpha>0$ {[}so that $t$ in Eq. (\ref{eq:MetricKomar}) is
a temporal coordinate{]}, the metric can be put in the form (\ref{eq:AxistatMetric}),
with 
\begin{eqnarray}
e^{2\Phi} & = & \frac{r^{4\lambda_{{\rm m}}}}{\alpha}\quad\Rightarrow\quad\Phi\ =\ 2\lambda_{{\rm m}}\ln(r)+K;\label{eq:CanonicalQM1}\\
\mathcal{A}_{\bar{\phi}} & = & \frac{j}{\lambda_{{\rm m}}-1/4}\ ;\qquad h_{rr}=h_{zz}=r^{4\lambda_{{\rm m}}(2\lambda_{{\rm m}}-1)};\qquad h_{\bar{\phi}\bar{\phi}}=\alpha r^{2(1-2\lambda_{{\rm m}})}\ ,\label{eq:CanonicalQM2}
\end{eqnarray}
$h_{ij}|_{i\ne j}=0$ and $K\equiv-\ln(\alpha)/2$. The gravitoelectric
and gravitomagnetic fields read, cf. Eqs. (\ref{eq:GEMFieldsQM}),
\begin{equation}
G_{i}=-\frac{2\lambda_{{\rm m}}}{r}\delta_{i}^{r};\qquad\vec{G}=-2\lambda_{{\rm m}}r^{-(1-4\lambda_{{\rm m}})^{2}/2-1/2}\partial_{r};\qquad\vec{H}=0\ .\label{eq:GEMCanonical}
\end{equation}
Thus, the gravitoelectric potential $\Phi$ and 1-form $G_{i}$ match
\emph{minus} their electric counterparts in Eqs. (\ref{eq:EMphi_and_A})-(\ref{eq:EB})
for a rotating charged cylinder (as viewed from the inertial rest
frame) identifying $\lambda_{{\rm m}}\leftrightarrow\lambda$. This
supports the interpretation of the Komar integral $\lambda_{{\rm m}}$
as the ``active'' gravitational mass per unit length. The gravitomagnetic
potential 1-form $\bm{\mathcal{A}}=\mathcal{A}_{\bar{\phi}}\mathbf{d}\bar{\phi}$
also resembles the magnetic potential 1-form $\mathbf{A}=\mathfrak{m}\mathbf{d}\phi$.
More importantly, $\mathcal{A}_{\bar{\phi}}$ is constant and $\vec{H}$
vanishes, just like their magnetic counterparts in Eqs. (\ref{eq:EMphi_and_A})-(\ref{eq:EB}).
The inertial fields $\vec{G}$ and $\vec{H}$ also match exactly those
of the Levi-Civita static metric (\ref{eq:Levi-Civita}), cf. Eq.
(\ref{eq:LeviCivitaGEM}); this means that a family of observers at
rest in the coordinates of (\ref{eq:MetricKomar}) measure the same
inertial forces as those at rest in the static metric (\ref{eq:Levi-Civita}).
Namely, since the gravitomagnetic field $\vec{H}$ vanishes in the
reference frame associated to the coordinates of (\ref{eq:MetricKomar}),
the only inertial force acting on test particles is the gravitoelectric
(Newtonian-like) force $m\vec{G}$. Thus, particles dropped from rest
or in radial motion move along radial straight lines, cf. Eq. (\ref{eq:QMGeo});
and, again, the circular geodesics have a \emph{constant} speed given
by 
\begin{equation}
v_{{\rm geo}}=\sqrt{\frac{\lambda_{{\rm m}}}{1/2-\lambda_{{\rm m}}}}\ .\label{eq:vgeoLambda}
\end{equation}
They are thus possible when $0\le\lambda_{{\rm m}}<1/4$ (it is when
$\lambda_{{\rm m}}>0$ that $\vec{G}$ is attractive, and they become
null for $\lambda_{{\rm m}}=1/4$).\textcolor{blue}{{} }Since $\vec{G}\stackrel{r\rightarrow\infty}{\rightarrow}\vec{0}$,
it follows moreover that the reference frame associated to the coordinate
system in (\ref{eq:MetricKomar}) is \emph{asymptotically} inertial,
and that the ``distant stars'' are at rest in such frame; that is,
it is a ``star-fixed'' frame. We notice also that the observers
at rest in such frame are, among the stationary observers, those measuring
a maximum $\vec{G}$, as can be seen from e.g. Eq. (9) of \cite{SemerakExtremally1996};
they are said to be ``extremely accelerated'' (for a brief review
of the privileged properties of such observers, we refer to \cite{Semerak_ClockEffect1999}).

Further consequences of the vanishing of $\vec{H}$ include: the vanishing
second term of Eq. (\ref{eq:GMClock}), which means that the gravitomagnetic
time delay for particles in geodesic motion around the cylinder, $\Delta t_{{\rm geo}}$,
equals precisely the Sagnac time delay for photons, Eq. (\ref{eq:DtBigLoop})
(this is a property inherent to extremely accelerated observers, see
\cite{BiniJantzenMashhoon_Clock1}); that gyroscopes at rest in the
coordinates of (\ref{eq:MetricKomar}) do not precess, the components
of their spin vector $\vec{S}$ remaining constant, cf. Eq. (\ref{eq:SpinPrec});
that no Sagnac effect arises in an optical gyroscope {[}not enclosing
the axis $r=0$, as depicted in Fig. \ref{fig:Sagnac}(b){]}, cf.
Eqs. (\ref{eq:DeltatH}).

As for the tidal tensors as measured by the observers at rest in the
coordinates of (\ref{eq:MetricKomar}), the gravitomagnetic tensor
vanishes (by construction), $\mathbb{H}_{\alpha\beta}=0$, and the
gravitoelectric tensor has non-vanishing components 
\begin{eqnarray}
\mathbb{E}_{rr} & = & -\frac{2\lambda_{{\rm m}}(1-2\lambda_{{\rm m}})^{2}}{r^{2}}\ ;\qquad\mathbb{E}_{zz}\ =\ \frac{4\lambda_{{\rm m}}^{2}(2\lambda_{{\rm m}}-1)}{r^{2}}\ ;\label{eq:CanonicalEab1}\\
\mathbb{E}_{\bar{\phi}\bar{\phi}} & = & -2\alpha r^{-8\lambda_{{\rm m}}^{2}}\lambda_{{\rm m}}(2\lambda_{{\rm m}}-1)\ .\label{eq:CanonicalEab2}
\end{eqnarray}
This is in fact the same as the gravitoelectric tidal tensor of the
static Levi-Civita metric. In order to see that, first notice that
Eqs. (\ref{eq:CanonicalEab1})-(\ref{eq:CanonicalEab2}) do not depend
on $j$; since the Levi-Civita limit is obtained by making $j\rightarrow0$,
the components $\mathbb{E}_{\alpha\beta}$ remain formally the same.
Now, since $\mathbb{E}_{\alpha\beta}$ is spatial with respect to
$u^{\alpha}$ ($\mathbb{E}_{\alpha\beta}u^{\beta}=\mathbb{E}_{\alpha\beta}u^{\alpha}=0$),
it can be identified with a tensor living on the space manifold $(\Sigma,h)$,
in which $\{r,\bar{\phi},z\}$ is a coordinate chart. The spatial
metric $h_{ij}$ depends only on $\lambda_{{\rm m}}$ and $\alpha$,
so it remains the same as well. We can then say that the tensor $\mathbb{E}_{\alpha\beta}$
is the same in both cases, i.e., the tidal effects as measured by
observers at rest in (\ref{eq:MetricKomar}) are the same as those
in the static metric (\ref{eq:Levi-Civita}) (with the identification
$\alpha\rightarrow1/a$).

Notice, on the one hand, that the congruence of observers at rest
in (\ref{eq:MetricKomar}) is the only one with respect to which $\mathbb{H}_{\alpha\beta}$
vanishes (since observers measuring $\mathbb{H}_{\alpha\beta}=0$
are, at each point, unique in a Petrov type I spacetime, see Sec.
\ref{sub:The-canonical-form}). On the other hand, observe that a
vanishing $\bm{\mathcal{A}}$, as well as a vanishing $\vec{H}$,
imply, via Eqs. (\ref{eq:GEMFieldsQM}), (\ref{eq:HijGEM}) {[}valid
for any stationary line element (\ref{eq:StatMetric}){]}, that $\mathbb{H}_{\alpha\beta}=0$;
that is: $\bm{\mathcal{A}}=0\Rightarrow\mathbb{H}_{\alpha\beta}=0$,
and $\vec{H}=0\Rightarrow\mathbb{H}_{\alpha\beta}=0$. This tells
us that (i) the gravitomagnetic potential 1-form $\bm{\mathcal{A}}$
in (\ref{eq:MetricKomar}) cannot be made to vanish in any coordinate
system where the metric is time-independent; (ii) Eq. (\ref{eq:MetricKomar})
is the only stationary form of the metric in which $\vec{H}=0$. Since
$\vec{H}=2\vec{\omega}$, cf. Eq. (\ref{eq:GEM Fields Cov}), this
amounts to saying that the observers $u^{\alpha}=(-g_{00})^{-1/2}\partial_{t}^{\alpha}$,
at rest in (\ref{eq:MetricKomar}), are the only vorticity-free (i.e.,
hypersurface orthogonal) congruence among all observer congruences
tangent to a Killing vector field. This implies that (iii) $\partial_{t}$,
in the coordinates of (\ref{eq:MetricKomar}), is the only hypersurface
orthogonal time-like Killing vector field in the Lewis metrics of
the Weyl class. In the range $0\le\lambda_{{\rm m}}<1/4$ (where,
as seen above, $\vec{G}$ is attractive and circular geodesics are
possible, and the metric has moreover a clear interpretation as the
external field of a cylindrical source, cf. \cite{GautreauHoffman69,GriffithsPodolsky2009,BonnorMartins1991,Bonnor1992,SantosLC,Bronnikov:2019clf}),
it is actually the only Killing vector field of the form $\xi^{\alpha}=\partial_{t}^{\alpha}+\varpi\partial_{\bar{\phi}}^{\alpha}$,
with $\varpi$ constant, which is time-like when\footnote{Any time-like Killing vector field in the Weyl class metric can, up
to a global constant factor, be written as $\xi^{\alpha}=\partial_{t}^{\alpha}+\varpi\partial_{\bar{\phi}}^{\alpha}+\mathcal{Z}\partial_{z}^{\alpha}$,
with $\varpi$ and $\mathcal{Z}$ constants. The time-like condition
$\xi^{\alpha}\xi^{\beta}g_{\alpha\beta}<0$ amounts, in the metric
(\ref{eq:MetricKomar}), to 
\[
\left[1-\frac{\varpi j}{\lambda_{{\rm m}}-1/4}\right]^{2}>\varpi^{2}\alpha^{2}r^{2(1-4\lambda_{{\rm m}})}+\alpha\mathcal{Z}^{2}r^{8\lambda_{{\rm m}}(\lambda_{{\rm m}}-1)}\ ,
\]
which, for $0\le\lambda_{{\rm m}}<1/4$, can be satisfied for all
$r$ only if $\varpi=0$ (since ${\rm lim}_{r\rightarrow\infty}r^{2(1-4\lambda_{{\rm m}})}=\infty$).} $r\rightarrow\infty$.

\subsubsection{Cosmic strings\label{sub:Cosmic-strings}}

In the limit $\lambda_{{\rm m}}=0$, Eq. (\ref{eq:MetricKomar}) yields
the exterior metric of a spinning cosmic string \cite{SantosGRG1995,Barros_Bezerra_Romero2003,JensenSoleng,Bezerra:1990ag,MenaNatarioTod}
of Komar angular momentum per unit length $j$ and angle deficit $2\pi(1-\alpha^{1/2})\equiv2\pi\delta$
(cf. also \cite{Linet1986,KibbleCosmicStrings,Nouri-Zonoz:2013rfa,Bronnikov:2019clf}).
In this case, for $r\ne0$, the spacetime is locally flat everywhere,
$R_{\alpha\beta\gamma\delta}=0$. All the GEM inertial and tidal fields
vanish, $\vec{G}=\vec{H}=0$, $\mathbb{E}_{\alpha\beta}=\mathbb{H}_{\alpha\beta}=0$,
thus there are no gravitational forces of any kind. This supports
the interpretation of the Komar mass as ``active gravitational mass'';
its vanishing here arises from an exact cancellation\footnote{\textcolor{black}{For a static string, this consists of the cancellation
\cite{Linet1986,KibbleCosmicStrings} between the energy density and
the string's tension, }$R_{\ 0}^{0}/(4\pi)=T_{\ 0}^{0}-T_{\ z}^{z}=0$\textcolor{black}{,
causing the integrand in Eq. (\ref{eq:KomarRicci}), for }$\xi^{\mu}=\partial_{t}^{\mu}$\textcolor{black}{{}
and }$n^{\mu}=\alpha^{1/2}\partial_{t}^{\mu}$\textcolor{black}{,
to vanish.}} \cite{Linet1986,KibbleCosmicStrings,JensenSoleng}, \emph{within
}the string, between the contributions of the energy density and the
stresses to the \textcolor{black}{integral in Eq. (\ref{eq:KomarRicci}).}
One consequence is that bound orbits for test particles are not possible.
Global gravitational effects however subsist, governed by the angle
deficit and by the gravitomagnetic potential 1-form $\bm{\mathcal{A}}=-4j\mathbf{d}\bar{\phi}$.
An example of the former are the double images of objects located
behind the strings \cite{KibbleCosmicStrings,Ford_Vilenkin_1981}.
Another is that a vector $V^{\alpha}$ parallel transported along
a closed loop enclosing the axis $r=0$ does not return to itself,
but to a new vector $V_{{\rm f}}^{\alpha}={\rm Hol}_{\ \beta}^{\alpha}V_{{\rm in}}^{\beta}\ne V_{{\rm in}}^{\alpha}$,
where ${\rm Hol}_{\ \beta}^{\alpha}$ is the holonomy matrix. In order
to determine it, one observes that, since\footnote{This holonomy implies, however, that $R_{\alpha\beta\gamma\delta}\ne0$
\emph{within} the string \cite{Ford_Vilenkin_1981,Nouri-Zonoz:2013rfa}
(a Dirac delta for infinitely thin strings). One can thus cast the
effect as a non-local manifestation, in a curvature-free region, of
the existence of a region with non-zero curvature. Parallelisms with
the Aharonov-Bohm effect have been drawn \cite{Dowker1967,Ford_Vilenkin_1981,SantosGRG1995,Nouri-Zonoz:2013rfa,Bronnikov:2019clf},
since the latter can likewise be cast as a manifestation, in a field
free region, of the existence of a region where the given field (e.g.
$\vec{B}$) is non-zero. This is not, however, as close an analogy
as that for the Sagnac effect, discussed in Secs. \ref{sub:Physical-distinction}
and \ref{sub:Aharonov-Bohm-effect}.} $R_{\alpha\beta\gamma\delta}=0$, it is invariant under continuous
deformations of the loop. Hence, it suffices to consider a circular
one in the form $t=z=0$, $r=const.$ Introducing the orthonormal
tetrad $e_{\hat{\alpha}}$ adapted to the laboratory observers (\ref{eq:uLab}):
$\mathbf{e}_{\hat{0}}=\alpha^{1/2}\partial_{t}$, $\mathbf{e}_{\hat{r}}=\partial_{r}$,
$\mathbf{e}_{\hat{\phi}}=r^{-1}\alpha^{-1/2}(\partial_{\phi}-4j\partial_{t})$,
$\mathbf{e}_{\hat{z}}=\partial_{z}$, we have $V_{{\rm f}}^{\hat{\alpha}}={\rm Hol}_{\ \hat{\beta}}^{\hat{\alpha}}V_{{\rm in}}^{\hat{\beta}}$,
with 
\[
\left[{\rm Hol}_{\ \hat{\beta}}^{\hat{\alpha}}\right]=\left[\begin{array}{cccc}
1 & 0 & 0 & 0\\
0 & \cos(2\pi\sqrt{\alpha}) & \sin(2\pi\sqrt{\alpha})\\
 & -\sin(2\pi\sqrt{\alpha}) & \cos(2\pi\sqrt{\alpha})\\
0 & 0 & 0 & 1
\end{array}\right]\ .
\]
This is a rotation about the $z-$axis by an angle $-2\pi\alpha^{1/2}$,
that is, $2\pi\delta$. The holonomy is actually the same along curves
that are only spatially closed, and is invariant under continuous
deformations of its projection $C$ on the space manifold $\Sigma$,
since $\Sigma$ is also flat. It is also the same as for a static
string ($j=0$), cf. \cite{Nouri-Zonoz:2013rfa,Ford_Vilenkin_1981,Dowker1967},
as one might expect from it having the same spatial metric $h_{ij}dx^{i}dx^{j}=dr^{2}+dz^{2}+\alpha r^{2}d\bar{\phi}^{2}$,
describing a conical geometry of angle deficit $2\pi\delta$.

Manifestations of $\bm{\mathcal{A}}$ are the Sagnac effect and the
synchronization holonomy, to be discussed next.

\subsection{The distinction between the rotating Weyl class and the static Levi-Civita
field\label{sub:The-distinction-between}}

The Levi-Civita metric (\ref{eq:Levi-Civita}) for the exterior field
of a static cylinder follows from the canonical form (\ref{eq:MetricKomar})
of the Weyl class metric by making $j=0$ (and identifying $\{\alpha,\lambda_{{\rm m}},\bar{\phi}\}\leftrightarrow\{a^{-1},(1-n)/4,\phi\}$).
Hence, in the notation of Eq. (\ref{eq:AxistatMetric}), they differ
only in the gravitomagnetic potential 1-form $\bm{\mathcal{A}}=j/(\lambda_{{\rm m}}-1/4)\mathbf{d}\bar{\phi}$,
which, as shown above, cannot be made to vanish in any coordinate
system where the metric is time-independent in the case of a rotating
cylinder. Therefore, the comparison between the two cases, both on
physical and mathematical grounds, amounts to investigating the implications
of $\bm{\mathcal{A}}$.

\subsubsection{Physical distinction\label{sub:Physical-distinction}}

As we have seen in Sec. \ref{sub:GEM-fields-Canonical}, the only
surviving gravitomagnetic object from Table \ref{tab:Levels} in the
canonical metric (\ref{eq:MetricKomar}) is the 1-form $\bm{\mathcal{A}}$
(or, equivalently, $\vec{\mathcal{A}}$) itself. Hence, the physical
distinction from the Levi-Civita metric lies only at that first level
of gravitomagnetism.

One physical effect that distinguishes the two metrics is thus the
Sagnac effect. Consider optical fiber loops fixed with respect to
the distant stars, i.e., at rest in the coordinate system of (\ref{eq:MetricKomar}).
In the Levi-Civita case, $j=0\Rightarrow\bm{{\rm \mathcal{A}}}=0$,
so it follows from Eq. (\ref{eq:SagnacDiffForm}) that no Sagnac effect
arises in any loop, and light beams propagating in the positive and
negative directions take the same time to complete the loop. For a
rotating cylinder ($j\ne0$), we have $\bm{\mathcal{A}}=\mathcal{A}_{\bar{\phi}}\mathbf{d}\bar{\phi}\ne0$
with $\mathcal{A}_{\bar{\phi}}$ constant; hence $\bm{\mathcal{A}}$
is a \emph{closed} ($\mathbf{d}\bm{\mathcal{A}}=0$) but \emph{non-exact}
form (since $\mathbf{d}\bar{\phi}$ is non-exact), defined in a space
manifold $\Sigma$ homeomorphic to $\mathbb{R}^{3}\backslash\{r=0\}$.
This means (see Sec. \ref{sub:Closed-forms,-exact}) that $\oint_{C}\bm{\mathcal{A}}$,
and thus the Sagnac time delay (\ref{eq:SagnacDiffForm}), vanish
along any loop which does not enclose the central cylinder, such as
the small loop in Fig. \ref{fig:Sagnac} (b), but has the \emph{same}
nonzero value 
\begin{equation}
\Delta t=4\pi\mathcal{A}_{\bar{\phi}}=-\frac{4\pi j}{1/4-\lambda_{{\rm m}}}\label{eq:DtCanonical}
\end{equation}
along any loop enclosing the cylinder, regardless of its shape {[}for
instance the circular loop depicted in Fig. \ref{fig:Sagnac} (b){]},
cf. Eq. (\ref{eq:DtBigLoop}).\setcounter{footnote}{0}

Notice the analogy with the situation in electromagnetism, in the
distinction between the field generated by static and rotating cylinders
(Sec. \ref{sec:The-electromagnetic-analogue:}): they likewise only
differ in the magnetic potential 1-form $\mathbf{A}$, which (in quantum
electrodynamics) manifests itself in the Aharonov-Bohm effect. Such
effect plays a role analogous to the Sagnac effect in the gravitational
setting; in fact, it is given by the formally analogous expression
(\ref{eq:dphiAharonov}), which is likewise independent of the particular
shape of the paths, as long as they enclose the cylinder. Earlier
works have already hinted at some qualitative\footnote{These works, however, do not compare directly analogous settings,
none of them considering the gravitational field of rotating cylinders.
In \cite{RuggieroAharonov2005} the parallelism drawn is between the
Aharonov-Bohm effect and the Sagnac effect in Kerr and Gödel spacetimes;
these fields are, however, of a different nature (from both that of
a cylinder and of the Aharonov-Bohm electromagnetic setting), since
$\mathbf{d}\bm{\mathcal{A}}\ne0\Leftrightarrow\vec{H}\ne0$, and so
$\Delta t=2\oint_{C}\bm{\mathcal{A}}$ therein is not invariant under
continuous deformations of the loop $C$. In \cite{RizziRuggieroAharonovI,RizziRuggieroAharonovII,RuggieroTartaglia2014NoteSagnac}
the Sagnac effect is that of a rotating frame in flat spacetime, where,
again, $\mathbf{d}\bm{\mathcal{A}}\ne0$. In \cite{Stachel:1981fg},
the metric of a static cylinder is considered, and it is suggested
that the effect would arise in a rotating cylinder, without actually
discussing the Lewis solutions explicitly. In \cite{RuggieroTartaglia2014NoteSagnac,RuggieroTartagliaSagnac2015}
it was concluded that the analogy holds only at lowest order; that
is due to the fact that therein (i) the effect is cast (via the Stokes
theorem) in terms of the flux of a ``gravitomagnetic field''; (ii)
a different (less usual) definition of such field is then used, {[}$\vec{H}=\tilde{\nabla}\times(e^{2\phi}\vec{\mathcal{A}})$,
instead of (\ref{eq:GEMFieldsQM}){]}, thereby obscuring the analogy
shown herein.} analogy between the Aharonov-Bohm effect and the Sagnac effect \cite{AshtekarMagnon,Stachel:1981fg,RizziRuggieroAharonovI,RizziRuggieroAharonovII,RuggieroAharonov2005,Barros_Bezerra_Romero2003},
or the global non-staticity of a locally static gravitational field
\cite{Stachel:1981fg}; on the other hand, it has been suggested \cite{SantosGRG1995}
that the Lewis metrics posses some kind of ``topological'' analogue
of the Aharonov-Bohm effect. Here we substantiate such suggestions
with concrete results for directly analogous settings, exposing a
striking one to one correspondence.

It is also worth mentioning the similarity with the situation for
PP waves \cite{PodolskyGyratons}, where the distinction between the
field produced by non-spinning and spinning sources (``gyratons'')
likewise boils down to a 1-form ($\bm{a}$, in the notation of \cite{PodolskyGyratons}),
associated to the off-diagonal part of the metric, vanishing in the
first case, and being a closed non-exact form in the second.

\subsubsection*{Coil of optical loops}

The apparatus above makes use of a star-fixed reference frame, which
is physically realized by aiming telescopes at the distant stars (e.g.
\cite{CiufoliniWheeler,CiufoliniNature2007}). It is possible, however,
still based on the Sagnac effect, to distinguish between the fields
of rotating and static cylinders without the need of setting up a
specific frame. The price to pay is that one must use more than one
loop, since the effect along a single loop can always be eliminated
by spinning it. In particular, we have seen in Sec. \ref{sub:circular-loop-around}
that it vanishes on circular loops whose angular momentum is zero;
that is, those comoving with the zero angular momentum observers (ZAMOs),
which have angular velocity {[}cf. Eq. (\ref{eq:OmegaZamo}){]} 
\begin{figure}
\includegraphics[width=1\textwidth]{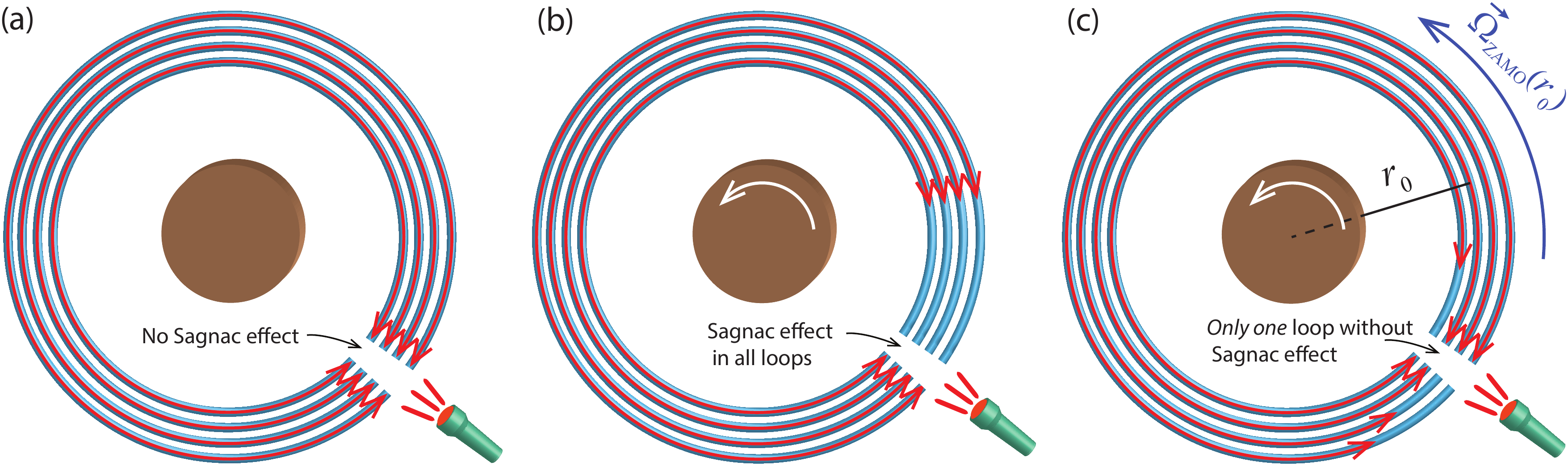}

\protect\protect\protect\caption{\label{fig:CoilLewis}Apparatus for physically distinguishing between
the static Levi-Civita metric and the Lewis metrics of the Weyl class,
based on the Sagnac effect: a set (``coil'') of optical fiber loops
around the central cylinder, in which counterpropagating light beams
are injected. (a) Levi-Civita static cylinder, coil at rest with respect
to the distant stars: the Sagnac effect vanishes in every loop. (b)
Rotating cylinder of the Weyl class, coil at rest with respect to
the distant stars: a Sagnac effect arises in every loop. (c) Lewis
cylinder of the Weyl class, coil rotating {[}with respect to the distant
stars{]} with the angular velocity of the ZAMO at $r_{0}$: the Sagnac
effect vanishes only for the loop of radius $r=r_{0}$; for $r>r_{0}$
($<r_{0}$) the beams co-rotating (counter-rotating) with the cylinder
take longer to complete the loop.}
\end{figure}

\begin{equation}
\Omega_{{\rm ZAMO}}(r)=-\frac{\mathcal{A}_{\bar{\phi}}e^{2\Phi}}{g_{\bar{\phi}\bar{\phi}}}=-\left[\frac{j}{1/4-\lambda_{{\rm m}}}-\frac{1/4-\lambda_{{\rm m}}}{j}\alpha^{2}r^{2(1-4\lambda_{{\rm m}})}\right]^{-1}\ .\label{eq:OmegaZAMOCanonical}
\end{equation}
Consider then a set (``coil'') of circular optical fiber loops concentric
with the cylinder, as depicted in Fig. \ref{fig:CoilLewis}. For a
static cylinder ($j=0$), and a coil at rest in the star-fixed coordinates
of (\ref{eq:MetricKomar}), the Sagnac effect vanishes in every loop.
When the metric is given in a different coordinate system, rotating
with respect to (\ref{eq:MetricKomar}), a Sagnac effect arises in
a coil at rest therein; such effect is however globally eliminated
by simply spinning the coil with some angular velocity. For a rotating
cylinder ($j\ne0$), and a coil at rest in the coordinates of (\ref{eq:MetricKomar})
{[}see Fig. \ref{fig:CoilLewis}(b){]}, a Sagnac effect arises in
\emph{every} loop, given by Eq. (\ref{eq:DtCanonical}). Now, along
\emph{one single loop} of radius $r_{0}$ {[}Fig. \ref{fig:CoilLewis}
(c){]}, the effect can always be eliminated, by spinning the coil
with an angular velocity equaling that of the ZAMO on site, $\Omega_{{\rm ZAMO}}(r_{0})$.
However, due to the $r-$dependence of $\Omega_{{\rm ZAMO}}(r)$,
in all other loops of radius $r\ne r_{0}$ a Sagnac effect arises.
Hence, given a Lewis metric in an arbitrary coordinate system, a physical
experiment to determine whether it corresponds to a static or rotating
cylinder would be to consider a coil of concentric optical fiber loops,
as illustrated in Fig. \ref{fig:CoilLewis}, and checking whether
one can globally eliminate the Sagnac effect along the whole coil
by spinning it with some angular velocity. This reflects the basic
fact that, contrary to the case around a static cylinder, in the rotating
case it is not possible to \emph{globally} eliminate $\bm{\mathcal{A}}$
through any rigid rotation (in fact, through any globally valid coordinate
transformation, cf. Secs. \ref{sub:GEM-fields-Canonical} and \ref{sub:Mathematical-distinction}).

It is worth observing that, for $\lambda_{{\rm m}}<1/4$ (case of
the range where circular geodesics are allowed, and the metric clearly
represents the field of a cylindrical source, see Sec. \ref{sub:GEM-fields-Canonical}),
$\mathcal{A}_{\bar{\phi}}$ has opposite sign to $j$ {[}cf. Eq. (\ref{eq:CanonicalQM2}){]},
and so, by Eq. (\ref{eq:DtCanonical}), for loops fixed with respect
to the distant stars, it is the light beams propagating in the sense
opposite to the cylinder's rotation that take longer to complete the
loop. Moreover, for spacelike $\partial_{\bar{\phi}}$ (i.e., $g_{\bar{\phi}\bar{\phi}}>0$),
$\Omega_{{\rm ZAMO}}(r)$ has the same sign of $j$, so that the ZAMOs
rotate (with respect to the distant stars) in the same sense as the
cylinder. Both effects are thus in agreement with the intuitive notion
that the cylinder's rotation ``drags'' the ``local spacetime geometry''
with it, and consequently with the physical interpretation in Sec.
\ref{sub:circular-loop-around}. 
\begin{figure}
\includegraphics[width=0.75\textwidth]{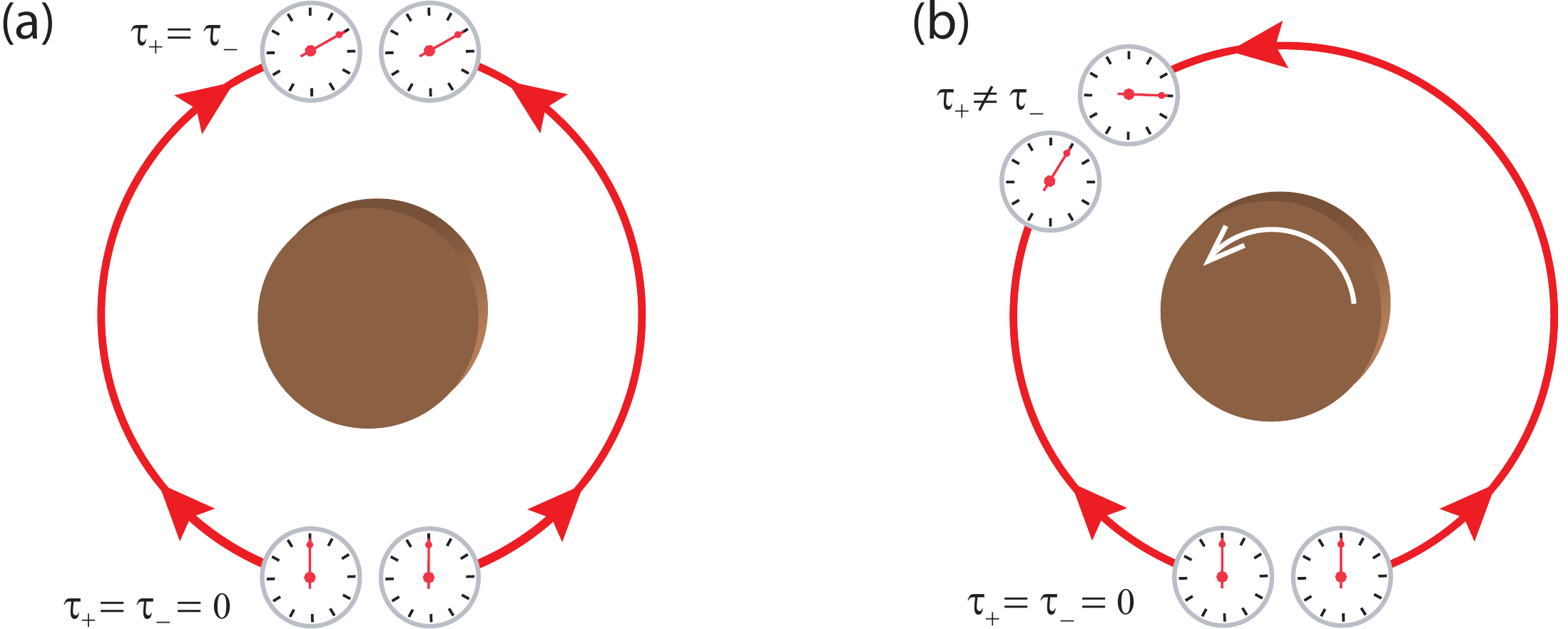}

\protect\protect\protect\caption{\label{fig:TwoCock}Apparatus for distinguishing between the field
of static and rotating cylinders, based on the observer invariant
gravitomagnetic clock effect: a pair of clocks in oppositely rotating
circular geodesics. (a) In the static case, both clocks measure the
same proper time between the events where they meet, $\tau_{+}=\tau_{-}$;
(b) in the rotating case, the proper times differ, $\tau_{+}>\tau_{-}$.}
\end{figure}

Finally, we notice that in the limit $\lambda_{{\rm m}}=0$, corresponding
to cosmic strings (Sec. \ref{sub:Cosmic-strings}), the Sagnac effect
subsists, and so all the above applies for the distinction between
the fields of spinning and non-spinning strings.

\subsubsection*{Gravitomagnetic clock effect}

Another effect that allows to distinguish between the fields of static
and rotating Weyl class cylinders is the gravitomagnetic clock effect.
As seen in Sec. \ref{sub:GEM-fields-Canonical}, the difference in
orbital periods for pairs of particles in oppositely rotating geodesics,
as measured in the star-fixed coordinate system of (\ref{eq:MetricKomar}),
reduces to the Sagnac time delay. Hence, one could replace the optical
fiber loops in Fig. \ref{fig:CoilLewis} by pairs of particles in
geodesic motion, with analogous results: in the case of the static
cylinder, the effect globally vanishes, the periods of circular geodesics
being independent of their rotation sense. In the case of the rotating
cylinder, the geodesics co-rotating with the cylinder have shorter
periods than the counter-rotating ones. (Notice that this is opposite
to the situation in the Kerr spacetime, cf. e.g. \cite{BiniJantzenMashhoon_Clock1,Tartaglia2000ClockEffect};
that is down to the dominance therein of the second term of (\ref{eq:GMClock}),
which vanishes herein). It is possible, by a transformation to a rotating
frame, to eliminate the delay for orbits of a given radius $r_{0}$;
but it is not possible to do so globally, i.e. for all $r$. It is
possible, however, to physically distinguish between the two metrics
using only \emph{one pair} of particles, through the observer invariant
two-clock effect discussed in Sec. \ref{sub:The-gravitomagnetic-clock}:
consider a pair of clocks in oppositely rotating circular geodesics,
as illustrated in Fig. \ref{fig:TwoCock}. For the Levi-Civita static
cylinder ($j=0$), the proper time measured between the events where
they meet is the same for both clocks ($\Delta\tau=0$). For the rotating
cylinder, by contrast, the proper times measured by each clock between
meeting events differ ($\Delta\tau\ne0$), being longer for the co-rotating
clock: $\tau_{+}>\tau_{-}$. Their values are computed from Eqs (\ref{eq:OmegaGeo}),
(\ref{eq:gammaPM}), (\ref{eq:InvClockEffect}), using the metric
components in (\ref{eq:MetricKomar}) {[}or, equivalently, in (\ref{eq:LewisMetric})-(\ref{eq:LewisFunctions}),
since the effect does not depend on the reference frame{]}.

\subsubsection{Local vs global staticity\label{sub:Local-vs-global}}

According to the usual definition in the literature (e.g. \cite{Barnes_1972,EhlersKundt,Hawking:1973uf,Bonnor1980,Stachel:1981fg,StephaniExact,WyllemanBeke2010}),
a spacetime is static \emph{iff} it admits a hypersurface-orthogonal
timelike Killing vector field $\xi^{\alpha}$. The hypersurface orthogonal
condition amounts to demanding its dual 1-form $\xi_{\alpha}$ to
be \emph{locally} \cite{Stachel:1981fg} of the form 
\begin{equation}
\xi_{\alpha}=\eta\partial_{\alpha}\psi\ ,\label{eq:HOcondition}
\end{equation}
where $\eta$ and $\psi$ are two smooth functions. This condition
is equivalent to the vanishing of the vorticity (\ref{eq:Vorticity})
of the integral curves of $\xi^{\alpha}$. One can show \cite{EhlersKundt}
that if this condition is satisfied then a coordinate system can be
found in which the metric takes a diagonal form. In such coordinates,
the hypersurfaces orthogonal to $\xi^{\alpha}$ are the level surfaces
of the time coordinate \cite{Bonnor1980}. This is, however, a local
notion, since such coordinates may not be globally satisfactory \cite{Bonnor1980,Tipler:1974gt}
(as exemplified in Sec. \ref{sub:Mathematical-distinction} below).

A distinction should thus be made between local and \emph{global}
staticity. Notions of global staticity have been put forth in different,
but equivalent formulations, by Stachel \cite{Stachel:1981fg} and
Bonnor \cite{Bonnor1980}, both amounting to demanding (\ref{eq:HOcondition})
to hold globally in the region under consideration, for some (single
valued) \emph{function} $\psi$. In \cite{Stachel:1981fg}, an enlightening
formulation is devised, in terms of the 1-form $\bm{\chi}$ ``inverse''
to $\xi^{\alpha}$, defined by $\chi_{\alpha}\propto\xi_{\alpha}$
and $\chi_{\alpha}\xi^{\alpha}=1\Rightarrow\chi_{\alpha}\equiv\xi_{\alpha}/\xi^{2}$:
it is noted that the condition that (\ref{eq:HOcondition}) is locally
obeyed is equivalent to $\bm{\chi}$ being closed, $\mathbf{d}\bm{\chi}=0$,
in which case $\xi^{\alpha}$ is dubbed a locally static Killing vector
field; and that the condition that (\ref{eq:HOcondition}) holds \emph{globally}
amounts to demanding $\bm{\chi}$ to be moreover \emph{exact}, i.e.,
$\bm{\chi}=\mathbf{d}\psi$ ($\Leftrightarrow\chi_{\alpha}=\partial_{\alpha}\psi$),
for some some global function\footnote{Therefore $\xi_{\alpha}=\xi^{2}\partial_{\alpha}\psi$, and (\ref{eq:HOcondition})
holds with $\eta=\xi^{2}$.} $\psi$. In this case $\xi^{\alpha}$ is dubbed globally static.
A spacetime is then classified as locally static \emph{iff} it admits
a locally static time-like Killing vector field $\xi^{\alpha}$, and
globally static \emph{iff} it admits a globally static $\xi^{\alpha}$.

Consider now a stationary metric in the form (\ref{eq:StatMetric}).
For the time-like Killing vector field $\xi^{\alpha}=\partial_{t}^{\alpha}$,
we have $\bm{\chi}=\mathbf{d}t-\bm{\mathcal{A}}$; thus, the condition
for $\xi^{\alpha}$ being locally static reduces to $\mathbf{d}\bm{\mathcal{A}}=0$,
i.e., to the spatial 1-form $\bm{{\rm \mathcal{A}}}$ being closed;
and it being globally static amounts to $\bm{\mathcal{A}}$ being
exact. It follows that

\begin{proposition}A spacetime is locally static iff it is possible
to find a coordinate system where the metric takes the form (\ref{eq:StatMetric})
with $\mathbf{d}\bm{\mathcal{A}}=0$. The spacetime is globally static
if $\bm{{\rm \mathcal{A}}}$ is moreover exact, i.e, if $\bm{{\rm \mathcal{A}}}=\mathbf{d}\varphi$,
for some globally defined (single valued) function $\varphi$. \end{proposition}
In the case of axistationary metrics, Eq. (\ref{eq:AxistatMetric}),
$\bm{\mathcal{A}}=\mathcal{A}_{\phi}\mathbf{d\phi}$ with $\mathcal{A}_{\phi}$
independent of $\phi$, so the closedness condition $0=\mathbf{d}\bm{\mathcal{A}}=\mathbf{d}\mathcal{A}_{\phi}\wedge\mathbf{d\phi}$
amounts to $\mathcal{A}_{\phi}=constant$ \cite{Bergh_Wils1983},
and the exactness condition to $\bm{{\rm \mathcal{A}}}=0$, since
$\oint_{C}\mathbf{d}\phi\ne0$ for any closed loop $C$ enclosing
the axis $r=0$.

The Levi-Civita static metric (\ref{eq:Levi-Civita}) is clearly locally
\emph{and} globally static, since $\bm{{\rm \mathcal{A}}}=0$ therein.
The Lewis metric of the Weyl class, as its canonical form (\ref{eq:MetricKomar})
reveals, is an example of a metric which is locally but \emph{not
globally} static.

We propose yet another equivalent definition of global staticity,
based on the hypersurfaces $\Sigma$ orthogonal to the Killing vector
field $\xi^{\alpha}$, which proves enlightening in this context.
Such hypersurfaces are the level surfaces $\psi=const.$ of the function
$\psi(t,r,\phi,z)$ in Eq. (\ref{eq:HOcondition}). Choosing, without
loss of generality, coordinates such that $\xi^{\alpha}=\partial_{t}^{\alpha}$,
it follows that $\partial_{\alpha}\psi=\chi_{\alpha}=g_{0\alpha}/g_{00}$,
i.e., by (\ref{eq:StatMetric}), 
\[
\mathbf{d}\psi=\mathbf{d}t-\mathcal{A}_{i}\mathbf{d}x^{i}\ \Leftrightarrow\ \psi=t-f(x^{i})\ ,
\]
with $\mathbf{d}f=\mathcal{A}_{i}\mathbf{d}x^{i}$. Thus, $\psi$
is a (single-valued) function \emph{iff} that is true for $f(x^{i})$,
which amounts to the level surfaces $t=f(x^{i})+const$ ($\Leftrightarrow\psi=const.$)
intersecting each integral line of $\partial_{t}$ \emph{exactly once}.
Such hypersurfaces are time slices. One can then say that a spacetime
is globally static \emph{iff} it admits a hypersurface orthogonal
Killing vector field, whose hypersurfaces intersect each worldline
of the congruence exactly once. Now, by definition, locally these
hypersurfaces consist of the events that are simultaneous with respect
to the laboratory observers (\ref{eq:uLab}) (whose worldlines are
tangent to $\partial_{t}$); if they intersect each worldline of the
congruence exactly once, they are global simultaneity hypersurfaces.
(This is immediately seen by defining a new time coordinate $t'=\psi$,
which is constant along the hypersurfaces $\Sigma$ orthogonal to
$\partial_{t'}=\partial_{t}$). Hence, 
\begin{figure}
\includegraphics[width=0.75\textwidth]{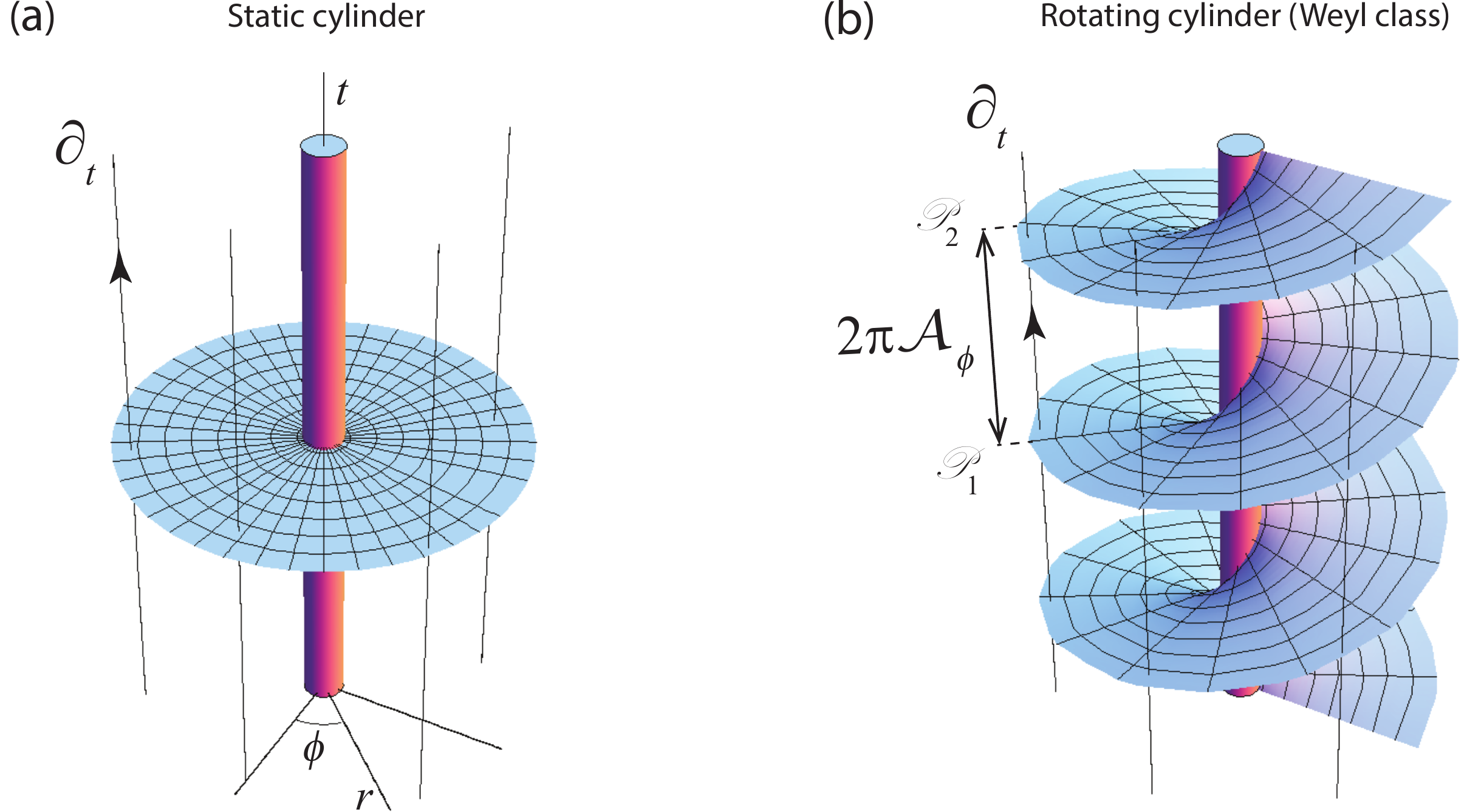} \protect\protect\protect\caption{\label{fig:OrthogonalHypersurfaces}$t,r,\phi$ plot of the hypersurfaces
orthogonal to the Killing vector field $\partial_{t}$ in: (a) the
Levi-Civita static metric; (b) the canonical form (\ref{eq:MetricKomar})
of the Lewis metric for a Weyl class rotating cylinder. The redundant
$z$ coordinate has been suppressed, and the bar in $\bar{\phi}$
in Eq. (\ref{eq:MetricKomar}) omitted. In (a) $\partial_{t}$ is
orthogonal to hypersurfaces of global simultaneity (the planes $t=const.$),
signaling that the spacetime is globally static. In (b) the orthogonal
hypersurfaces are helicoids, described by $t-\mathcal{A}_{\phi}\phi=const.$,
which are not hypersurfaces of global simultaneity, intersecting each
integral curve of $\partial_{t}$ infinitely many times. The spacetime
is thus locally, but \emph{not globally} static. Each $2\pi$ turn
along $\phi$ leads to a different event in time; the jump between
turns is the synchronization gap $2\pi\mathcal{A}_{\phi}$. }
\end{figure}

\begin{proposition}A spacetime is locally static iff it admits a
hypersurface orthogonal Killing vector $\xi^{\alpha}$; it is moreover
globally static iff such hypersurfaces are of global simultaneity,
i.e, if they intersect each integral line of $\xi^{\alpha}$ exactly
once.\end{proposition}

In Fig. \ref{fig:OrthogonalHypersurfaces}, the hypersurfaces orthogonal
to the Killing field $\partial_{t}$ in the Levi-Civita metric (\ref{eq:Levi-Civita})
and in the canonical form (\ref{eq:MetricKomar}) for Lewis-Weyl metric
are plotted, in a 3-D chart $\{t,r,\phi\}$ that omits the $z$ coordinate
{[}and the bar in $\bar{\phi}$ in Eq. (\ref{eq:MetricKomar}){]}.
In the former these are the planes $t=const.$, which are hypersurfaces
of global simultaneity, along which all clocks can be synchronized.
For the rotating Lewis-Weyl metric such hypersurfaces are helicoids,
described by $t-\mathcal{A}_{\bar{\phi}}\bar{\phi}=const.$, which
intersect each integral curve of $\partial_{t}$ \emph{infinitely
many times}, signaling that the spacetime is \emph{not} globally static.
Each $2\pi$ turn in the $\bar{\phi}$ coordinate does not lead back
to the same event $\mathscr{P}_{1}$, but to another ($\mathscr{P}_{2}$)
at a different coordinate time ($\Delta t=2\pi\mathcal{A}_{\bar{\phi}}$),
hence they are clearly \emph{not} global simultaneity hypersurfaces.
Consequently, a global clock synchronization between the hypersurface
orthogonal Killing observers is not possible in the Lewis-Weyl rotating
metric. In other words, observers at rest with respect to the distant
stars can globally synchronize their clocks in the Levi-Civita, but
not in the Lewis-Weyl rotating metric. This is another physical difference,
to be added to those discussed in Sec. \ref{sub:Physical-distinction}.

The global non-staticity of the Lewis-Weyl metric can also be seen
from the fact that the hypersurfaces $\psi=t-\mathcal{A}_{\bar{\phi}}\bar{\phi}=const.$
form a foliation whose space of leaves is the circle rather than the
real line; in other words, leaves given by $\psi=2n\pi\mathcal{A}_{\bar{\phi}}$
coincide for integer $n$, implying that $\psi$ is not single valued.
Indeed, $\psi$ is a function \emph{only locally}, for $\bar{\phi}\in[0,2\pi[$;
otherwise it takes multiple values for the same point: $\psi(t,r,\bar{\phi},z)\ne\psi(t,r,\bar{\phi}+2n\pi,z)$.

The locally static and globally stationary character of the Lewis-Weyl
metric is thus transparent in the canonical form (\ref{eq:MetricKomar})
{[}though not in the usual form (\ref{eq:LewisMetric})-(\ref{eq:LewisFunctions}){]},
and it is physically manifest in the setups in Figs. \ref{fig:CoilLewis}
(b)-(c) and \ref{fig:TwoCock} (b). The setups in Figs. \ref{fig:CoilLewis}-\ref{fig:TwoCock}
are also examples that Stachel's criteria for global staticity is
well posed and sound on physical grounds.

\subsubsection{Global staticity and holonomy\label{sub:Holonomy}}

A stationary spacetime is a principal bundle over the space manifold
$\Sigma$, since this manifold is simply the quotient of the spacetime
by the integral lines of the time-like Killing vector field $\xi^{\alpha}$,
that is, by the $\mathbb{R}$-action corresponding to the flow of
$\xi^{\alpha}$ \cite{Harris_1992,Anderson:1999xz}. A local trivialization
of this bundle is simply a choice of a time coordinate $t$ such that
$\partial_{t}^{\alpha}=\xi^{\alpha}$, and the structure group is
the additive group $(\mathbb{R},+)$. Choosing instead the parameterization
$s=e^{t}$ changes sums to products and allows us to see the stationary
spacetime as a principal bundle with the more familiar multiplicative
structure group $(\mathbb{R}^{+},\cdot)=GL^{+}(1,\mathbb{R})$. The
distribution of hyperplanes orthogonal to $\xi^{\alpha}$ defines
a connection on this bundle, whose parallel transport corresponds
to the synchronization of the clocks carried by the observers tangent
to $\xi^{\alpha}$, using the Einstein procedure \cite{LandauLifshitz,Bazanski1997}.
Indeed, the synchronization equation along some curve $x^{i}(\lambda)$,
which amounts to the condition that the curve be orthogonal (at every
point) to $\xi^{\alpha}$, reads 
\begin{equation}
\frac{dt}{d\lambda}-\mathcal{A}_{i}\frac{dx^{i}}{d\lambda}=0\quad\Leftrightarrow\quad\frac{ds}{d\lambda}-\mathcal{A}_{i}\frac{dx^{i}}{d\lambda}s=0\ ,\label{transport}
\end{equation}
and so the connection 1-form is $\bm{\mathcal{A}}$. The curvature
2-form is therefore $\bm{\mathcal{F}}=\mathbf{d}\bm{\mathcal{A}}$,
and so (cf. Sec. \ref{sub:Local-vs-global}) the condition for $\xi^{\alpha}$
to be hypersurface orthogonal is that this connection be flat.

To compute the holonomy of this connection along a closed curve $C$
on $\Sigma$ we integrate Eq. (\ref{transport}) along the curve:
\begin{equation}
\frac{1}{s}\frac{ds}{d\lambda}=\mathcal{A}_{i}\frac{dx^{i}}{d\lambda}\quad\Leftrightarrow\quad\ln\left(\frac{s_{{\rm final}}}{s_{{\rm initial}}}\right)=\oint_{C}\mathcal{A}_{i}dx^{i}\ .
\end{equation}
Therefore the initial and final values of $s$ under parallel transport
along $C$ are related by 
\begin{equation}
s_{{\rm final}}={\rm Hol}(C)\,s_{{\rm initial}}\ ,
\end{equation}
where the holonomy of the connection along $C$, ${\rm Hol}(C)$,
is the group element 
\begin{equation}
{\rm Hol}(C)=e^{\oint_{C}\mathcal{A}_{i}dx^{i}}\in\mathbb{R}^{+}.
\end{equation}
If the connection is flat then the holonomy depends only on the homotopy
class of $C$, that is, it is invariant under continuous deformations
of $C$. Moreover, the holonomy is trivial, that is, ${\rm Hol}(C)=1$
for all closed curves $C$, if and only if $\oint_{C}\bm{\mathcal{A}}=0$
for all closed curves $C$, i.e., if and only if $\bm{\mathcal{A}}$
is exact. It follows from Sec. \ref{sub:Local-vs-global} that the
local staticity of a spacetime is equivalent to the existence of a
time-like Killing vector field $\xi^{\alpha}$ whose synchronization
connection is flat (i.e., a hypersurface orthogonal $\xi^{\alpha}$),
and global staticity to it having moreover a trivial holonomy. Hence,
another way of phrasing the distinction between the Levi-Civita (\ref{eq:Levi-Civita})
and the rotating Weyl class metrics (\ref{eq:MetricKomar}) is that
in the former, but not in the latter, the hypersurface orthogonal
Killing observers have a synchronization connection with trivial holonomy.

\subsubsection{Geometrical distinction\label{sub:Mathematical-distinction}}

It is well known (e.g. \cite{SantosGRG1995}) that the transformation
\begin{equation}
t'=(t+b\phi)\ ;\qquad\phi'=\frac{n-bc}{n}\left[\phi-\Omega t\right]\ ;\qquad\Omega=\frac{c}{n-bc}\label{eq:StaticTransform}
\end{equation}
puts the Weyl class Lewis metric (\ref{eq:LewisMetric})-(\ref{eq:LewisFunctions})
into a form similar to the Levi-Civita line element (\ref{eq:Levi-Civita}),
with $\{t',\phi'\}$ in the place of $\{t,\phi\}$. Hence, \emph{locally},
they are isometric (i.e., locally indistinguishable). On the other
hand, it is also known that this transformation is not globally satisfactory
\cite{Bonnor1980,Tipler:1974gt}, and that the two solutions globally
differ, which is sometimes (inaccurately) assigned to topological
differences. Their distinction, from a mathematical point of view,
is indeed a subtle and not so well understood issue in the literature.
It is however a realization of the mathematical relationship between
globally, and locally but non-globally static spacetimes established
by Stachel \cite{Stachel:1981fg}, as we shall now show.

We start by observing that the topology of the underlying manifolds
is in fact the same: $\mathbb{R}^{1}\times\mathbb{R}^{3}\backslash\{r=0\}$.
Therefore, it must be at the level of the metric that the differences
arise. Let us then dissect the nature of transformation (\ref{eq:StaticTransform}).
In what pertains to the angular coordinate, it consists of a rotation
$\bar{\phi}=\phi-\Omega t$ with the angular velocity $\Omega$ that
leads to the star-fixed coordinates $\{x^{\bar{\alpha}}\}$ of Eqs.
(\ref{eq:Canonical})-(\ref{eq:CanonicalFunctions}), composed with
the ``re-scaling'' $\phi'=\bar{\phi}(n-bc)/n$, which accounts for
the different angular deficits of the spatial metrics $h_{ij}$ {[}Eq.
(\ref{eq:AxistatMetric}){]} that occur \emph{when} \emph{one identifies}
the parameter $a$ in Eq. (\ref{eq:Levi-Civita}) with that in (\ref{eq:Canonical})-(\ref{eq:CanonicalFunctions}).
The latter step is actually not necessary {[}one can instead identify
$a$ in (\ref{eq:Levi-Civita}) with $\alpha^{-1}${]}, which is clear
from the canonical form (\ref{eq:MetricKomar}) of the metric. The
transformation can actually be much simplified starting from the latter,
which is immediately diagonalized (since $\mathcal{A}_{\bar{\phi}}$
is constant) through the transformation 
\begin{equation}
t'=t-\mathcal{A}_{\bar{\phi}}\bar{\phi}\equiv t-\frac{j}{\lambda_{{\rm m}}-1/4}\bar{\phi}\ ;\qquad\phi'=\bar{\phi}\ ,\label{eq:StaticTransformCanonical}
\end{equation}
leading to 
\begin{equation}
ds^{2}=-\frac{r^{4\lambda_{{\rm m}}}}{\alpha}dt'^{2}+r^{4\lambda_{{\rm m}}(2\lambda_{{\rm m}}-1)}(dr^{2}+dz^{2})+\alpha r^{2(1-2\lambda_{{\rm m}})}d\phi'^{2}\ ,\label{eq:LCTransmute}
\end{equation}
which is \emph{locally} the Levi-Civita line element. One may check
{[}substituting, in (\ref{eq:StaticTransformCanonical}), $\bar{\phi}=\phi-\Omega t${]}
that it diagonalizes the original form (\ref{eq:LewisMetric})-(\ref{eq:LewisFunctions})
of the metric as well, yielding (\ref{eq:LCTransmute}). Transformation
(\ref{eq:StaticTransformCanonical}) amounts to redefining the time
coordinate so that it is constant along the hypersurfaces orthogonal
to the Killing vector field $\partial_{t}$, plotted in Fig. \ref{fig:OrthogonalHypersurfaces}
(b). That is, $t'$ is the function $\psi$ as defined in Sec. \ref{sub:Local-vs-global}
above. Since, in the original coordinates in (\ref{eq:MetricKomar}),
$\bar{\phi}$ is a periodic coordinate, with the identification $(t,\bar{\phi})=(t,\bar{\phi}+2\pi)$,
transformation (\ref{eq:StaticTransformCanonical}) leads to a coordinate
system where the events $(t',\phi')$ and $(t'-2\pi\mathcal{A}_{\bar{\phi}},\,\phi'+2\pi)$
are identified, and neither $\phi'$ or $t'$ are periodic\footnote{Sometimes \cite{Tipler:1974gt,Bonnor1980,Stachel:1981fg} it is asserted
that $t'$ is periodic; in rigor this is not correct (for the coordinate
lines of $t'$ are not closed), it is the identification above for
the \emph{pair} $(t',\phi')$ that is generated by transformation
(\ref{eq:StaticTransform}).}. In the Levi-Civita static metric, however, the periodic quantity
is the angular coordinate {[}$\phi$, in the notation in (\ref{eq:Levi-Civita}){]},
which is a requirement of the matching to the interior solution \cite{Bonnor1979_linemass}.
Therefore, to effectively convert the metric (\ref{eq:MetricKomar})
into the Levi-Civita metric, one must, in addition to the coordinate
transformation (\ref{eq:StaticTransformCanonical}), discard the original
identifications and force instead, in (\ref{eq:LCTransmute}), $\phi'$
to be periodic, through the identification $(t',\phi')=(t',\,\phi'+2\pi)$.
Such prescription, however, is not a \emph{global} diffeomorphism.
Namely, the map is neither injective nor single-valued: for instance,
events $\mathscr{P}_{1}$: $(t,\bar{\phi})=(0,\bar{\phi}_{1})$ and
$\mathscr{P}_{2}$: $(t,\bar{\phi})=(2\pi\mathcal{A}_{\bar{\phi}},\,\bar{\phi}_{1}+2\pi)$
in Fig. \ref{fig:OrthogonalHypersurfaces}, which are \emph{distinct}
in the original manifold, would be mapped into the \emph{same} event
$(t',\phi')=(-\mathcal{A}_{\bar{\phi}}\bar{\phi}_{1},\,\bar{\phi}_{1})=(-\mathcal{A}_{\bar{\phi}}\bar{\phi}_{1},\,\bar{\phi}_{1}+2\pi)$
in the static solution; conversely, the ordered pairs $\mathscr{P}{}_{3}$:
$(t,\bar{\phi})=(0,0)$ and $\mathscr{P}{}_{4}$: $(t,\bar{\phi})=(0,2\pi)$,
which yield the \emph{same} event in the original manifold, would
be mapped into the two \emph{distinct} events $\mathscr{P}'_{3}$:
$(t',\phi')=(0,0)$ and $\mathscr{P}'_{4}$: $(t',\phi')=(-2\pi\mathcal{A}_{\bar{\phi}},\,2\pi)$
in the static solution. Only \emph{locally} is the map bijective.
Since only through such a map is it possible to obtain one from the
other, that means that no global identification between the two metrics
exists, thus they are \emph{not globally isometric}.

It is worth noting that, in spite of the fact that the underlying
manifolds are topologically indistinguishable, topology still plays
an important role in the relationship between the exterior field of
static and rotating cylinders of the Weyl class, in that, as explained
in Sec. \ref{sub:Closed-forms,-exact}, it is the cylindrical ``hole''
along the axis $r=0$ that allows the existence of closed but non
exact forms, i.e., curl-free forms $\bm{\mathcal{\sigma}}$ with non-vanishing
circulation $\oint_{C}\bm{\mathcal{\sigma}}$ along closed loops $C$.
Now, when a local but non-global diffeomorphism, such as the prescription
above, exists between two manifolds, a closed but non-exact 1-form
in one manifold can be mapped into an exact one in the other manifold
\cite{Stachel:1981fg}. On the other hand, as discussed in Sec. \ref{sub:Local-vs-global},
global staticity consists of the exact character of the 1-form $\bm{\chi}$,
inverse to the hypersurface orthogonal time-like Killing vector field
($\partial_{t}$, in this case). Consequently, globally static and
locally but non-globally static metrics, connected by local diffeomorphisms,
can coexist on such underlying topology. This is precisely the situation
between the rotating and static Lewis metrics of the Weyl class: the
1-form inverse to the Killing vector field $\partial_{t}$ on the
metric (\ref{eq:MetricKomar}), $\bm{\chi}=\mathbf{d}t-\bm{\mathcal{A}}$,
which is \emph{not exact} (manifesting the global non-staticity of
$\partial_{t}$), is mapped, via (\ref{eq:StaticTransformCanonical}),
into the \emph{exact} 1-form $\mathbf{d}t'$, inverse of the globally
static Killing vector field $\partial_{t'}$, on the target manifold
{[}the Levi-Civita spacetime, described by (\ref{eq:LCTransmute})
under the identification $(t',\phi')=(t',\,\phi'+2\pi)$, with $t'$
assumed a single valued function{]}.

\subsection{Matching to the van Stockum cylinder\label{sub:Matching Stockum}}

It was shown by van Stockum \cite{Stockum1938} that the Lewis metric
has a smooth matching with the interior solution corresponding to
an infinite, rigidly rotating cylinder of dust. In order to address
the matching problem, we first establish the connection between the
Lewis metric and van Stockum's form for the exterior solution. The
latter can be written as \cite{BonnorCQG1995} 
\begin{equation}
ds_{*}^{2}=-Fdt_{*}^{2}+2Mdt_{*}d\phi+\mathcal{H}(dr_{*}^{2}+dz_{*}^{2})+Ld\phi^{2}\ ,\label{eq:Stockum}
\end{equation}
with
\begin{eqnarray}
F & = & \frac{(2N-1)(r_{*}/R)^{2N+1}+(2N+1)(r_{*}/R)^{1-2N}}{4N}\ ;\label{eq:FStockum}\\
M & = & wR^{2}\frac{(2N+1)(r_{*}/R)^{2N+1}+(2N-1)(r_{*}/R)^{1-2N}}{4N}\ ;\label{eq:MStockum}\\
L & = & R^{2}\frac{(2N+1)^{3}(r_{*}/R)^{2N+1}+(2N-1)^{3}(r_{*}/R)^{1-2N}}{16N}\ ;\label{eq:LStockum}\\
\mathcal{H} & = & e^{-w^{2}R^{2}}(r_{*}/R)^{-2w^{2}R^{2}};\qquad\ N=\sqrt{1/4-w^{2}R^{2}}\ .\label{eq:HStockum}
\end{eqnarray}
There are thus only two independent, positive parameters $w$ and
$R$, the latter being the cylinder's radius. The line element $ds_{*}$
in Eqs. (\ref{eq:Stockum})-(\ref{eq:HStockum}), as well as the coordinates
$t_{*}$, $r_{*}$, $z_{*}$, have the (usual) dimensions of length;
this contrasts with the usual Lewis line element in (\ref{eq:LewisMetric})-(\ref{eq:LewisFunctions}),
where $ds$ is dimensionless, and written in terms of \emph{dimensionless
coordinates} $t$, $r$ and $z$. Hence, in order to compare the two,
we must first write, for the Lewis metric, a line element in the form
$ds_{*}^{2}=\mathcal{R}^{2}ds^{2}$, where $\mathcal{R}$ is a constant
with dimensions of length. Through the parameter redefinition $a=a_{*}\mathcal{R}^{1-n}$,
$b=b_{*}/\mathcal{R}$, $c=\mathcal{R}c_{*}$, this line element becomes
\begin{equation}
ds_{*}^{2}=-f(r_{*})dt_{*}^{2}+2k(r_{*})dt_{*}d\phi+\left[\frac{r_{*}}{\mathcal{R}}\right]^{(n^{2}-1)/2}(dr_{*}^{2}+dz_{*}^{2})+l(r_{*})d\phi^{2}\ ,\label{eq:LewisLenght}
\end{equation}
where $(t_{*},r_{*},z_{*})\equiv(\mathcal{R}t,\mathcal{R}r,\mathcal{R}z)$
are coordinates with dimensions of length, $f(r_{*})\equiv f(r_{*},a_{*},c_{*},n)$,
$k(r_{*})\equiv k(r_{*},a_{*},b_{*},c_{*},n)$, and $l(r_{*})\equiv l(r_{*},a_{*},b_{*},c_{*},n)$.
By comparing the expressions for $g_{r_{*}r_{*}}$, and matching terms
with the same powers in $r_{*}$ in the remainder of the metric components,
we find that the metric (\ref{eq:Stockum})-(\ref{eq:HStockum}) follows
from (\ref{eq:LewisLenght}) and (\ref{eq:LewisFunctions}) through
the substitutions\footnote{There have been previous approaches \cite{SantosCQG1995,GriffithsPodolsky2009}
at establishing this connection. The expressions for $b_{*}$, $c_{*}$
and $n$ agree with those in Eqs. (5.17)-(5.20) of \cite{SantosCQG1995},
but $a_{*}$ differs. This is because Eqs. (5.1)-(5.4) therein actually
do not correspond to van Stockum's exterior solution in the usual
coordinates (Eqs. (10.11)-(10.15) of \cite{Stockum1938}), which stems
from the omission, in Eqs. (5.1)-(5.4) of \cite{SantosCQG1995}, of
the \emph{dependent} parameter $r_{0}\equiv r_{0}(w,R)$ showing up
in Eqs. (9.7) and (10.1) of \cite{Stockum1938}. The resulting metric
is consequently one in a special system of units where $r_{0}=1$,
and $w$ and $R$ are \emph{not independent}, being related by Eqs.
(10.3) and (10.9) of \cite{Stockum1938} --- an implicit relation
which can only be solved numerically. On the other hand, $a_{*}$
and $n$ match the result in \cite{GriffithsPodolsky2009} p. 244,
but $b_{*}$ and $c_{*}$ have opposite signs, due to $g_{0\phi}$
therein having opposite sign to van Stockum's in Eqs. (\ref{eq:Stockum}),
(\ref{eq:MStockum}). }
\begin{eqnarray}
\mathcal{R} & = & R/\sqrt{e}\ ;\qquad n\ =\ 2N\ ;\qquad a_{*}\ =\ \frac{2N+1}{4N}R^{2N-1}\ ;\label{eq:LewistoStockum}\\
b_{*} & = & \frac{1-2N}{1+2N}wR^{2}\ ;\qquad c_{*}=-\frac{\sqrt{1-4N^{2}}}{2R}=-w\ .\label{eq:LewistoStockum2}
\end{eqnarray}
Notice that parameters $n$, $a_{*}$, $b_{*}$, $c_{*}$ are real
\emph{iff} $wR<1/2$; hence the van Stockum cylinder belongs to the
Weyl class for $wR<1/2$, and to the Lewis class for $wR>1/2$. The
metric can be put in the form (\ref{eq:AxistatMetric}), with
\begin{equation}
e^{2\Phi}=F\ ;\qquad\mathcal{A}_{\phi}=\frac{M}{F}\ ;\qquad h_{rr}=h_{zz}=\mathcal{H};\qquad h_{\phi\phi}=r_{*}^{2}e^{-2\Phi}\ .\label{eq:StockumExtQM}
\end{equation}
The corresponding gravitoelectric and gravitomagnetic fields are
\begin{equation}
\vec{G}=\frac{2w^{2}Re^{w^{2}R^{2}}\left[(r_{*}/R)^{4N}-1\right](r_{*}/R)^{2w^{2}R^{2}-1}}{2N+1+(r/R)^{4N}(2N-1)};\qquad\ \vec{H}=\frac{8wNe^{w^{2}R^{2}}(r_{*}/R)^{2N-1+2w^{2}R^{2}}}{2N+1+(r_{*}/R)^{4N}(2N-1)}.\label{eq:GEMStockumExt}
\end{equation}
Observe that $\vec{G}=0$ for $r_{*}=R$; by virtue of (\ref{eq:3DAccel}),
this means that a test particle dropped from rest therein remains
at rest (i.e., particles at rest are geodesic). Again, this hints
at the fact that the metric is written in a rotating coordinate system,
the centrifugal inertial force exactly canceling out the gravitational
attraction. Observe moreover that $g_{00}$ becomes positive (i.e.,
the Killing vector $\partial_{t_{*}}$ ceases to be time-like) for
$r_{*}^{4N}>R^{4N}(2N+1)/(1-2N)$, which, as discussed in Sec. \ref{sub:The-canonical-form}
(see also Sec. \ref{sub:EMRotating-frame}), is typical of a rotating
frame.

\subsubsection{Interior solution\label{sub:Interior-solution}}

The interior solution is given by Eq. (\ref{eq:Stockum}), with \cite{Stockum1938,BonnorCQG1995}
\begin{equation}
F=1;\qquad M=wr_{*}^{2};\qquad L=r_{*}^{2}-w^{2}r_{*}^{4};\qquad\mathcal{H}=e^{-w^{2}r_{*}^{2}}\ ,\label{eq:StockumInt}
\end{equation}
depending on the single parameter\footnote{The constant $w$ yields the cylinder's angular velocity with respect
to a rigid spatial frame which, \emph{at the cylinder's axis} $r_{*}=0$,
undergoes Fermi-Walker transport \cite{Stockum1938} (i.e., a rigid
frame such that $\vec{H}=0$ at the axis).} $w$. It can be put in the form (\ref{eq:AxistatMetric}), with 
\begin{equation}
\Phi=0;\qquad\mathcal{A}_{\phi}=wr_{*}^{2};\qquad h_{r_{*}r_{*}}=h_{z_{*}z_{*}}=e^{-w^{2}r_{*}^{2}};\qquad h_{\phi\phi}=r_{*}^{2}\ .\label{eq:QMStockumInt}
\end{equation}
The corresponding gravitoelectric and gravitomagnetic fields are 
\begin{equation}
\vec{G}=0\ ;\qquad\ \vec{H}=2we^{w^{2}r_{*}^{2}}\partial_{z_{*}}\ ,\label{eq:GEMStockumInt}
\end{equation}
and the gravitomagnetic tidal tensor as measured by the rest observers
has the only non-vanishing components $\mathbb{H}_{rz}=\mathbb{H}_{zr}=-w^{3}r_{*}$.
Thus $\mathbb{H}_{\alpha\beta}$ is \emph{symmetric}; since $\mathbb{H}_{[\alpha\beta]}=-4\pi\epsilon_{\alpha\beta\mu\nu}J^{\mu}u^{\nu}$
\cite{CostaHerdeiro}, where $J^{\mu}\equiv-T^{\mu\sigma}u_{\sigma}$
is the mass-energy current as measured by the rest observers of 4-velocity
$u^{\alpha}$, this means that no spatial mass currents {[}$h_{\ \nu}^{\mu}J^{\nu}$,
see Eq. (\ref{eq:SpaceProjector}){]} are measured by $u^{\alpha}$,
i.e., the metric is written in a coordinate system co-rotating with
the dust, cf. \cite{Stockum1938}. Observe moreover that $\vec{G}=0$
everywhere inside the cylinder; this is just the condition that the
circular motion of the dust particles is solely driven by gravity
(i.e., geodesic), so in the dust rest frame a centrifugal inertial
force arises that exactly balances the gravitational attraction.

Let $\sigma^{(3)}$ be a stationary 3-D hypersurface which is the
common boundary of two stationary spacetimes, and $\sigma$ the projected
2-D surface on the corresponding space manifolds $\Sigma$, as defined
in Sec. \ref{sec:Preliminaries}. Let $\vec{n}$ be the unit vector
normal to $\sigma$. The matching of the two solutions along $\sigma^{(3)}$
amounts to matching the induced metric on $\sigma^{(3)}$, $g_{\alpha\beta}|_{\sigma^{(3)}}$,
plus the extrinsic curvature of $\sigma^{(3)}$. In the GEM formalism,
and when $\sigma$ is connected, this is guaranteed (see \cite{MenaNatario2008}
and footnote 3 therein) by the continuity across $\sigma$ of the
GEM fields $\vec{G}$ and $\vec{H}$, gravitomagnetic potential 1-form\footnote{When $\sigma$ is \emph{simply} connected (which is \emph{not} the
case herein), the continuity of the restriction of $\bm{\mathcal{A}}$
to $\sigma$ (up to $\mathbf{d}f$) is equivalent to the continuity
of the normal component of $\vec{H}$, hence the matching conditions
reduce to the continuity of $\vec{G}$, $\vec{H}$, $h_{ij}$ and
$K_{ij}$ \cite{MenaNatario2008}.} $\bm{\mathcal{A}}$ (up to an \emph{exact} form $\mathbf{d}f$, for
some function $f$ on $\sigma$, corresponding to the freedom associated
to the choice of $t$), spatial metric $h_{ij}$, and extrinsic curvature
$K_{ij}\equiv\mathcal{L}_{n}h_{ij}$ of the spatial 2-surface $\sigma$:
\begin{eqnarray*}
 &  & \vec{G}_{{\rm int}}=\vec{G}_{{\rm ext}}\ ;\qquad\vec{H}_{{\rm int}}=\vec{H}_{{\rm ext}}\ ;\qquad\bm{\mathcal{A}}_{{\rm int}}=\bm{\mathcal{A}}_{{\rm ext}}+\mathbf{d}f\ ;\\
 &  & (h_{{\rm int}})_{ij}=(h_{{\rm ext}})_{ij}\ ;\qquad(K_{ij})_{{\rm int}}=(K_{ij})_{{\rm ext}}\ .
\end{eqnarray*}
It follows from Eqs. (\ref{eq:StockumExtQM})-(\ref{eq:GEMStockumExt})
and (\ref{eq:QMStockumInt})-(\ref{eq:GEMStockumInt}) that these
conditions (with $\bm{\mathcal{A}}_{{\rm int}}=\bm{\mathcal{A}}_{{\rm ext}}\Rightarrow\mathbf{d}f=0$)
are satisfied across the cylinder's surface $r_{*}=R$ with unit normal
$\vec{n}=(h_{r_{*}r_{*}})^{-1/2}\vec{\partial}_{r_{*}}$, and so indeed
the interior solution (\ref{eq:StockumInt}) smoothly matches the
exterior (\ref{eq:FStockum})-(\ref{eq:HStockum}). The rotation of
coordinates that we noticed (Sec. \ref{sub:The-canonical-form}) in
the usual form of the Lewis-Weyl metric has thus a simple interpretation
here: the coordinate system in (\ref{eq:LewisMetric})-(\ref{eq:LewisFunctions})
{[}or equivalently, in (\ref{eq:Stockum})-(\ref{eq:HStockum}){]},
is \emph{rigidly co-rotating} with the interior cylinder.

\subsubsection{Matching in canonical form\label{sub:Canonical-form-Van-Stockum}}

We have seen in Sec. \ref{sub:The-canonical-form} that the star-fixed
(``canonical'') coordinates for the Lewis metric of the Weyl class
are obtained from the usual coordinates in (\ref{eq:LewisMetric})-(\ref{eq:LewisFunctions})
by the transformation (\ref{eq:TransformCanonical}), with $\Omega\equiv d\phi/dt$
one of the \emph{dimensionless} angular velocities in (\ref{eq:PrincipalObs})
(depending on the sign of $a$). Since here $a_{*}>0$, cf. Eq. (\ref{eq:LewistoStockum}),
the star-fixed coordinates for the Weyl class van Stockum exterior
metric analogously follows by applying to (\ref{eq:Stockum})-(\ref{eq:HStockum})
the transformation 
\begin{equation}
\bar{\phi}=\phi-\Omega_{*}t_{*}\ ;\qquad\Omega_{*}\equiv\frac{d\phi}{dt_{*}}=\frac{\Omega}{\mathcal{R}}=\frac{c_{*}}{n-b_{*}c_{*}}=-\frac{4w}{(1+2N)^{2}}\ ,\label{eq:OmegaCanStockum}
\end{equation}
where the angular velocity $\Omega_{*}$ now has the (usual) dimensions
of inverse length, and, in the last equality, we substituted Eqs.
(\ref{eq:LewistoStockum})-(\ref{eq:LewistoStockum2}). This yields
the line element 
\begin{equation}
ds^{2}=-\bar{F}dt_{*}^{2}+2\bar{M}dt_{*}d\bar{\phi}+\mathcal{H}(dr_{*}^{2}+dz_{*}^{2})+Ld\bar{\phi}^{2}\ ,\label{eq:CanonicalStockumExt}
\end{equation}
with $\mathcal{H}$ and $L$ given by Eqs. (\ref{eq:LStockum})-(\ref{eq:HStockum}),
and
\begin{equation}
\bar{F}=\frac{16N}{(1+2N)^{3}}\left[\frac{r_{*}}{R}\right]^{1-2N}\ ;\qquad\ \bar{M}=-\frac{R^{4}w^{3}}{2N}\bar{F}\ .\label{eq:FMCanonical}
\end{equation}
One can show, after some algebra, that (\ref{eq:OmegaCanStockum})
indeed corresponds to the transformation to the star-fixed coordinate
system obtained in \cite{Stockum1938} {[}Eqs. (4.7) and (10.16) therein{]},
and Eqs. (\ref{eq:CanonicalStockumExt})-(\ref{eq:FMCanonical}) to
the exterior metric as written in such coordinate system {[}Eqs. (10.17)
therein, apart from a typo in the expression for $F$, where an extra
$2wR$ factor is present{]}. Observe that $g_{00}=-\bar{F}$ is now
negative for all $r_{*}$, so that the Killing vector field $\partial_{t_{*}}$
is time-like everywhere, contrary to the situation in (\ref{eq:Stockum})-(\ref{eq:HStockum}).
The Komar mass and angular momentum per unit length for the metric
(\ref{eq:CanonicalStockumExt})-(\ref{eq:FMCanonical}) can be obtained
by applying the integrals (\ref{eq:KomarCylinder}) to any tube of
unity $z_{*}-$length enclosing the cylinder {[}or by substituting
(\ref{eq:LewistoStockum})-(\ref{eq:LewistoStockum2}) in (\ref{eq:MassCylinder})-(\ref{eq:AngularMomentumCylinder}),
recalling that $a=a_{*}\mathcal{R}^{1-n}$, $b=b_{*}/\mathcal{R}$,
$c=\mathcal{R}c_{*}$, and observing that $j_{*}=j\mathcal{R}${]};
they read, respectively, 
\begin{equation}
\lambda_{{\rm m}}=\frac{1-2N}{4}=\frac{1-\sqrt{1-4w^{2}R^{2}}}{4}\ ;\qquad j_{*}=\frac{R^{4}w^{3}}{4}\ .\label{eq:Lambdam}
\end{equation}
Notice that $j_{*}$ has the usual dimensions of length. The metric
(\ref{eq:CanonicalStockumExt})-(\ref{eq:FMCanonical}) can be written
in a canonical form akin to that in Sec. \ref{sub:The-metric-in-3-parameters}.
For that, we first observe that {[}similarly to the usual form of
the Lewis metric (\ref{eq:LewisMetric})-(\ref{eq:LewisFunctions}){]}
the line element $ds$ in (\ref{eq:MetricKomar}), as well as the
coordinates $t$, $r$, $z$ therein, are dimensionless; hence we
need to write, for the same metric, a line element $ds_{*}^{2}=\mathcal{R}^{2}ds^{2}$
with the dimensions of length:\footnote{One may thus argue that the most general (dimensional) canonical form
of the metric contains four parameters ($\alpha_{*}$, $\lambda_{{\rm m}}$,
$j_{*}$, and $\mathcal{R}$); and that, likewise, the usual form
(\ref{eq:LewisMetric})-(\ref{eq:LewisFunctions}) of the Lewis metric
actually implicitly contains five (not four) parameters \cite{GriffithsPodolsky2009}:
$a_{*}$, $b_{*}$, $c_{*}$, $n$, and $\mathcal{R}$, since a parameter
$\mathcal{R}$, defining a length-scale, must be introduced in order
to yield a line element (\ref{eq:LewisLenght}) with the usual dimensions
of length.}
\begin{equation}
ds_{*}^{2}=-\frac{r_{*}^{4\lambda_{{\rm m}}}}{\alpha_{*}}\left(dt_{*}-\frac{j_{*}}{\lambda_{{\rm m}}-1/4}d\bar{\phi}\right)^{2}+\left[\frac{r_{*}}{\mathcal{R}}\right]^{4\lambda_{{\rm m}}(2\lambda_{{\rm m}}-1)}(dr_{*}^{2}+dz_{*}^{2})+\alpha_{*}r_{*}^{2(1-2\lambda_{{\rm m}})}d\bar{\phi}^{2},\label{eq:CanonicalLenght}
\end{equation}
where $\alpha_{*}=\alpha\mathcal{R}^{4\lambda_{{\rm m}}}$, $\mathcal{R}$
is, again, a constant with dimensions of length, and $(t_{*},r_{*},z_{*})\equiv(\mathcal{R}t,\mathcal{R}r,\mathcal{R}z)$
are coordinates with dimensions of length. The canonical form of the
van Stockum exterior solution then follows from using, in (\ref{eq:CanonicalLenght}),
$\lambda_{{\rm m}}$ and $j_{*}$ as given by (\ref{eq:Lambdam}),
and
\begin{equation}
\alpha_{*}=\frac{R^{4\lambda_{{\rm m}}}(1-2\lambda_{{\rm m}})^{3}}{1-4\lambda_{{\rm m}}}\ ;\qquad\;\mathcal{R}=R/\sqrt{e}\ .\label{eq:alphaStockum}
\end{equation}
It naturally possesses all the ``canonical'' properties listed in
Sec \ref{sub:The-metric-in-3-parameters}. In this special case, however,
$\lambda_{{\rm m}}$, $j_{*}$, and $\alpha_{*}$ are not independent
parameters, as is clear from Eqs. (\ref{eq:Lambdam}) and (\ref{eq:alphaStockum});
the metric has only two independent parameters (which boil down to
$R$ and $w$), just like in the original coordinate system in (\ref{eq:Stockum})-(\ref{eq:HStockum}).
It is also useful to write the metric in the form (\ref{eq:AxistatMetric}),
with 
\begin{eqnarray}
e^{2\Phi} & = & \ \bar{F}\ =\alpha_{*}^{-1}R^{4\lambda_{{\rm m}}}(r_{*}/R)^{4\lambda_{{\rm m}}}\quad\Rightarrow\quad\Phi\ =\ 2\lambda_{{\rm m}}\ln(r_{*}/R)+const.\ ;\label{eq:StockumExtQMCan}\\
\mathcal{A}_{\bar{\phi}} & = & \frac{j_{*}}{\lambda_{{\rm m}}-1/4}=-\frac{R^{4}w^{3}}{2N}\ ;\label{eq:StockumExtQMCan2}\\
h_{r_{*}r_{*}} & = & h_{z_{*}z_{*}}=\left[\frac{r_{*}}{R}e^{1/2}\right]^{4\lambda_{{\rm m}}(2\lambda_{{\rm m}}-1)};\qquad h_{\bar{\phi}\bar{\phi}}=r_{*}^{2}e^{-2\Phi}.\label{eq:StockumExtQMCan3}
\end{eqnarray}
Since $\Omega_{*}$ in Eq. (\ref{eq:OmegaCanStockum}) is the angular
velocity of the star-fixed frame with respect to a frame co-rotating
with the interior cylinder, then the cylinder rotates with angular
velocity $-\Omega_{*}$ with respect to the star-fixed frame (cf.
\cite{Stockum1938}). Observe that $\Omega_{*}$ is \emph{negative};
this means that the cylinder is rotating in the \emph{positive} $\bar{\phi}$
direction. Observe moreover that $\mathcal{A}_{\bar{\phi}}$ is negative;
this implies, via Eqs. (\ref{eq:AngMomentumLab}) and (\ref{eq:OmegaZamo}),
that the star-fixed ``laboratory'' observers have negative angular
momentum, and that the zero angular momentum observers rotate in the
same sense of the cylinder (i.e., are ``dragged'' around by the
cylinder's rotation), as occurs e.g. in the Kerr spacetime, and in
agreement with an intuitive notion of frame-dragging. The GEM fields
read 
\begin{equation}
G_{i}=-\frac{2\lambda_{{\rm m}}}{r_{*}}\delta_{i}^{r}\ ;\qquad\ \vec{H}=0\ ,\label{eq:GEMStockumCan}
\end{equation}
the discussion of their physical effects in Sec. \ref{sub:GEM-fields-Canonical}
applying herein.

The interior solution written in star-fixed coordinates is likewise
obtained from (\ref{eq:Stockum}), (\ref{eq:StockumInt}) (the metric
written in coordinates comoving with the cylinder) by the transformation
(\ref{eq:OmegaCanStockum}), yielding a metric of the form (\ref{eq:CanonicalStockumExt}),
with $\mathcal{H}$ and $L$ still given by Eqs. (\ref{eq:StockumInt})
and
\begin{eqnarray}
\bar{F} & = & 1+4\frac{r_{*}^{4}}{R^{4}}\frac{\lambda_{{\rm m}}^{2}}{(1-2\lambda_{{\rm m}})^{2}}+2\frac{r_{*}^{2}}{R^{2}}\frac{\lambda_{{\rm m}}\left[2(1-2\lambda_{{\rm m}})^{2}-1\right]}{(1-2\lambda_{{\rm m}})^{3}}\ ;\label{eq:FCanonicalInt}\\
\bar{M} & = & wr_{*}^{2}\frac{r_{*}^{2}w^{2}-4(1-\lambda_{{\rm m}})\lambda_{{\rm m}}}{(1-2\lambda_{{\rm m}})^{2}}\ .\label{eq:MCanonicalInt}
\end{eqnarray}
Observe from Eqs. (\ref{eq:FCanonicalInt}) and (\ref{eq:Lambdam})
that $\bar{F}$ depends only on the (dimensionless) quantities $r_{*}/R$
and $\lambda_{{\rm m}}$. Since $0<r_{*}<R$ within the cylinder,
and $wR<1/2\Rightarrow0<\lambda_{{\rm m}}<1/4$ for the Weyl class,
it follows that $\bar{F}>0\Rightarrow g_{00}<0$ everywhere inside
the cylinder, and so the Killing vector field field $\partial_{t_{*}}$
is everywhere time-like therein. Moreover, it follows from the expressions
for $L$ and $\mathcal{H}$ in Eqs. (\ref{eq:StockumInt}) that the
coordinate basis vectors $\partial_{r_{*}}$, $\partial_{\bar{\phi}}$,
and $\partial_{z_{*}}$ are everywhere spacelike. This tells us that
the coordinate system fixed to the distant stars is well defined everywhere
within the cylinder. Writing the metric in the form (\ref{eq:AxistatMetric})
yields the GEM fields and spatial metric:
\begin{eqnarray*}
G_{i} & = & \frac{4\lambda_{{\rm m}}r_{*}\left[\lambda_{{\rm m}}-r_{*}^{2}w^{2}+\Delta(2\lambda_{{\rm m}}-1)w^{2}\right]}{r_{*}^{4}w^{2}-4\lambda_{{\rm m}}^{2}r_{*}^{2}+\Delta\left[r_{*}^{2}\left(2-4\lambda_{{\rm m}}\right)+\Delta(4\lambda_{{\rm m}}^{2}-4\lambda_{{\rm m}}+1)\right]w^{2}}\delta_{i}^{r_{*}}\ ;\\
\vec{H} & = & -\frac{2\Delta w^{3}e^{r_{*}^{2}w^{2}}(\Delta^{2}w^{2}+2r_{*}^{2}+2R^{2})}{(3r_{*}^{2}+R^{2})(1-4\lambda_{{\rm m}})+\Delta-2w^{2}(R^{4}-r_{*}^{4})-2\Delta^{3}w^{4}}\partial_{z_{*}}\ ;\\
h_{r_{*}r_{*}} & = & h_{z_{*}z_{*}}=e^{-w^{2}r_{*}^{2}}\ ;\qquad\ h_{\bar{\phi}\bar{\phi}}=r_{*}^{2}e^{-2\Phi}=\frac{r_{*}^{2}}{\bar{F}}\ ,
\end{eqnarray*}
where $\Delta\equiv R^{2}-r_{*}^{2}$. At the cylinder's surface $r_{*}=R$
($\Rightarrow\Delta=0$), and so we have 
\begin{eqnarray}
 &  & (G_{{\rm int}})_{i}=(G_{{\rm ext}})_{i}=-\frac{2\lambda_{{\rm m}}}{R}\delta_{i}^{r_{*}}\ ;\qquad\ \vec{H}_{{\rm int}}=\vec{H}_{{\rm ext}}=0\ ;\nonumber \\
 &  & (\mathcal{A}_{\phi})_{{\rm int}}=(\mathcal{A}_{\phi})_{{\rm ext}}=\frac{j_{*}}{\lambda_{{\rm m}}-1/4}\ ;\qquad\ (h_{{\rm int}})_{\bar{\phi}\bar{\phi}}=(h_{{\rm ext}})_{\bar{\phi}\bar{\phi}}=\alpha_{*}R^{2(1-2\lambda_{{\rm m}})}\ ;\label{eq:MatchingCanonical}\\
 &  & (h_{{\rm int}})_{z_{*}z_{*}}=(h_{{\rm int}})_{r_{*}r_{*}}=(h_{{\rm ext}})_{z_{*}z_{*}}=(h_{{\rm ext}})_{r_{*}r_{*}}=e^{2\lambda_{{\rm m}}(2\lambda_{{\rm m}}-1)}\ .\nonumber 
\end{eqnarray}
The extrinsic curvature ($K_{ij}\equiv\mathcal{L}_{n}h_{ij}$) of
that surface, with unit normal $\vec{n}=(h_{r_{*}r_{*}})^{-1/2}\vec{\partial}_{r_{*}}$,
has non-vanishing components
\begin{eqnarray}
(K_{{\rm int}})_{\bar{\phi}\bar{\phi}} & = & (K_{{\rm ext}})_{\bar{\phi}\bar{\phi}}=\frac{2R(1-2\lambda_{{\rm m}})^{4}}{1-4\lambda_{{\rm m}}}e^{\lambda_{{\rm m}}(1-2\lambda_{{\rm m}})}\ ;\label{eq:Kphiphi}\\
(K_{{\rm int}})_{z_{*}z_{*}} & = & (K_{{\rm ext}})_{z_{*}z_{*}}=-\frac{4\lambda_{{\rm m}}(1-2\lambda_{{\rm m}})}{R}e^{-\lambda_{{\rm m}}(1-2\lambda_{{\rm m}})}\ .\label{eq:Kzz}
\end{eqnarray}
Thus, indeed there is a smooth matching between the interior metric
in star-fixed coordinates and the exterior metric in canonical (star-fixed)
form. This is the expected result, for we knew that the matching is
possible in the more usual coordinates employed in Sec. \ref{sub:Interior-solution}.

The Komar mass per unit length can be computed from the interior solution
by using Eq. (\ref{eq:KomarRicci}), $Q_{\xi}(\mathcal{V})=-K/(8\pi)\int_{\mathcal{V}}R_{\ \beta}^{\alpha}\xi^{\beta}n_{\alpha}d\mathcal{V}$,
with $\mathcal{V}$ the cylinder of radius $r_{*}=R$ and unit $z_{*}-$length
on the hypersurface $\Sigma_{t_{0}}$ of constant time $t_{*}=t_{0}$,
$n_{\alpha}=-(1-w^{2}r_{*}^{2})^{-1/2}\nabla_{\alpha}t_{*}$ the unit
covector normal to $\Sigma_{t_{0}}$, $\xi^{\alpha}=\partial_{t_{*}}^{\alpha}$,
$d\mathcal{V}=\sqrt{g_{\Sigma}}dr_{*}d\bar{\phi}dz_{*}$, where $g_{\Sigma}=e^{-2w^{2}r_{*}^{2}}(r_{*}^{2}-w^{2}r_{*}^{4})$
is the determinant of the metric induced on $\Sigma_{t_{0}}$, and
(again) $K=-2$. It yields\footnote{We note that different values have been obtained in \cite{Bonnor1980}
by using Hansen-Winicour integrals (which are approximations to Komar
integrals \cite{HansenWinicour1975}), for different choices of the
time-like Killing vector field --- namely, the vector $\partial_{t_{*}}$
of the coordinate system in (\ref{eq:StockumInt}), co-rotating with
the cylinder, and another one tangent to the ZAMOS near the axis.
Such fields are not time-like at infinity, and so, as discussed in
Secs. \ref{sub:Komar-Integrals} and \ref{sub:Komar-Canonical}, the
corresponding integrals should not be interpreted as the cylinder's
mass per unit length. The different definitions match only for small
$w^{2}R^{2}$, yielding $\lambda_{{\rm m}}\approx w^{2}R^{2}/2$.}, as expected, the same result (\ref{eq:Lambdam}) obtained from the
exterior solution. The same is true for the angular momentum per unit
length $j_{*}$.

\subsection{The Lewis class\label{sub:The-Lewis-class}}

When $n$ is imaginary, the structure of the curvature invariants,
Eqs. (\ref{eq:QuadInvariants})-(\ref{eq:BInv}) and (\ref{eq:MInvariant}),
is the following: 
\[
\star\!\mathbf{R}\cdot\mathbf{R}=0;\qquad\mathbf{R}\cdot\mathbf{R}\ge0\ \ (<0)\ \ {\rm for}\ \ |n|\le\sqrt{3}\ \ (>\sqrt{3});\qquad\mathbb{M}<0\ \mbox{(real)}\ .
\]
These conditions mean that there are no observers, at any point, for
which $\mathbb{H}_{\alpha\beta}=0$ \cite{McIntosh_1994,Invariants,StephaniExact}.
This in turn implies, via Eq. (\ref{eq:HijGEM}), that $\vec{H}$
cannot vanish in any coordinate system where the metric is time-independent.
Therefore the metric possesses (locally and globally) intrinsic gravitomagnetic
tidal tensor $\mathbb{H}_{\alpha\beta}$ and globally intrinsic gravitomagnetic
field $\vec{H}$, in the classification scheme of \cite{Invariants}.
Since $\vec{H}$ is proportional to the vorticity of the observer
congruence {[}$\vec{H}=2\vec{\omega}$, cf. Eq. (\ref{eq:GEM Fields Cov}){]},
this amounts to saying that hypersurface orthogonal time-like Killing
vector fields do not exist. Hence, contrary to the Weyl class case,
the metric is not locally static (as is well known, e.g. \cite{SantosCQG1995}).
Thus these are fundamentally very different gravitational fields.

The fact that $\vec{H}\ne0$ in any coordinate system where the metric
is time-independent implies, e.g., that radial geodesics are not possible,
and gyroscopes (with $\vec{S}\nparallel\vec{H}$) will always be seen
to precess therein, cf. Eq. (\ref{eq:SpinPrec}). The fact that $\mathbb{H}_{\alpha\beta}\ne0$
for all observers means that spinning bodies in this spacetime are
always acted by a force (\ref{eq:SpinCurvature}).

\section{Conclusion}

In this work we investigated the exterior gravitational fields produced
by infinite cylinders, described by the Lewis metrics, focusing on
a class of them --- the Weyl class --- whose metrics are known to
be locally static, and to encompass the field of both static (the
Levi-Civita solution) and rotating cylinders. We aimed at establishing
the distinction between the two cases, both in terms of the physical
effects and of the geometrical properties where the rotation imprints
itself. We started by observing that gravitomagnetism has three levels
(corresponding to three different orders of differentiation of $\vec{\mathcal{A}}$),
described by the three mathematical objects: the gravitomagnetic vector
potential $\vec{\mathcal{A}}$, the gravitomagnetic field $\vec{H}$,
and the gravitomagnetic tidal tensor $\mathbb{H}_{\alpha\beta}$.
Then we unveiled a hitherto unnoticed feature of the Weyl class metric:
that by a simple coordinate rotation it can be put into an especially
simple form, where (by contrast with the usual form in the literature)
the Killing vector field $\partial_{t}$ is time-like everywhere,
and the associated coordinate system is fixed to the distant stars.
In such a reference frame both $\vec{H}$ and $\mathbb{H}_{\alpha\beta}$
vanish everywhere, the vector $\vec{\mathcal{A}}$ being the only
surviving gravitomagnetic object, which, in the case of a rotating
cylinder, cannot be made to vanish by any \emph{global} coordinate
transformation. This perfectly mirrors the electromagnetic analogue
(Sec. \ref{sec:The-electromagnetic-analogue:}): in the exterior of
an infinitely long rotating charged cylinder both the magnetic field
$\vec{B}=\nabla\times\vec{A}$ and the magnetic tidal tensor $B_{\alpha\beta}$
vanish, just like for a static cylinder; only the magnetic vector
potential $\vec{A}$ is non-vanishing. (Reinforcing the analogy, the
gravitoelectric potential $\Phi$ in these coordinates also remarkably
matches its electromagnetic counterpart, if we identify charge with
mass.) The resulting metric, moreover, depends only on three parameters:
the Komar mass and angular momentum per unit length, plus the angle
deficit. We argue this to be the \emph{canonical form} of the Lewis
metrics of the Weyl class. It makes explicit, for the Weyl class,
and in terms of parameters with a clear physical meaning, the earlier
finding in \cite{MacCallumSantos1998} that there are only three independent
parameters in the Lewis metric. It also makes explicit that the exterior
metric of a rotating cylinder formally differs from that of a static
one only by the presence of a non-vanishing, but irrotational $\vec{\mathcal{A}}$
(i.e., of a closed 1-form $\bm{\mathcal{A}}$). By contrast with classical
electrodynamics, where a vector potential with vanishing curl $\nabla\times\vec{A}=\vec{B}=0$
is pure gauge, but similarly to quantum electrodynamics, where it
manifests itself in the Aharonov-Bohm effect (Sec. \ref{sub:Aharonov-Bohm-effect}),
the gravitomagnetic vector potential $\vec{\mathcal{A}}$ \emph{does
manifest itself physically}, in effects involving loops around the
central cylinder, namely in the Sagnac effect, clock synchronization,
and the gravitomagnetic clock effect. The Sagnac effect, in particular,
is seen to be described exactly by an equation formally analogous
to the Aharonov-Bohm effect in the exterior of an infinitely long
rotating charged cylinder (or of a long solenoid). This substantiates,
with a concrete result, earlier suggestions in the literature: the
suggestion in \cite{SantosGRG1995} that the Lewis metrics possess
some topological analogue of the Aharonov-Bohm effect (by showing
what it is); and the claim in \cite{AshtekarMagnon,Stachel:1981fg,RizziRuggieroAharonovI,RizziRuggieroAharonovII,RuggieroAharonov2005,Barros_Bezerra_Romero2003}
that the Sagnac effect can be seen as a gravitational analogue of
the Aharonov-Bohm effect (by revealing a one to one correspondence
using the gravitational setup that is physically analogous to the
Aharonov-Bohm electromagnetic setting \cite{AharonovBohm}).

The physical effects mentioned above are global, in that they arise
only on paths $C$ enclosing the central cylinder. The gravitomagnetic
clock effect is naturally so, as it is defined for circular orbits.
The Sagnac effect and synchronization gap, both given by the circulation
of the gravitomagnetic potential 1-form, $\oint_{C}\bm{\mathcal{A}}$,
vanish (in the canonical, star-fixed frame) along any loop not enclosing
the cylinder, and have the same value along any loop enclosing it,
regardless of its shape. Global effects are seen to actually be the
only physical differences between the metrics, since all local and
quasi-local dynamical fields (i.e., tidal and inertial fields, respectively)
are shown to be the same as for the static cylinder.

The difference between metrics of rotating and static Weyl class cylinders
turns out to be an archetype of the distinction between globally static,
and locally static but globally stationary spacetimes. We reformulated
the Stachel-Bonnor notions of local and global staticity into equivalent,
more enlightening forms in this context, by showing that: (i) local
staticity amounts to existence of a coordinate system (\ref{eq:StatMetric})
where the gravitomagnetic potential 1-form $\bm{\mathcal{A}}$ is
closed, and global staticity to it being moreover exact; (ii) equivalently,
while local staticity amounts to the existence of a hypersurface orthogonal
Killing time-like vector field, global staticity amounts to such hypersurface
being moreover a global simultaneity hypersurface. This distinction
can moreover be formulated in terms of a connection that describes
the clock synchronization for observers tangent to $\xi^{\alpha}$,
local staticity amounting to such connection being flat, and global
staticity to its holonomy being trivial. We also dissected the nature
of the well known transformation that takes the Weyl class metric
into the static Levi-Civita one, showing it not to be a global diffeomorphism
(thus not a globally valid coordinate transformation), and the two
metrics to be locally, but not globally isometric, in spite of the
underlying manifolds sharing same topology.

The distinction above, both on physical and geometrical grounds, was
made transparent by writing the Weyl class metrics in their ``canonical''
form, based on star-fixed coordinates, which therefore play a key
role in this work. In the ``real world'' such reference frame is
physically set up by pointing telescopes at the distant stars, and
used in various experiments (including the detection of gravitomagnetic
effects, such as gyroscope and orbital precessions \cite{CiufoliniWheeler,CiufoliniNature2007}).
It should be noted, however, that the underlying physical distinction
between the two fields is not an artifact, nor does it rely on the
use of any particular frame. In fact, in Sec. \ref{sub:Physical-distinction}
we propose (thought) physical apparatuses --- namely a coil of optical
loops, and the observer independent gravitomagnetic clock effect ---
that are frame independent.

\ack{}{We thank the anonymous Referees for useful suggestions.
This work was partially supported by FCT/Portugal through projects
UID/MAT/04459/2019 and UIDB/MAT/04459/2020.}

\section*{References}

 \bibliographystyle{utphys}
\bibliography{Ref_Lewis}

\end{document}